%% file: Bmodel.tex
\documentclass[12pt]{article}

\input{Preamble}

\newcommand*{\xrlab}{A}
{
  \makeatletter
  \newcommand{\hbs@tocstring}{}
  \newcommand{\hbs@bmstring}{}  
  \externaldocument[\xrlab-]{Amodel}  
}

\begin{document}

\begin{titlepage}
  \vspace*{-2cm}
  \VersionInformation
  \hfill
  \parbox[c]{5cm}{
    \begin{flushright}
      UPR 1178-T
      \\
      DISTA-2007
      \\
      TUW-07-08
    \end{flushright}
  }
  \vspace*{\stretch{1}}
  \begin{center}
     \Huge 
     Worldsheet Instantons and Torsion Curves\\
     Part B: Mirror Symmetry
  \end{center}
  \vspace*{\stretch{2}}
  \begin{center}
    \begin{minipage}{\textwidth}
      \begin{center}
        \large         
        Volker Braun${}^1$,
        Maximilian Kreuzer${}^2$,
        \\
        Burt A. Ovrut${}^1$, and
        Emanuel Scheidegger${}^3$
      \end{center}
    \end{minipage}
  \end{center}
  \vspace*{1mm}
  \begin{center}
    \begin{minipage}{\textwidth}
      \begin{center}
        ${}^1$ 
        Department of Physics,
        University of Pennsylvania,        
        \\
        209 S. 33rd Street, 
        Philadelphia, PA 19104--6395, USA
      \end{center}
      \begin{center}
        ${}^2$ 
        Institute for Theoretical Physics,
        Vienna University of Technology, 
        \\
        Wiedner Hauptstr. 8-10, 1040 Vienna, 
	Austria 		
      \end{center}
      \begin{center}
        ${}^3$
        Dipartimento di Scienze e Tecnologie Avanzate, 
        Universit\`a del Piemonte Orientale
        \\
        via Bellini 25/g, 15100 Alessandria, Italy, 
        and INFN - Sezione di Torino, Italy
      \end{center}
    \end{minipage}
  \end{center}
  \vspace*{\stretch{1}}
  \begin{abstract}
    \normalsize 
    We apply mirror symmetry to the problem of counting holomorphic
    rational curves in a Calabi-Yau threefold $X$ with
    $\Z_3\oplus\Z_3$ Wilson lines. As we found in
    Part~A~\cite{Braun:2007xh}, the integral homology group
    $H_2(X,\Z)=\Z^3\oplus\Z_3\oplus\Z_3$ contains \emph{torsion}
    curves. Using the B-model on the mirror of $X$ as well as its
    covering spaces, we compute the instanton numbers. We observe that
    $X$ is self-mirror even at the quantum level. Using the
    self-mirror property, we derive the complete prepotential on $X$,
    going beyond the results of Part~A. In particular, this yields the
    first example where the instanton number depends on the torsion
    part of its homology class. Another consequence is that the
    threefold $X$ provides a non-toric example for the conjectured
    exchange of torsion subgroups in mirror manifolds.
  \end{abstract}
  \vspace*{\stretch{5}}
  \begin{minipage}{\textwidth}
    \underline{\hspace{5cm}}
    \centering
    \\
    Email: 
    \texttt{vbraun}, \texttt{ovrut@physics.upenn.edu}, 
    \texttt{Maximilian.Kreuzer@tuwien.ac.at},
    \texttt{esche@mfn.unipmn.it}
  \end{minipage}
\end{titlepage}

\tableofcontents

\input{Bmodel-Intro}

\input{Bmodel-Cover}

\input{Bmodel-Quotient}


\input{Bmodel-Fin}

\bibliographystyle{utcaps} \renewcommand{\refname}{Bibliography}
\addcontentsline{toc}{section}{Bibliography} 
\bibliography{Volker,Emanuel}

\end{document}

%% file: Preamble.tex
\newcommand{\VersionInformation}{}  
\InputIfFileExists{Debug}{}{}

\usepackage{amsmath}
\usepackage{amsthm}
\usepackage{amsfonts}          
\usepackage{amssymb}           
\usepackage[mathscr]{eucal}
\usepackage{graphicx}
\usepackage{color}
\usepackage{hypbmsec}

\usepackage{rotating}
\usepackage{slashbox}
\usepackage[isu,bf]{caption}
\newlength{\xtrawidth}
\newlength{\xtraheight}
\usepackage[all]{xy}           

\usepackage{xr-hyper}

\newcommand{\partA}{Part~A}
\newcommand{\partB}{Part~B}

\newcommand*{\xeqref}[2][\xrlab]{\eqref{#1-#2}}
\newcommand*{\xautoref}[2][\xrlab]{\autoref{#1-#2}}

\ifx\NoBackreferences\EMPTYMACRO
  \usepackage[backref,linktocpage,bookmarks]{hyperref}
\else
  \usepackage[linktocpage,bookmarks]{hyperref}
\fi

\ifx\DEBUG\EMPTYMACRO
\else
  \usepackage{showlabels}
\fi

\def\clap#1{\hbox to 0pt{\hss#1\hss}}

\def\mathrlap{\mathpalette\mathrlapinternal}
\def\mathclap{\mathpalette\mathclapinternal}

\def\mathrlapinternal#1#2{%
\rlap{$\mathsurround=0pt#1{#2}$}}
\def\mathclapinternal#1#2{%
\clap{$\mathsurround=0pt#1{#2}$}}	

\makeatletter
  \def\adots{\mathinner{\mkern2mu\raise\p@\hbox{.}
      \mkern2mu\raise4\p@\hbox{.}\mkern1mu
      \raise7\p@\vbox{\kern7\p@\hbox{.}}\mkern1mu}}
\makeatother

\newcommand{\eqdef}{%
  \mathrel{\lower.1mm
    \hbox{$\stackrel{\lower.424ex\hbox{\scriptsize def}}{=}$}}
}
\newcommand{\Q}{\ensuremath{{\mathbb{Q}}}}
\newcommand{\R}{\ensuremath{{\mathbb{R}}}}
\newcommand{\C}{\ensuremath{{\mathbb{C}}}}

\newcommand{\Z}{\mathbb{Z}}
\newcommand{\CP}{\ensuremath{\mathop{\null {\mathbb{P}}}}\nolimits}

\newcommand{\CY}{Calabi-Yau}
\newcommand{\CYm}{\CY{} manifold}

\newcommand{\even}{\ensuremath{\mathrm{ev}}}

\newcommand{\iunit}{\ensuremath{\mathrm{i}}}
\newcommand{\Cunits}{\ensuremath{\C^\times}}
\newcommand{\free}{\ensuremath{\text{free}}}
\newcommand{\tors}{\ensuremath{\text{tors}}}

\newcommand{\Moduli}{\mathcal{M}}

\DeclareMathOperator{\diff}{d\!}
\DeclareMathOperator{\Span}{span}

\DeclareMathOperator{\Hom}{Hom}

\DeclareMathOperator{\Li}{Li}

\newcommand{\textdef}[1]{{\it #1}}

\newcommand{\Xt}{{\ensuremath{\widetilde{X}}}}
\newcommand{\Xb}{{\ensuremath{\overline{X}}}}

\newcommand{\ZZZ}{\ensuremath{{\Z_3\times\Z_3}}}

\newcommand{\Osheaf}{\ensuremath{\mathscr{O}}}

\newcommand{\dual}{\ensuremath{\vee}}

\newcommand{\CPambient}{\ensuremath{\CP^2\times \CP^1 \times \CP^2}}
\newcommand{\dP}[1]{\ensuremath{dP_{#1}}}
\newcommand{\FPtimes}{\underline{\times}}

\newcommand{\Fprepotential}{\mathscr{F}}

\newcommand{\Fprepot}[1]{\ensuremath{\Fprepotential_{{#1},0}}}
\newcommand{\FprepotNP}[1]{\ensuremath{\Fprepot{#1}^\text{np}}}

\newcommand{\FprepotXNP}{\FprepotNP{X}}

\newcommand{\FprepotXtNP}{\FprepotNP{\Xt}}

\newcommand{\Kahler}{K\"ahler}
\newcommand{\Kcone}{\ensuremath{\mathcal{K}}}

\newcommand{\mathemph}[1]{\textcolor{red}{\mbox{\boldmath $#1$}}}

\newcommand{\tmod}{~\mathrm{mod}~}

{\end{list}}

{\everymath{\displaystyle\everymath{}}\array}%
{\endarray}

\newtheorem{lemma}{Lemma}

\def\cO{\mathcal{O}}
\def\mC{\mathbb{C}}
\def\mP{\mathbb{P}}
\def\mQ{\mathbb{Q}}
\def\mR{\mathbb{R}}

\def\mZ{\mathbb{Z}}
\def\IZ{\mathbb{Z}}
\def\rhob{\bar{\rho}}
\def\nub{\bar{\nu}}

\def\spcheck{^\dual}
\DeclareMathOperator{\Vol}{Vol}
\DeclareMathOperator{\NE}{NE}
\DeclareMathOperator{\ch}{c}
\newcommand{\IdealSR}{I_\text{SR}}
\def\lb{\bar l}
\def\Db{\bar D}
\def\Kb{\bar K}
\newcommand{\Jb}{\Bar{J}}
\newcommand{\Jbstar}{\Bar{J}^\ast}
\newcommand{\Jt}{\Tilde{J}}
\newcommand{\FprepotentialB}{\ensuremath{\Fprepotential^B}}
\newcommand{\FprepotBY}{\ensuremath{\FprepotentialB_{Y^*,0}}}

\include{pst-plot}
\psset{unit=3 pt}
\def\putlab#1)#2#3{\put#1){\makebox(0,0)[#2]{\small #3}}} 
\def\putlin#1,#2,#3,#4,#5){\put#1,#2){\line(#3,#4){#5}}}
\def\putvec#1,#2,#3,#4,#5){\put#1,#2){\vector(#3,#4){#5}}}
\def\putcx#1,#2){\put#1,#2){\circle*{1.4}}}
\newcommand{\PMdos}[4]{{%
    \setlength{\fboxsep}{1pt}%
    \setlength{\fboxrule}{0.5pt}%
    \framebox{$\begin{smallmatrix}
        #1 & #2 \\ #3 & #4
      \end{smallmatrix}
      $}}}
\def\PMtres#1#2#3#4#5#6#7#8#9{
	\begin{picture}(11,11.3)(-1,-2)        
	\put(0,5){$#1$}\put(3,5){$#2$}\put(6,5){$#3$}
	\put(0,2){$#4$}\put(3,2){$#5$}\put(6,2){$#6$}
	\put(0,-1){$#7$}\put(3,-1){$#8$}\put(6,-1){$#9$}
	\end{picture}}
\def\PMplus{{\tiny\begin{picture}(5,5)(0,1)\put(0,0)+\put(0,2)+\put(0,4)+
        \put(2,0)+\put(2,2)+\put(2,4)+\put(4,0)+\put(4,2)+\put(4,4)+
        \end{picture}}}
\def\PMminus{{\tiny\begin{picture}(5,5)(0,1)\put(0,0){--}\put(0,2){--}
	\put(0,4){--}\put(2,0){--}\put(2,2){--}\put(2,4){--}\put(4,0){--}
        \put(4,2){--}\put(4,4){--}\end{picture}}}

\newcommand{\GrantsAcknowledgements}{This research was supported in
    part by the Department of Physics and the Math/Physics Research
    Group at the University of Pennsylvania under cooperative research
    agreement DE-FG02-95ER40893 with the U.~S.~Department of Energy
    and an NSF Focused Research Grant DMS0139799 for ``The Geometry of
    Superstrings'', in part by the Austrian Research Funds FWF grant
    number P18679-N16, in part by the European Union RTN contract
    MRTN-CT-2004-005104, in part by the Italian Ministry of University
    (MIUR) under the contract PRIN 2005-023102 ``Superstringhe, brane
    e interazioni fondamentali'', and in part by the Marie Curie Grant
    MERG-2004-006374.}


%% file: Bmodel-Intro.tex
\section{Introduction}
\label{sec:intro}

Counting world sheet instantons (that is, holomorphic curves) on a
Calabi-Yau threefold has had a large number of applications in
mathematics and physics, ever since it was essentially solved by
mirror symmetry several years ago~\cite{Candelas:1990rm}. The purpose
of this paper is to take into account an important subtlety that does
not appear in very simple Calabi-Yau manifolds like hypersurfaces in
smooth toric varieties.  This subtlety is the appearance of
\emph{torsion} curve classes. That is, the homology\footnote{In the
  following, $\Z_3\eqdef \Z/3\Z$ always denotes the integers mod $3$.
  Similarly, we write $(\Z_3)^n=\oplus_n \Z_3 =
  \Z_3\oplus\cdots\oplus\Z_3$ for the Abelian group generated by $n$
  generators of order $3$.} group
\begin{equation}
  H_2\big( X, \Z \big) = \Z^3 \oplus \Z_3 \oplus \Z_3
\end{equation}
contains the \emph{torsion\footnote{Not to be confused with the
    torsion tensor of a connection.} subgroup} $\Z_3\oplus\Z_3$.
Here, the manifold of interest $X$ is a quotient of one of Schoen's
Calabi-Yau manifolds~\cite{MR923487, Braun:2004xv} by a freely acting
symmetry group. There are already a few known examples of such
Calabi-Yau manifolds with torsion curves~\cite{Aspinwall:1995rb,
  Batyrev:2005jc, gross-2005, Ferrara:1995yx, Aspinwall:1995mh}, but
the proper instanton counting has never been done before.

The prime motivation for studying these curves is that one would like
to compute the superpotential for the vector bundle
moduli~\cite{Lima:2001nh, Lima:2001jc, Buchbinder:2002ic,
  Buchbinder:2002pr, Buchbinder:2002wz, Buchbinder:2003pi,
  Buchbinder:2002ji} in a heterotic MSSM~\cite{Braun:2005nv,
  Braun:2005ux, Braun:2005bw, Braun:2005zv, Braun:2005fk,
  Braun:2005xp, Braun:2006me, Braun:2006ae, Braun:2006th}. Our main
result will be that there exist smooth rigid rational curves in $X$
that are alone in their homology class. This proves that, in general,
no cancellation between contributions to the superpotential $W$ from
instantons in the same homology class can occur.

Therefore we would like to count rational curves on $X$. In physical
terms, we need to find the instanton correction $\FprepotXNP$ to the
genus zero prepotential of the (A-model) topological string on $X$.
This is usually written as a (convergent) power series in $h^{11}$
variables $q_a=e^{2\pi\iunit t^a}$. Each summand is the contribution
of an instanton, and the (integer) coefficients are the multiplicities
of instantons in each homology class. According
to~\cite{Aspinwall:1994uj, Braun:2007tp, Braun:2007xh} the novel
feature of the $3$-torsion curves on $X$ is that for each $3$-torsion
generator we need an additional variable $b_j$ such that $b_j^3=1$.
The Fourier series of the prepotential on $X$ becomes
\begin{equation}
  \FprepotXNP(p,q,r,\, b_1,b_2)
  = 
  \sum_{
    \begin{smallmatrix}
      n_1,n_2,n_3\in \Z \\ 
      m_1,m_2\in \Z_3
    \end{smallmatrix}  
  }
  n_{(n_1,n_2,n_3,m_1,m_2)} \Li_3
  \big( p^{n_1} q^{n_2} r^{n_3} b_1^{m_1} b_2^{m_2} \big)
  ,
\end{equation}
where $n_{(n_1,n_2,n_3,m_1,m_2)}$ is the instanton number in the curve
class $(n_1,n_2,n_3,m_1,m_2)$. 

For the purpose of computing the prepotential, we can either use
directly the A-model or start with the B-model and apply mirror
symmetry. The A-model calculation was carried out in the companion
paper~\cite{Braun:2007xh}, entitled \partA{}. The results were:
\begin{itemize}
\item A set of powerful techniques to compute the torsion subgroups in
  the integral homology and cohomology groups of $X$. They are
  spectral sequences starting with the so-called group (co)homology of
  the group action on the universal cover $\Xt$.
\item A closed formula for the genus zero prepotential
  \begin{equation}
    \label{eq:Aresult}
    \FprepotXNP
    (p,q,r,b_1,b_2)
    =
    \bigg( \sum_{i,j=0}^2 p b_1^i b_2^j \bigg)
    P(q)^4
    P(r)^4
    +O(p^2)
    = 
    \sum_{i,j=0}^2 \Li_3(p b_1^i b_2^j) 
    + \cdots
  \end{equation} 
  to linear order in $p$, extending the one computed
  in~\cite{Hosono:1997hp} for the universal cover $\Xt$. Here, if
  $p(k)$ is the number of partitions of $k\in\Z_\geq$, then $P(q)$ is
  the generating function for partitions,
  \begin{equation}
    \label{eq:GenfnPartitions}
    P(q) \eqdef
    \sum_{i=0}^\infty p(i) q^i =
    \frac{q^\frac{1}{24}}{\eta(\tfrac{1}{2\pi\iunit}\ln q)}
    .
  \end{equation}
\item Expanding eq.~(\ref{eq:Aresult}) as an instanton series we find 
  that the number of rational curves of degree $(1,0,0,m_1,m_2)$ is:
  \begin{equation}
    \label{eq:100curves}
    n_{(1,0,0,m_1,m_2)} = 1, \qquad \forall \; m_1,m_2 \in \mZ_3.  
  \end{equation}
  Furthermore, these curves have normal bundle 
  $\Osheaf_{\mP^1}(-1)\oplus\Osheaf_{\mP^1}(-1)$. 
  Hence, there are indeed 9 smooth rigid rational curves which are alone
  in their homology class.
\end{itemize}
Alternatively, one can start with the B-model topological string and
apply mirror symmetry, which is what we will do in this paper,
entitled \partB{}. This will allow us to obtain the higher order terms
in $p$. The order in $p$ up to which one wants to compute the
instanton numbers is only limited by computer power. We will again
find a closed formula at every order in $p$, however, this time by
guessing it from the instanton calculation, and hence only up to the
order given by this limitation. The way to arrive at this result is as
follows:
\begin{itemize}
\item The universal cover $\Xt$ admits a simple realization as a
  complete intersection in a toric variety. In this situation, mirror
  symmetry boils down to an algorithm to compute instanton numbers.
  Unfortunately, there are many non-toric divisors which cannot be
  treated this way. It turns out that, after descending to $X$,
  precisely the torsion information is lost. In this approach one can
  only compute $\FprepotXNP(q_1,q_2,q_3,\, 1,1)$.
\item As a pleasant surprise we find strong evidence that the manifold
  $X$ is self-mirror. In particular, we attempt to compute the
  instanton numbers on the mirror $X^\ast$ by descending from the
  covering space $\Xt^\ast$.  The embedding of $\Xt^\ast$ into a toric
  variety is such that all $19$ divisors are toric. In principle, this
  allows for a complete analysis including the full $\Z_3\oplus\Z_3$
  torsion information, but this is too demanding in view of current
  computer power.
\item Although the full quotient $X=\Xt/(\ZZZ)$ is not toric, it turns
  out that a certain partial quotient $\Xb=\Xt/\Z_3$ can be realized
  as a complete intersection in a toric variety. That way, one only
  has to deal with $h^{11}(\Xb)=7$ parameters, which is manageable on
  a desktop computer. Assuming the self-mirror property, we work with
  the mirror $\Xb^\ast$, for which again all divisors are toric, and
  we can compute the expansion of $\FprepotXNP(p,q,r,\, 1,b_2)$ to any
  desired degree.  A symmetry argument then allows one to recover the
  $b_1$ dependence as well.  Finally, we can extract the instanton
  numbers $n_{(n_1,n_2,n_3,m_1,m_2)}$ including the torsion
  information.
\item As can be seen from the A-model result eq.~\eqref{eq:Aresult},
  we observe that the prepotential $\FprepotXNP$ at order $p$ factors
  into $\sum_{i,j=0}^2 b_1^i b_2^j$ times a function of $p$, $q$, $r$
  only. This means that the instanton number does not depend on the
  torsion part of its homology class. We will explain the underlying
  reason for this factorization and show that it breaks down at order
  $p^3$. This fits nicely with the B-model computation at order $p^3$,
  where the instanton numbers \emph{do} depend on the torsion part.
\item Another consequence of the self-mirror property is that $X$ is a
  non-toric example for the conjecture of~\cite{Batyrev:2005jc}. By
  this conjecture, certain torsion subgroups of the integral homology
  groups are exchanged under mirror symmetry.
\end{itemize}
An easily readable overview and a discussion of the physical
consequences of our findings for superpotentials and moduli
stabilization of heterotic models was presented in~\cite{Braun:2007tp}.
The present \partB{} is self-contained and can be read independently
of \partA~\cite{Braun:2007xh}. All necessary results from \partA{} 
are reproduced in this part.

As a guide through this paper, we start in~\autoref{sec:CY} with a
brief overview of the topology of the various spaces involved as
determined in \partA~\cite{Braun:2007xh}. This is followed by a review
of the Batyrev-Borisov construction of mirror pairs of complete
intersections in toric varieties in~\autoref{sec:BmodelIntro}. We
illustrate this construction by means of the covering spaces $\Xt$ and
$\Xt^\ast$ of our example. The review includes the techniques to
compute the B-model prepotential and the mirror map. These are applied
in~\autoref{sec:Bquotient} to the partial quotients $\Xb$ and $\Xb^*$
yielding the main results stated above. This assumes that $X$ as well
as $\Xb$ are self-mirror, and evidence for this property is
recapitulated in~\autoref{sec:self-mirror}. Moreover, we show how the
torsion subgroups are exchanged. \autoref{sec:p3check} contains an
explanation for the breakdown of the factorization alluded to above.
Putting all the information together we try to guess a closed form for
the prepotential in~\autoref{sec:conjecture}. Finally, we present our
conclusions in~\autoref{sec:conclusion}.
In the course of this work we will notice that a certain flop of $X$
is very natural from the toric point of view, and we will present it
in~\autoref{sec:X*flop}.

\section
{Calabi-Yau Threefolds}
\label{sec:CY}

\subsection
[The Calabi-Yau Threefold X]
{The Calabi-Yau Threefold $\mathbf{X}$}
\label{sec:X}

The \CYm{} $X$ of interest is constructed as a free $G\eqdef \ZZZ$
quotient of its universal covering space $\Xt$. The latter is one of
Schoen's \CY{} threefolds~\cite{MR923487}. It is simply connected and
hence easier to study. Among its various descriptions are the fiber
product of two \dP9 surfaces, a resolution of a certain $T^6$
orbifold~\cite{Lust:2006zh}, or a complete intersection in a toric
variety. In the present \partB{}, we will mostly use the latter
viewpoint. The simplest way is to introduce the toric ambient variety
$\CPambient$ with homogeneous coordinates
\begin{equation}
  \Big( 
  [x_0:x_1:x_2],~ 
  [t_0:t_1],~
  [y_0:y_1:y_2]
  \Big)
  \in 
  \CPambient
  .
\end{equation}
The embedded \CY{} threefold $\Xt$ is then obtained as the complete
intersection of a degree $(0,1,3)$ and a degree $(3,1,0)$ hypersurface
in $\CPambient$. We restrict the coefficients of their defining 
equations $F_i=0$ to a particular set of three complex parameters 
$\lambda_1$, $\lambda_2$, $\lambda_3$, such that the polynomials $F_i$ read
\begin{subequations}
  \begin{align}
    \label{eq:B1}
    t_0 \Big( x_0^3+x_1^3+x_2^3 \Big) + 
    t_1 \Big( x_0 x_1 x_2 \Big)
    &\eqdef F_1 \\
    \label{eq:B2}
    \big( \lambda_1 t_0 + t_1\big)
    \Big( y_0^3+y_1^3+y_2^3 \Big) 
    +
    \big( \lambda_2 t_0 + \lambda_3 t_1\big)
    \Big( y_0 y_1 y_2 \Big)
    &\eqdef F_2 
    .
  \end{align}
\end{subequations}
For the special complex structure parametrized by $\lambda_1$,
$\lambda_2$, $\lambda_3$ the complete intersection is invariant under
the $G=\ZZZ$ action generated by ($\zeta\eqdef e^{\frac{2\pi i}{3}}$)
\begin{subequations}
\begin{equation}
  \label{eq:g1action}
  g_1:  
  \begin{cases}
    [x_0:x_1:x_2] \mapsto
    [x_0:\zeta x_1:\zeta^2 x_2]
    \\
    [t_0:t_1] \mapsto
    [t_0:t_1] 
    ~\text{(no action)}
    \\
    [y_0:y_1:y_2] \mapsto
    [y_0:\zeta y_1:\zeta^2 y_2]
  \end{cases}
\end{equation}
and
\begin{equation}
  \label{eq:g2action}
  g_2:  
  \begin{cases}
    [x_0:x_1:x_2] \mapsto
    [x_1:x_2:x_0]
    \\
    [t_0:t_1] \mapsto
    [t_0:t_1] 
    ~\text{(no action)}
    \\
    [y_0:y_1:y_2] \mapsto
    [y_1:y_2:y_0]
  \end{cases}
\end{equation}
\end{subequations}
One can show that the fixed points of this group action in $\CPambient$ do 
not satisfy eqns.~\eqref{eq:B1} and~\eqref{eq:B2}, hence the action on 
$\Xt$ is free.

\subsection
[The Intermediate Quotient $\Xb$]
[The Intermediate Quotient Xbar]
{The Intermediate Quotient $\mathbf{\Xb}$}
\label{sec:Xb}

The partial quotient 
\begin{equation}
  \label{eq:Xb}
  \Xb \eqdef \Xt \big/ G_1
\end{equation}
will be of particular interest in this paper because this quotient is
generated by phase symmetries, see eq.~\eqref{eq:g1action}, and hence
is toric. In particular, we will need a basis of \Kahler{} classes. As
usual, we will not distinguish degree-$2$ cohomology and degree-$4$
homology classes but identify them via Poincar\'e duality.
\partA~\cite{Braun:2007xh} \xautoref{sec:X3specseq} shows
that\footnote{The torsion in $H^2$ are just the Wilson lines, that is,
  first Chern classes of flat line bundles. They will play no role in
  the following. The torsion curves in $H_2$, on the other hand, are
  the focus of this paper.}
\begin{equation}
  \label{eq:H2Xb}
  H^2\big(\Xb,\Z\big)
  = 
  H^2(\Xt,\Z)^{G_1} 
  \oplus \Z_3
  =
  \Span_{\Z}
  \Big\{ 
  \phi,\,
  \tau_1,\, \upsilon_1,\, \psi_1,\,
  \tau_2,\, \upsilon_2,\, \psi_2
  \Big\}
  \oplus
  \Z_3
  .
\end{equation}
Hence, by abuse of notation, we can identify the free generators on
$\Xb$ with the $G_1$-invariant generators on $\Xt$, see \partA{}
eq.~\xeqref{eq:H2XtG1inv}, via the pull back by the quotient map. The
triple intersection numbers on $\Xb=\Xt/\Z_3$ are one-third of the
corresponding intersection numbers on $\Xt$ listed in
\partA{} eq.~\xeqref{eq:cubicformXtG1inv}. Hence, the intersection
numbers on $\Xb$ are
\begin{equation}
  \label{eq:cubicformXb}
  \begin{aligned}
    \phi \tau_1 \tau_2 =&\; 3 &
    \phi \tau_1 \upsilon_2 =&\; 3 &
    \phi \tau_1 \psi_2 =&\; 6 &
    \phi \upsilon_1 \tau_2 =&\; 3 &
    \phi \upsilon_1 \upsilon_2 =&\; 3 \\
    \phi \upsilon_1 \psi_2 =&\; 6 &
    \phi \psi_1 \tau_2 =&\; 6 &
    \phi \psi_1 \upsilon_2 =&\; 6 &
    \phi \psi_1 \psi_2 =&\; 12 &
    \tau_1^2 \tau_2 =&\; 1 \\
    \tau_1^2 \upsilon_2 =&\; 1 &
    \tau_1^2 \psi_2 =&\; 2 &
    \tau_1 \upsilon_1 \tau_2 =&\; 3 &
    \tau_1 \upsilon_1 \upsilon_2 =&\; 3 &
    \tau_1 \upsilon_1 \psi_2 =&\; 6 \\
    \tau_1 \psi_1 \tau_2 =&\; 3 &
    \tau_1 \psi_1 \upsilon_2 =&\; 3 &
    \tau_1 \psi_1 \psi_2 =&\; 6 &
    \tau_1 \tau_2^2 =&\; 1 &
    \tau_1 \tau_2 \upsilon_2 =&\; 3 \\
    \tau_1 \tau_2 \psi_2 =&\; 3 &
    \tau_1 \upsilon_2^2 =&\; 3 &
    \tau_1 \upsilon_2 \psi_2 =&\; 6 &
    \tau_1 \psi_2^2 =&\; 6 &
    \upsilon_1^2 \tau_2 =&\; 3 \\
    \upsilon_1^2 \upsilon_2 =&\; 3 &
    \upsilon_1^2 \psi_2 =&\; 6 &
    \upsilon_1 \psi_1 \tau_2 =&\; 6 &
    \upsilon_1 \psi_1 \upsilon_2 =&\; 6 &
    \upsilon_1 \psi_1 \psi_2 =&\; 12 \\
    \upsilon_1 \tau_2^2 =&\; 1 &
    \upsilon_1 \tau_2 \upsilon_2 =&\; 3 &
    \upsilon_1 \tau_2 \psi_2 =&\; 3 &
    \upsilon_1 \upsilon_2^2 =&\; 3 &
    \upsilon_1 \upsilon_2 \psi_2 =&\; 6 \\
    \upsilon_1 \psi_2^2 =&\; 6 &
    \psi_1^2 \tau_2 =&\; 6 &
    \psi_1^2 \upsilon_2 =&\; 6 &
    \psi_1^2 \psi_2 =&\; 12 &
    \psi_1 \tau_2^2 =&\; 2 \\
    \psi_1 \tau_2 \upsilon_2 =&\; 6 &
    \psi_1 \tau_2 \psi_2 =&\; 6 &
    \psi_1 \upsilon_2^2 =&\; 6 &
    \psi_1 \upsilon_2 \psi_2 =&\; 12 &
    \psi_1 \psi_2^2 =&\; 12
    .
  \end{aligned}
\end{equation}
Clearly, $G_2$ acts on the partial quotient $\Xb$. From \partA{} 
eq.~\xeqref{eq:Bg1invg2act} it follows that, of the $7$
non-torsion divisors above, only $3$ are $G_2$-invariant. This
invariant part is particularly manageable and will be important in the
following. We find 
\begin{equation}
  \label{eq:H2XGinv}
  H^2\big(\Xt,\Z\big)^G_\free 
  =
  H^2\big(\Xb,\Z\big)^{G_2}_\free
  =
  \Span_\Z \Big\{ \phi, \tau_1, \tau_2 \Big\}
\end{equation}
with products $3 \tau_i^2=\tau_i\phi$. In particular, the triple
intersection numbers on $\Xb$ are
\begin{equation}
  \label{eq:t1t2fcubic}
  \tau_1^2 \tau_2 = 1 
  ,\quad
  \tau_1 \phi \tau_2  = 3
  ,\quad
  \tau_1 \tau_2^2 = 1 
  ,\quad
\end{equation}
and $0$ otherwise. Finally, the second Chern class of $\Xb$ is
$c_2(\Xb)=12(\tau_1^2+\tau_2^2)$. Therefore,
\begin{equation}
  c_2\big(\Xb\big) \cdot \tau_1 = 12
  ,\quad
  c_2\big(\Xb\big) \cdot \phi = 0
  ,\quad
  c_2\big(\Xb\big) \cdot \tau_2 = 12
  .
\end{equation}

\subsection{Variables}
\label{sec:variables}

As we discussed in \partA{} \xautoref{sec:InstGeneral}, the
instanton-generated superpotential should be thought of as a series
with one variable for each generator in $H_2$.
\begin{table}[htbp]
  \centering
  \renewcommand{\arraystretch}{1.5}
  \begin{tabular}{c|cccc}
    \parbox[c][12mm]{25mm}{\centering Calabi-Yau \\ threefold}
    & 
    $H_2\big(-,\Z\big)$ & 
    \parbox{25mm}{\centering Free \\ generators} & 
    \parbox{25mm}{\centering Torsion \\ generators}  
    \\ \hline
    $\Xt$ &
    $\Z^{19}$ & $\big\{p_0,q_0,\dots,q_8,r_0,\dots,r_8\big\}$ &
    $\varnothing$
    \\
    $\Xb=\Xt/G_1$ &
    $\Z^{7}\oplus\Z_3$ & $\big\{P,Q_1,Q_2,Q_3,R_1,R_2,R_3\big\}$ &
    $\big\{b_1\big\}$
    \\
    $X=\Xt/G$ &
    $\Z^3\oplus\Z_3\oplus\Z_3$ & $\big\{p,q,r\big\}$ &
    $\big\{b_1,b_2\big\}$
  \end{tabular}
  \caption{The different Calabi-Yau threefolds, curve classes, 
    and variables used to expand the prepotential.}
  \label{tab:variables}
\end{table}
In particular, we will be interested in the Calabi-Yau threefolds
$\Xt$, $\Xb$, and $X$. For these, the degree-$2$ integral homology and
the variables used (see
\partA{}~\cite{Braun:2007xh} for precise definitions) are in
summarized \autoref{tab:variables}. Pushing down the curves by the
respective quotients lets us express the prepotential on the quotient
in terms of the prepotential on the covering space. We found in
\partA{} that
\begin{equation}
  \label{eq:Xt2Xb}
  \begin{split}
    \FprepotNP{\Xb}&\big(P,Q_1,Q_2,Q_3,R_1,R_2,R_3,b_1) 
    \\
    =&
    \frac{1}{|G_1|}    
    \FprepotXtNP\Big(\! 
    \begin{array}[t]{l}
      P Q_1^5 Q_2^6 R_1^5 R_2^6 ,
      \\
      ~
      Q_1^5 Q_2^6 ,\,
      Q_1^{-2} Q_2^{-2} Q_3^{-3} b_1 ,\,
      Q_1^{-1} Q_2^{-1} ,\,
      Q_3^3 ,\,
      Q_3^2 b_1 ,\,
      Q_3 ,\,
      1 ,\,
      b_1 ,\,
      Q_1 Q_3^3 ,
      \\
      ~~
      R_1^5 R_2^6 ,\,
      R_1^{-2} R_2^{-2} R_3^{-3} b_1^2 ,\,
      R_1^{-1} R_2^{-1} ,\,
      R_3^3 ,\,
      R_3^2 b_1^2 ,\,
      R_3 ,\,
      1 ,\,
      b_1^2 ,\,
      R_1 R_3^3                    
      ~\, \mathrlap{\smash{\Big)}}
    \end{array}
  \end{split}
\end{equation}
and
\begin{equation}
  \label{eq:Xb2X}
  \FprepotXNP\big(p,q,r,b_1,b_2) 
  =
  \frac{1}{|G_2|}    
  \FprepotNP{\Xb}\big(
    p ,\,
    q ,\,
    b_2 ,\,
    b_2 ,\,
    r ,\,
    b_2^2 ,\,
    b_2^2 ,\,
    b_1
  \big)
  .
\end{equation}


%% file: Bmodel-Cover.tex
\section{Toric Geometry and Mirror Symmetry}
\label{sec:BmodelIntro}

In this section we review mirror symmetry and the construction of the
B-model for the mirror of the covering space $\Xt$. Since $\Xt$ is a
complete intersection in a toric variety, we can use the standard
constructions. Because we expect the model to be self-mirror, we will
analyze the B-model for $\Xt$ and its mirror $\Xt^\ast$. The toric
geometry for $\Xt$ is much simpler\footnote{Meaning that $\Xt$ is a
  complete intersection in the very simple toric variety $\CPambient$,
  whereas $\Xt^\ast$ is embedded in a complicated toric ambient
  variety.} than for $\Xt^\ast$, but contains less information.  In
this section we will start with the simpler model in order to review
the Batyrev-Borisov construction for the mirror of a complete
intersection in a toric variety. Then we will apply this construction
to the more complicated model, now without going into too many
details.  We will see that, on the simpler side, not all parameters
are toric and no torsion is visible. However, on the more complicated
mirror side, all parameters are toric which will allow us, in
principle, to perform the B-model computation of the complete
prepotential. As $\Xt\cong\Xt^\ast$ is expected to be self-mirror,
this determines the complete prepotential
$\FprepotNP{\Xt}=\FprepotNP{\Xt^\ast}$ as well. In practice, however,
the analysis is computationally too involved.

Fortunately, the space $\Xb=\Xt/G_1$ and its mirror will turn out to
be both tractable with toric methods and sufficiently informative.
This quotient will be the subject of~\autoref{sec:Bquotient}. Finally,
this is also the starting point for arguing
in~\autoref{sec:self-mirror} that the self-mirror property persists at
the level of instanton corrections.

Recall that, in \autoref{sec:X} we defined our Calabi-Yau manifold as
the complete intersection
\begin{equation}
  \label{eq:defXtilde} 
  \Xt ~\eqdef~ 
  \Big\{ F_1=0 ,\, F_2=0\big\} 
  ~\subset~ 
  \CPambient
\end{equation} 
with the two polynomials $F_1$, $F_2$ as in eqns.~\eqref{eq:B1}
and~\eqref{eq:B2}, respectively. In order to construct the mirror
manifold following Batyrev and Borisov, we need to reformulate this
definition in terms of toric geometry. We review here some essential
ingredients of toric geometry, for details we refer
to~\cite{Kreuzer:2006ax, Klemm:2004km} and references therein. We will
give the abstract definitions and concepts step by step, and at each
step illustrate them with the example $\Xt$ and its mirror manifold
$\Xt^\ast$.

\subsection{Toric Varieties}
\label{sec:toric}

Given a lattice $N$ of dimension $d$, a toric variety $V_{\Sigma}$ is 
defined in terms of a fan $\Sigma$ which is a collection of rational 
polyhedral\footnote{Here, the tip of the cone is always 
  the origin of $N$. A cone is rational if it is spanned
  by rays which pass through lattice points (other than the origin),
  that is, have rational slopes. A cone is polyhedral if it is the
  cone over an $(d-1)$-dimensional polytope.  In other words, curved
  faces are not allowed.} cones $\sigma \subset N$ such that it
contains all faces and intersections of its elements. $V_\Sigma$ is
compact if the support of $\Sigma$ covers all of the real extension
$N_\mR$ of the lattice $N$. The resulting $d$-dimensional variety
$V_\Sigma$ is smooth if all cones are simplicial and if all maximal
cones are generated by a lattice basis.

Let $\Sigma^{(1)}$ denote the set of one-dimensional cones (rays) with
primitive generators $\rho_i$, $i=1,\dots,n$. The simplest description
of $V_\Sigma$ introduces $n$ homogeneous coordinates $z_i$
corresponding to the generators $\rho_i$ of the rays in
$\Sigma^{(1)}$. These homogeneous coordinates are then subjected to
weighted projective identifications
\begin{equation}
  \label{eq:weights} 
  \big[z_1:\dots:z_n \big] =
  \big[\lambda^{q_1^{(a)}}z_1:\dots:\lambda^{q_n^{(a)}}z_n \big]
  \qquad 
  a = 1,\dots,h
\end{equation} 
for any nonzero complex number $\lambda\in\mC^\times$, where the integer
$n$-vectors $q_i^{(a)}$ are generators of the linear relations $\sum
q_i^{(a)}\rho_i=0$ among the primitive lattice vectors\footnote{We
  will use the same symbol $\rho_\ast$ to denote the generators in
  $\Sigma^{(1)}$ and the corresponding primitive lattice vectors in
  $N$.} $\rho_i$.  In order to obtain a well-behaved quotient, we must
exclude an exceptional set $Z(\Sigma)\subset\mC^n$ that is defined in
terms of the fan, as will be explained below. Hence, the quotient is
\begin{equation}
  \label{eq:toricvariety} 
  V_\Sigma=
    \Big(\mC^n-Z(\Sigma)\Big)
    \Big/
    \Big( \big(\mC^\times\big)^h \times \Gamma \Big)
    ,
\end{equation} 
where $\Gamma \simeq N/\Span\{\rho_i\}$ is a finite abelian group.
There are $h=n-d$ independent $\mC^\times$ identifications, therefore
the complex dimension of $V_\Sigma$ equals the rank $d$ of the lattice
$N$. The identifications by $\Gamma$ are only non-trivial if the $\rho_i$ 
do not span the lattice $N$. Refinements of the lattice $N$ with fixed 
$\rho_i$ can hence be used to construct quotients of toric varieties 
$V_\Sigma$ by discrete phase symmetries such as $\mZ_3$. Such quotients
will be discussed in~\autoref{sec:Bquotient}.
Note that the rays $\rho_i$ are in 1--to--1 correspondence with the
$\left(\C^\times\right)$-invariant divisors $D_i$ on $V_\Sigma$, which
are defined as
\begin{equation}
  D_i = \big\{z_i = 0 \big\} ~\subset V_\Sigma .
\end{equation}
Conversely, the homogeneous coordinate $z_i$ is a section of the line
bundle $\cO(D_i)$.

For example, consider the simplest compact toric variety, the
projective space $\mP^d$. Its fan $\Sigma=\Sigma(\Delta)$ is generated
by the $n=d+1$ vectors
\begin{equation}
  \rho_1=e_1,
  \quad
  \rho_2=e_2,
  \quad
  \dots,
  \quad
  \rho_{n-1}=e_d,
  \quad
  \rho_{n}=-\sum_{i=1}^d e_i
\end{equation}
of a $d$-dimensional simplex $\Delta$. They satisfy a single linear
relation, $\sum_{i=1}^{n} \rho_i =0$. Therefore $q_i=1$ for all $i$,
and the homogeneous coordinates in eq.~\eqref{eq:weights} are the
usual homogeneous coordinates on $\mP^d$.

For products of toric varieties we simply extend the relations for any
single factor by zeros and take the union of them. Hence, the fan of
the polyhedron $\Delta^*$ describing the $5$-dimensional toric variety
$\mP^2\times \mP^1 \times \mP^2$ in eq.~\eqref{eq:defXtilde} is
generated by the $n=5+3=8$ vectors
\begin{equation}
  \label{eq:Delta}
  \begin{aligned}
    \rho_1&=e_1,&
    \rho_2&=e_2,&
    \rho_3&=-e_1-e_2,&
    \rho_4&=e_3,
    \\
    \rho_5&=-e_3,&
    \rho_6&=e_4,&
    \rho_7&=e_5,&
    \rho_8&=-e_4-e_5
  \end{aligned}
\end{equation} 
satisfying the linear relations
\begin{equation}
  \label{eq:linreldelta}
  \sum_{i=1}^3\rho_i=\sum_{i=4}^5\rho_i=\sum_{i=6}^8\rho_i=0 
  .
\end{equation}
Except for the origin, there are no other lattice points in the
interior of $\Delta^*$. The corresponding homogeneous coordinates will
be denoted by 
\begin{equation}
  \label{eq:varchange}
  \begin{gathered}
    z_1 = x_0
    ,\quad
    z_2 = x_1
    ,\quad
    z_3 = x_2
    , \\
    z_4 = t_0
    ,\quad
    z_5 = t_1
    , \\
    z_6 = y_0
    ,\quad
    z_7 = y_1
    ,\quad
    z_8 = y_2
    .
  \end{gathered}
\end{equation}
In more general situations, given a
polytope $\Delta^* \subset N$ we will denote the resulting toric
variety by $\mP_{\Delta^*}=V_{\Sigma(\Delta^*)}$.

\subsection{The Batyrev-Borisov Construction}
\label{sec:batyrev-borisov}

Batyrev showed that a generic section of $K_{\mP_{\Delta^*}}^{-1}$,
the anticanonical bundle of $\mP_{\Delta^*}$, defines a Calabi-Yau
hypersurface if $\Delta^*$ is reflexive, which means, by definition,
that $\Delta^*$ and its dual
\begin{equation}
  \label{eq:dual} 
  \Delta=
  \Big\{
  x\in M_\mR
  \,\Big|\,
  (x,y)\ge-1 
  ~\forall y\in\Delta^*
  \Big\}
\end{equation} 
are both lattice polytopes. Here, $M=\mathrm{Hom}(N,\mZ)$ is the
lattice dual to $N$ and $M_\mR$ is its real extension. Mirror symmetry
corresponds to the exchange of $\Delta$ and
$\Delta^*$~\cite{Batyrev:1994hm}. The generalization of this
construction to complete intersections of codimension $r>1$ is due to
Batyrev and Borisov~\cite{Borisov:1993ab, Batyrev:1994pg}. For that
purpose, they introduced the notion of a \textdef{nef partition}.
Consider a dual pair of $d$-dimensional reflexive polytopes
$\Delta\subset M_\mR,\Delta^*\subset N_\mR$. In that context, a
partition $E=E_1\cup\dots\cup E_r$ of the set of vertices of
$\Delta^*$ into disjoint subsets $E_1,\dots,E_r$ is called a
nef-partition if there exist $r$ integral upper convex
$\Sigma(\Delta^*)$-piecewise linear support functions
$\phi_l:N_\mR\rightarrow \mR$, $l=1,\dots,r$ such that
\begin{equation} 
  \phi_l(\rho)=
  \begin{cases} 
    1 & \text{if } \rho\in E_l, \\ 
    0 & \text{otherwise.}
  \end{cases}
\end{equation} 
Each $\phi_l$ corresponds to a divisor
\begin{equation} 
  \label{eq:CICY}
  D_{0,l}=\sum_{\rho\in E_l} D_{\rho}
\end{equation} 
on $\mP_{\Delta^*}$, and their intersection 
\begin{equation}
  Y=D_{0,1}\cap\dots\cap D_{0,r}
\end{equation}
defines a family $Y$ of Calabi-Yau complete intersections of
codimension $r$.  Moreover, each $\phi_l$ corresponds to a lattice
polyhedron $\Delta_l$ defined as
\begin{equation}
  \label{eq:di} 
  \Delta_l=
  \Big\{
  x\in M_\mR
  \,\Big|\,
  (x,y)\geq -\phi_l(y)
  \quad \forall y\in N_\mR
  \Big\}
  .
\end{equation} 
The lattice points $m \in \Delta_l$ correspond to monomials 
\begin{equation}
  z^m =
  \prod_{i=1}^{n} z_i^{\langle m, \rho_i\rangle}
  \quad\in \Gamma\left(\mP_{\Delta^*},\cO(D_{0,l})\right)
  .
\end{equation}
One can show that the sum of the functions $\phi_l$ is equal to the
support function of $K_{\mP_{\Delta^*}}^{-1}$ and, therefore, the
corresponding Minkowski sum is $\Delta_1 + \dots + \Delta_r = \Delta$.
Moreover, the knowledge of the decomposition $E=E_1\cup\dots\cup E_r$
is equivalent to that of the set of supporting polyhedra
$\Pi(\Delta)=\{\Delta_1,\dots,\Delta_r\}$, and therefore this data is
often also called a nef partition. 

It can be shown that given a nef partition $\Pi(\Delta)$ the
polytopes\footnote{The brackets $\big< \cdots \big>$ denote the convex
  hull.}
\begin{equation}
  \nabla_l=
  \big\langle
  \{0\}\cup E_l
  \big\rangle
  ~\subset N_\mR
\end{equation}
define again a nef partition
$\Pi^\ast(\nabla)=\{\nabla_1,\dots,\nabla_r\}$ such that the Minkowski
sum $\nabla=\nabla_1+\dots+\nabla_r$ is a reflexive polytope. This is
the combinatorial manifestation of mirror symmetry in terms of dual
pairs of nef partitions of $\Delta^*$ and $\nabla^*$, which we
summarize in the diagram
\begin{equation}
  \label{eq:polyDN}
  \vcenter{\xymatrix{
      \strut
      \Delta=\Delta_1+\ldots+\Delta_r
      &&
      \strut
      \Delta^*=\big\langle\nabla_1,\ldots,\nabla_r\big\rangle
      \ar@{<->}[lldd]|{\parbox{20mm}{\centering\small Mirror\\ Symmetry}}
      \\ 
      M_\mR 
      &
      & 
      N_\mR
      \\
      \nabla^*=\big\langle\Delta_1,\ldots,\Delta_r\big\rangle
      &
      (\Delta_l,\nabla_{l'})\ge-\delta_{l\,l'} 
      & 
      \nabla=\nabla_1+\ldots+\nabla_r
      \save 
      "1,1"."3,1"*+=[F]\frm{}
      \restore
      \save 
      "1,3"."3,3"*+=[F]\frm{}
      \restore
    }}
  .
\end{equation} 
In the horizontal direction, we have the duality between the lattices
$M$ and $N$ and mirror symmetry goes from the upper right to the lower
left. The other diagonal has also a meaning in terms of mirror
symmetry as we will explain below.  The complete intersections $Y
\subset \mP_{\Delta^*}$ and $Y^* \subset \mP_{\nabla^*}$ associated to
the dual nef partitions are then mirror Calabi-Yau varieties.

Let us now apply the Batyrev-Borisov construction to the complete
intersection eq.~\eqref{eq:defXtilde}, hence $r=2$. There exist
several nef-partitions of $\Delta^*$. The one which has the correct
degrees $(3,1,0)$ and $(0,1,3)$ is, up to exchange of $t_0$ and $t_1$,
$E_1=\{\rho_i\, |i=1,\dots,4\}$ and $E_2=\{\rho_i\, |i=5,\dots,8\}$.
Adding the origin and taking the convex hull yields the polytopes
\begin{align}
  \label{eq:nefcover} 
  \nabla_1&=\big\langle \rho_1,\dots,\rho_4, 0 \big\rangle
  , & 
  \nabla_2& = \big\langle \rho_5,\dots,\rho_8 , 0 \big\rangle
  ,
\end{align} 
where the $\rho_i$ are defined in eq.~\eqref{eq:Delta}. The two
divisors cutting out the Calabi-Yau threefold are, according to
eq.~\eqref{eq:CICY},
\begin{equation}
  \label{eq:defXt}
  D_{0,1}=\sum_{i=1}^4 D_{i}
  ,\quad
  D_{0,2}=\sum_{i=5}^8 D_{i}  
  \qquad\Rightarrow\quad
  \Xt = D_{0,1} \cap D_{0,2} ~\subset \mP_{\Delta^*}
\end{equation}
Note that, while $\Delta^*$ has no further lattice points, its dual
$\Delta$ has $18$ vertices and $300$ lattice points. Using the
computer package PALP~\cite{Kreuzer:2002uu}, we determine the
associated polytopes $\Delta_1$ and $\Delta_2$ of the global sections
of $\cO(D_{0,1})$ and $\cO(D_{0,2})$, respectively. In an appropriate
lattice basis there is, up to symmetry, a unique nef partition
consisting of
\begin{align}
  \label{eq:nefmirror} 
  \Delta_1&=\big\langle \nu_1,\dots,\nu_6,0\big\rangle
  ,    &
  \Delta_2&=\big\langle \nu_7,\dots,\nu_{12},0\big\rangle
  ,
\end{align} 
where
\begin{equation}
  \begin{aligned}
    \label{eq:Nabla}
    \nu_1&=2e_1-e_2,&\nu_2&=-e_1+2e_2,&\nu_3&=-e_1-e_2,
    \\
    \nu_4&=2e_1-e_2-e_3,&\nu_5&=-e_1+2e_2-e_3,&\nu_6&=-e_1-e_2-e_3,
    \\
    \nu_7&=2e_4-e_5,&\nu_8&=-e_4+2e_5,&\nu_9&=-e_4-e_5,
    \\
    \nu_{10}&=e_3+2e_4-e_5,&\nu_{11}&=e_3-e_4+2e_5,&\nu_{12}&=e_3-e_4-e_5
    .
  \end{aligned} 
\end{equation}
Among these $12$ vectors there are the $7$ independent linear
relations
\begin{equation}
  \begin{aligned}
    \label{eq:linrelnabla} 
    3 \nu_3+\nu_4+\nu_5-2\nu_6 &=0, &\qquad
    3 \nu_9+\nu_{10}+\nu_{11}-2\nu_{12} &=0, 
    \\
    \nu_1-\nu_3-\nu_4+\nu_6 &=0, & 
    -\nu_1+\nu_2+\nu_4-\nu_5 &=0, 
    \\
    \nu_7-\nu_9-\nu_{10}+\nu_{12} &= 0, & 
    -\nu_7+\nu_8+\nu_{10}-\nu_{11} &=0, 
    \\
    -\nu_2+\nu_5-\nu_8+\nu_{11}&=0
    . 
  \end{aligned} 
\end{equation}
The convex hull $\nabla^* = \langle \Delta_1,\Delta_2 \rangle$ yields
the fan $\Sigma(\nabla^*)$ and, consequently, the toric variety
$\mP_{\nabla^*}$. Let $D^*_i, i=1,\dots,12$ be the divisors associated
to the vertices $\nu_i$. Then, by eq.~\eqref{eq:CICY}, the nef
partition eq.~\eqref{eq:nefmirror} defines the divisors 
\begin{equation}
  D^*_{0,1}=\sum_{i=1}^6 D^*_{i}
  ,\quad
  D^*_{0,2}=\sum_{i=7}^{12} D^*_{i},  
  \qquad\Rightarrow\quad
  \Xt^*= D^*_{0,1} \cap D^*_{0,2} ~\subset \mP_{\nabla^*}
\end{equation}
cutting out the mirror complete intersection $\Xt^\ast$. In contrast
to $\Delta^*$, the polytope $\nabla^*$ contains extra integral points.
We find that it contains, in addition to the origin and the vertices
in eq.~\eqref{eq:Nabla}, the $26$ points
\begin{equation}
  \label{eq:points}
  \begin{gathered}
    \begin{aligned}
      \nu_{13}&=\frac{1}{3}(\nu_4+\nu_5+\nu_6)=-e_3,
      &
      \nu_{12+6k+i+j}&=\frac{1}{3}(\nu_{3k+i}+2\nu_{3k+j}),
      \\
      \nu_{14}&=\frac{1}{3}(\nu_{10}+\nu_{11}+\nu_{12})=e_3,
      &
      \nu_{15+6k+i+j}&=\frac{1}{3}(\nu_{3k+j}+2\nu_{3k+i})
    \end{aligned}   
    \\
    \forall ~ 
    k\in\{0,\dots,3\} 
    ,~
    (i,j)\in\big\{(1,2),(1,3),(2,3)\big\}
    .
  \end{gathered}
\end{equation}
For completeness, note that the dual polytope $\nabla$ has $15$
vertices and $24$ lattice points. Running PALP to compute the Hodge
numbers using the formula of~\cite{Batyrev:1995ca}, we obtain
\begin{equation}
  \label{eq:HodgeXt}
  h^{1,1}\big(\Xt\big)=
  h^{1,2}\big(\Xt\big)=
  h^{1,1}\big(\Xt^*\big)=
  h^{1,2}\big(\Xt^*\big)=19, 
\end{equation}
in agreement with \partA~\cite{Braun:2007xh},
eq.~\xeqref{eq:Hodgecover}.

So far, we have mainly focused on the information contained in the
reflexive polytopes $\Delta^*$ and $\nabla^*$ and ignored their duals.
We have already mentioned that in the reflexive case a generic section
of $K^{-1}_{\mP_{\Delta^*}}$ defines a Calabi-Yau manifold, and that
such sections are provided by the lattice points of $\Delta$.  In
other words, $\Delta$ and $\nabla$ are the Newton polytopes of $Y$ and
$Y^*$, respectively. That is, the complete intersection $Y$ ($Y^\ast$)
is defined by $r$ polynomial equations, and the exponents of the
monomials in each equation are the lattice points in $\Delta$
($\nabla$). More precisely, the Minkowski sum for, say,
$\Delta=\Delta_1+\dots+\Delta_r$ defines $r$ homogeneous polynomials
\begin{equation}
  \label{eq:CIeqs} 
  F_l(z) = 
  \sum_{\substack{m \in\\ \Delta_l\cap M}} a_{l,m} 
  \prod_{l'=1}^r
  \prod_{\substack{\rho_i \in\\ \nabla_{l'}\cap N}} 
  z_i^{\langle m, \rho_i \rangle+\delta_{l\,l'}}
  , \qquad 
  l = 1,\dots,r
\end{equation} 
with coefficients $a_{l,m} \in \mC$. The simultaneous vanishing of
$F_1,\dots,F_r$ then defines the complete intersection Calabi-Yau
manifold $Y \subset \mP_{\Delta^*}$.  Exchanging $\Delta_l$ and
$\nabla_{l'}$ in eq.~\eqref{eq:CIeqs} yields the equations $F_l^*$
defining the mirror manifold $Y^*$. It is in this sense that the map
from the upper left to the lower right in eq.~\eqref{eq:polyDN} is
also a manifestation of mirror symmetry. Since we will not need the
actual polynomials for $\Xt$ and $\Xt^*$, we refrain from writing them
explicitly. Instead, we refer the reader to \autoref{sec:Bquotient},
where we determine the equations in a simpler situation.

\subsection{Toric Intersection Ring}
\label{sec:intersection-ring}

Up to now we have only considered one of the ingredients in the fan
$\Sigma$, namely, the generators $\rho \in \Sigma^{(1)}$ which defined
the $\mC^\times$ action in eq.~\eqref{eq:toricvariety}. The second
ingredient is the exceptional set $Z(\Sigma)$. It corresponds to fixed
loci of a continuous subgroup of $\left(\mC^\times\right)^h$ for which
the quotient eq.~\eqref{eq:toricvariety} is not well defined.
Therefore, these loci have to be removed. In terms of the homogeneous
coordinates $z_i$, this happens precisely when a subset $\{z_i\,| i
\in I\}$, $I \subseteq \{1,\dots,n\}$, of the coordinates vanishes
simultaneously such that there is no cone $\sigma \in \Sigma$
containing all of the $\rho_i \subseteq \sigma$, $i \in I$. Hence, the
set $Z(\Sigma)$ is the union of the sets $Z_I = \{[z_1:\dots:z_n]\,|
z_i=0\, \forall i \in I\}$. Minimal index sets $I$ with this property
are called primitive collections~\cite{Batyrev:1991ab}. In order to
determine the index sets $I$ we need a coherent\footnote{Coherent
  triangulations, sometimes also called regular triangulations,
  satisfy some technical property that is equivalent to the associated
  toric quotient being \Kahler.} triangulation $T=T(\Delta^*)$ of the
polytope $\Delta^*$ for which all simplices contain the origin.
Different triangulations will yield different exceptional sets and,
hence, different toric varieties. However, for simplicity, we will
mostly suppress the choice of a triangulation in the notation.  In the
case of complete intersections, only those triangulations of
$\Delta^*$ are compatible with a given nef partition that can be
lifted to a triangulation of the corresponding Gorenstein cone,
see~\cite{Stienstra:1998ab}.

The polytope defining projective space $\mP^d$ admits a unique
triangulation with the required properties, and this triangulation
consists of $n=d+1$ simplices. The only primitive collection is $I=
\{1,\dots,n\}$. This is well-known from the definition of projective
space, where we have to remove the origin $z_1=\dots=z_{d+1}=0$ from
$\mC^{d+1}$. Similarly, the polyhedron $\Delta^*$ for the ambient
space $\mP_{\Delta^*}$ of $\Xt$ admits a unique triangulation, and the
primitive collections are those of its factors, that is,
\begin{align}
  \label{eq:primitive} 
  I_1 &= \{1,2,3\}
  ,\, & 
  I_2& =\{4,5\}
  ,\, &
  I_3&=\{6,7,8\}
  .
\end{align} 
The mirror polyhedron $\nabla^*$, on the other hand, admits a huge
number of triangulations. We will discuss particularly interesting
triangulations of the mirror polyhedron at the end of
\autoref{sec:nablatriangs}.

The primitive collections determine the cohomology ring of toric
varieties and, together with the nef partition, complete intersections. 
Recall that if the collection
$\rho_{i_1},\dots,\rho_{i_k}$ of rays is not contained in at least one
cone, then the corresponding homogeneous coordinates $z_{i_l}$ are not
allowed to vanish simultaneously. Therefore, the corresponding
divisors $D_{i_l}$ have no common intersection. Hence, we obtain
non-linear relations $R_I=D_{i_1}\cdot\ldots\cdot D_{i_k}=0$ in the
intersection ring. It can be shown that all such relations are
generated by the primitive collections $I=\{i_1,\ldots, i_k\}$ defined
above. The ideal generated by these $R_I$ is called Stanley-Reisner
ideal
\begin{equation}
  \IdealSR
  = 
  \big< 
  R_I 
  ,\, 
  I \text{ primitive collection}
  \big>
  \quad
  \subset \mZ[D_1,\ldots, D_n]
  ,
\end{equation}
and $\mZ[D_1,\dots,D_n]/\IdealSR$ is the Stanley-Reisner ring.
The intersection ring of a non-singular compact toric variety
$\mP_\Sigma$ is~\cite{Danilov:1978ab}
\begin{equation} 
  H^*\big(\mP_\Sigma,\Z\big)=
  \mZ\left[D_1,\ldots, D_n\right]
  \Big/
  \big\langle 
    \IdealSR,\,
    \sum_i (m,\rho_i)D_i
  \big\rangle
  .
\end{equation} 
In other words, the intersection ring can be obtained from the
Stanley-Reisner ring by adding the linear relations $\sum_i( m,\rho_i
) D_i=0$, where it is sufficient to take a set of basis vectors for
$m\in M$. In particular, the intersection number of the divisors
spanning a maximal-dimensional simplicial cone
$\sigma=\Span_{\R\geq}\{\rho_{i_1},\ldots, \rho_{i_d}\}$ is
\begin{equation}
  \label{eq:intersect} 
  D_{i_1}\cdot\ldots\cdot D_{i_d}=\frac{1}{\Vol(\sigma)}
  ,
\end{equation} 
where $\Vol(\sigma)$ is the lattice-volume, that is, the geometric
volume divided by the volume $\tfrac{1}{d!}$ of a basic simplex. For
practical purposes it is sufficient to compute one of these volumes,
the remaining intersections can be obtained using the linear and
non-linear relations.

Having found the intersection ring of the ambient toric variety, we
now turn to the complete intersection $Y \subset \mP_{\Delta^*}$. The
toric part of its even-degree intersection ring
is~\cite{Batyrev:1994rs}
\begin{equation}
  \label{eq:CYring} 
  H^\even_\text{toric}\big(Y,\mQ\big) =
  \mQ\left[D_1,\ldots, D_n\right] 
  \big/ 
  I_Y
  ,
\end{equation}
where $I_Y$ is the ideal quotient
\begin{equation}
  \label{eq:CYideal}
  I_Y = 
  \Big\langle 
  \IdealSR,\sum_i (m,\rho_i) D_i
  \Big\rangle 
  \,:\, 
  \prod_{l=1}^r D_{0,l} 
  .
\end{equation}
Note that it can happen that some of the $D_i$ appear as generators of
$I_Y$. This means that they can be set to zero in the intersection
ring. Geometrically, this means that these divisors do not intersect a
generic complete intersection $Y$. While the intersection ring depends
on the triangulation $T(\Delta^*)$ through the primitive collections
defining the Stanley-Reisner ideal, we conjecture that the divisors
$D_i$ not intersecting $Y$ are independent of the choice of
triangulation. This conjecture is proven for $r=1$ and supported by a
large amount of empirical evidence for $r>1$. We conclude that the
dimension $\dim H^2_\text{toric}(Y)$ is in general smaller than
$h^{1,1}(Y)$ for the following two reasons: Only $h=n-d = \dim
H^2(\mP_{\Delta^*},\mZ)$ divisors are realized in the ambient toric
variety $\mP_{\Delta^*}$, and some of them may not descend to the
complete intersection $Y$.  Using the adjunction formula we can
compute the the Chern classes of $Y$ by expanding
\begin{equation}
  \label{eq:ChernY}
  \ch(Y) = \frac{\prod\limits_{i=1}^n(1+D_i)}{\prod\limits_{l=1}^r (1+D_{0,l})}.
\end{equation}
The intersection ring together with the second Chern class determine
the diffeomorphism type of a simply-connected Calabi-Yau
manifold~\cite{Wall:1966ab}. If we consider the cohomology with
integral coefficients there can be torsion and, in fact, this is what
this paper is all about. Unfortunately, a combinatorial formula in
terms of the fan $\Sigma(\Delta)$ for the torsion in the integral
cohomology of a toric Calabi-Yau manifold is only known in the
hypersurface case~\cite{Batyrev:2005jc}. 

We now illustrate these concepts in the example of the complete
intersection $\Xt \subset \mP_{\Delta^*}=\mP^2 \times \mP^1 \times
\mP^2$ and its mirror manifold $\Xt^*$. In eq.~\eqref{eq:primitive} we
already determined the primitive collections, hence the corresponding
Stanley-Reisner ideal is
\begin{equation} 
  \label{eq:SRI1}
  \IdealSR=\big\langle D_1D_2D_3, D_4D_5,D_6D_7D_8\big\rangle
  .
\end{equation} 
The linear equivalences are $D_1=D_2,\,D_1=D_3,\,D_4=D_5,\,
D_6=D_7,\,D_6=D_8$ and, hence, we can choose $K_1=D_4,\, K_2=D_1,\,
K_3=D_6$ as a basis for $H^2(\mP_{\Delta^*})$. In terms of this basis,
we obtain $D_{0,1}=K_1+3K_2$ and $D_{0,2}=K_1+3K_3$, see
eq.~\eqref{eq:CICY}.  Therefore, the ideal $I_{\Xt}$ in
eq.~(\ref{eq:defXt}) is
\begin{equation}
  \label{eq:idealXt} 
  I_{\Xt}=
  \left\langle
    {K_{{3}}}^{2}K_{{2}}-{K_{{2}}}^{2}K_{{3}},\,
    K_{{1}}K_{{2}}-3\,{K_{{2}}}^{2},\,
    K_{{1}}K_{{3}}-3\,{K_{{3}}}^{2},\,
    {K_{{1}}}^{2},\,{K_{{2}}}^{3},\,{K_{{3}}}^{3} 
  \right\rangle
  .
\end{equation} 
Next, we define the restriction of the $K_i$ to $\Xt$ to be the divisors
\begin{equation}
  \Jt_i =
  K_i\cdot \Xt = K_i(K_1+3K_2)(K_1+3K_3)  
  .
\end{equation}
We need to compute one of the intersection numbers directly from the
volume of a cone, say, $\Jt_1\Jt_2\Jt_3 =
K_1K_2K_3(K_1+3K_2)(K_1+3K_3) = 9K_1K_2^2K_3^2$, where we made use of
the relations in $I_{\Xt}$. Using eq.~\eqref{eq:intersect}, this
intersection can be evaluated to be
\begin{equation}
  \label{eq:normalization}
  \begin{split}
    9K_1K_2^2K_3^2 =&\;
    9D_1D_2D_4D_6D_7 = 
    9 / \Vol\big(
    \langle\rho_1,\rho_2,\rho_4,\rho_6,\rho_7\rangle
    \big) 
    \\
    =&\;
    9 / \Vol\big(\langle e_1,e_2,e_3,e_4,e_5 \rangle\big) 
    = 
    9
    .
  \end{split}
\end{equation}
Then, again using eq.~\eqref{eq:idealXt}, we see that the only
non-vanishing intersection numbers and the second Chern class are
\begin{equation}
  \label{eq:ringXt}
  \begin{gathered} 
    \Jt_2^2\Jt_3 = 3    , \quad 
    \Jt_1\Jt_2\Jt_3 = 9   , \quad
    \Jt_2\Jt_3^2  =3    ,\\
    \ch_2\big(\Xt\big) \cdot  \Jt_1 = 0  ,\quad
    \ch_2\big(\Xt\big) \cdot  \Jt_2 = 36  ,\quad
    \ch_2\big(\Xt\big) \cdot  \Jt_3 = 36
    .
  \end{gathered}
\end{equation}
Note that only $h^{1,1}_{\mathrm{toric}}(\Xt)=3$ of the
$h^{1,1}(\Xt)=19$ parameters are realized torically. Comparing the
triple intersection numbers with eq.~\eqref{eq:t1t2fcubic}, it is
clear that these $3$ toric divisors are precisely the $G$-invariant
divisors on $\Xt$.

A similar, though much more complicated, calculation can be done for
$\Xt^* \subset \mP_{\nabla^*}$. Using the results
of~\autoref{sec:nablatriangs} one can show that, among the points in
eq.~\eqref{eq:points}, the $14$ divisors $D^*_{13}, D^*_{14},
D^*_{12+6k+i+j}, D^*_{15+6k+i+j},\,k=0,2$ appear as generators of
eq.~\eqref{eq:CYideal} and, therefore, do not intersect $\Xt^*$.
Subtracting from the remaining $24$ divisors in eqns.~\eqref{eq:Nabla}
and~\eqref{eq:points} the remaining $5$ linear relations in
eq.~\eqref{eq:linrelnabla}, we find that all
$h^{1,1}_{\mathrm{toric}}(\Xt^*)=h^{1,1}(\Xt^*)=19$ moduli are
realized torically.

\subsection{Mori Cone}
\label{sec:mori-cone}

As we have just seen, the cohomology classes $D_i$ span
$H^2(\mP_{\Sigma},\mR) = H^{1,1}(\mP_{\Sigma})$. The K\"ahler classes
of a smooth projective toric variety $\mP_{\Sigma}$ form an open cone
in $H^{1,1}(\mP_{\Sigma})$ called the K\"ahler cone
$\Kcone(\mP_{\Sigma})$. This cone has a combinatorial description in
terms of the fan $\Sigma$, which we now review. 

First, define a support function to be a continuous function $\psi:
N_{\mR} \to \mR$ given on each cone $\sigma\in\Sigma$ by an
$m_{\sigma} \in M_{\mR}$ via
\begin{equation}
  \psi(\rho) = (m_\sigma,\rho)
  \quad \forall \rho\in \sigma \subset N_\R
  .
\end{equation}
A support function determines a divisor $D = \sum_i \psi(\rho_i) D_i$.
We say that $D$ is convex if $\psi$ is a convex function on $N_{\mR}$.
The convex classes form a non-empty strongly convex polyhedral cone in
$H^{1,1}(\mP_{\Sigma})$ whose interior is the \Kahler{} cone
$\Kcone(\mP_{\Sigma})$. Such a support function is strictly convex if
and only if 
\begin{equation}
  \label{eq:convex} 
  \psi(\rho_{i_1} + \dots + \rho_{i_k}) >
  \psi(\rho_{i_1}) + \dots + \psi(\rho_{i_k})
\end{equation} 
for every primitive collection $I =
\{i_1,\dots,i_k\}$~\cite{Batyrev:1994rs}.  The dual of the \Kahler{}
cone $\Kcone(\mP_{\Sigma})$ is called the Mori cone or the cone of
numerically effective curves $\NE(\mP_{\Sigma})$. Its generators
can be described by vectors $l^{(a)}$ of the corresponding linear
relations $\sum_i l^{(a)}_i \rho_i = 0$.  Each face of the \Kahler{}
cone $\Kcone(\mP_{\Sigma})$ is dual to an edge of $\NE(\mP_{\Sigma})$.
These edges are generated by curves $c^{(a)}$, and the entries of the
vector $l^{(a)}$ are
\begin{equation}
  \big( l^{(a)} \big)_i = c^{(a)} \cdot D_i
  .
\end{equation}
A practical algorithm to find the generators for $l^{(a)}$ in terms of
the triangulation $T(\Delta^*)$ is described
in~\cite{Berglund:1995gd}.
%
%
Of course, we are not interested in the ambient space but in a
complete intersection $Y \subset \mP_{\Delta^*}$. The restriction of a
\Kahler{} class on the ambient space yields a \Kahler{} class on $Y$,
but not every \Kahler{} class on $Y$ arises that way. We define the
toric part of the \Kahler{} cone on $Y$ as the restriction~\cite{Cox:1999ab}
\begin{equation}
  \Kcone(Y)_\text{toric} = 
  \Kcone(\mP_{\Sigma})\big|_Y
  ~\subset
  \Kcone(Y)
  .
\end{equation}
In the simplicial case, we can always take the basis 
$J_i$ of $H^2_\text{toric}(Y,\Q)$ to be edges
of the \Kahler{} cone. The dual of the toric \Kahler{} cone of $Y$ is
the (toric) Mori cone $\NE(Y)_\text{toric}$. This is sufficient for
mirror symmetry purposes, however, it can be larger than the actual
cone of effective curves.  Once the generators $l^{(a)}$ of
$\NE(\mP_{\Delta^*})$ are determined, we need to add the information
about the nef partition. For this purpose, we define 
\begin{equation}
  l^{(a)}_{0,m} 
  \eqdef
  -D_{0,m} \cdot c^{(a)} 
  \quad
  m=1,\dots,r
  .
\end{equation}
Finally, it is customary to write the generators of the Mori cone
$\NE(Y)_\text{toric}$ as
\begin{equation} 
  \label{eq:genmori}
  l^{(a)}=
  \big(
  l^{(a)}_{0,1},\ldots,l^{(a)}_{0,r}; 
  l^{(a)}_1,\ldots,l^{(a)}_n
  \big),
\end{equation} 
which are, by abuse of notation, again denoted by $l^{(a)}$. The
knowledge of the (toric) Mori cone is important for several reasons.
It defines the local coordinates on the complex structure moduli space
of the mirror $Y^*$ near the point of maximal unipotent monodromy.
Moreover, the generators enter the coefficients of the fundamental
period which is a solution of the Picard-Fuchs equations as we
will review in~\autoref{sec:b-model-prepotential}.

For example, using the unique primitive collections in
eq.~\eqref{eq:primitive}, the Mori cone for $\mP_{\Delta^*}$ is
generated\footnote{We sort the Mori cone generators such that the
  first one corresponds to the $\CP^1$ of the ambient space, and the
  second and third generator are the hyperplane sections of the two
  $\CP^2$. In other words, we have $\Jt_a\cdot c^{(b)} = \delta_a^b$. 
  This is the basis of curves that we used for the A-model
  computation.} by
\begin{equation}
  \label{eq:MoriDelta}
  \renewcommand{\arraystretch}{1.3}
  \begin{array}{r@{(}r@{,~}r@{,~}r@{,~}r@{,~}r@{,~}r@{,~}r@{,~}r@{)}l} 
    l^{(1)}= &0&0&0&1&1&0&0&0  \\
    l^{(2)}= &1&1&1&0&0&0&0&0  \\ 
    l^{(3)}= &0&0&0&0&0&1&1&1  &.
  \end{array} 
\end{equation}
Recalling the nef partition
$D_{0,1}=D_1+\dots+D_4,\,D_{0,2}=D_5+\dots+D_8$, we prepend
$(-D_{0,1}\cdot c^{(a)},-D_{0,2}\cdot c^{(a)}) = (-3,0),\, (-1,-1),\, (0,-3)$,
$a=1,2,3$, to obtain the generators 
\begin{equation}
  \label{eq:MoriXt}
  \renewcommand{\arraystretch}{1.3}
  \begin{array}{r@{(}r@{,}r@{;~}
      r@{,~}r@{,~}r@{,~}r@{,~}r@{,~}r@{,~}r@{,~}r@{)}l} 
    l^{(1)}= & -1&-1&0&0&0&1&1&0&0&0   \\
    l^{(2)}= & -3& 0&1&1&1&0&0&0&0&0   \\
    l^{(3)}= &  0&-3&0&0&0&0&0&1&1&1   &
  \end{array}
\end{equation}
of the Mori cone $\NE(\Xt)_{\text{toric}}$. Due to the large number of 
toric moduli, the calculation for the Mori cone $\NE(\mP_{\nabla^*})$ 
of the ambient toric variety of the mirror $\Xt^*$ is much more complex.

\subsection{B-Model Prepotential}
\label{sec:b-model-prepotential}

Mirror symmetry identifies the quantum corrected K\"ahler moduli space
of $Y$ with the classical complex structure moduli space of $Y^*$, see
the excellent treatise in~\cite{Cox:1999ab} for details. The
deformations of the complex structure of $Y^*$ are encoded in the
periods $\varpi=\int_{\gamma} \Omega$ and the latter can be computed
from the equations $F^*_l$ that cut out $Y^*\subset \mP_{\nabla^*}$.
Given the Mori cone eq.~\eqref{eq:genmori} and the classical
intersections numbers $\kappa_{abc} = J_a\cdot J_b\cdot J_c$ we
follow~\cite{Givental:1996ab, Givental:1998ab, Stienstra:1998ab,
  Cox:1999ab} to write down a local expansion of the periods,
convergent near the large complex structure point, which is
characterized by its maximal unipotent monodromy. In the following, we
will review just the bare essentials. 

The coefficients $a_i$ in the polynomial constraints $F^*_l$ of the
complete intersection $Y^*$, see eq.~\eqref{eq:CIeqs}, define the
complex structure of $Y^\ast$. A particular set of local coordinates
$u_a$ on the complex structure moduli space on $Y^*$ is defined by
\begin{equation}
  u_b=\prod_{m=1}^r a_{m,0}^{l^{(b)}_{0,m}}\prod_{i=1}^n a_i^{l^{(b)}_i}
  \quad 
  b=1,\dots,h
\end{equation}
where $h \eqdef h^{1,1}_{\mathrm{toric}}(Y)$ and $a_{m,0}$ is the coefficient
in~(\ref{eq:CIeqs}) corresponding to the origin in $\nabla_l$. 
In these coordinates,
the point of maximal unipotent monodromy is at $u_b=0$. We define the
cohomology-valued period
\begin{equation}
  \label{eq:period}
  \varpi(u,J) = \sum_{\{n_a\geq=0\}} 
  \frac{ 
    \prod\limits_{m=1}^r 
    \big(1-\sum\limits_{a=1}^{h}  
    l^{(a)}_{0,m}J_a\big)_{-\sum_{a=1}^{h}  l^{(a)}_{0,m}n_a}
  }{
    \prod\limits_{i=1}^n 
    \big(1+\sum\limits_{a=1}^{h} 
    l^{(a)}_i J_a\big)_{\sum_{a=1}^{h} l^{(a)}_i n_a}
  }
  \prod_{a=1}^h u_a^{n_a+J_a} 
  .
\end{equation}
where $(x)_n = \Gamma(x+n)/\Gamma(x)$ is the Pochhammer symbol. Note
that the choice of triangulation is implicit in the generators
$l^{(a)}$ of the Mori cone. Expanding $\varpi(u,J)$ by cohomology
degree yields
\begin{equation}
  \label{eq:periodexpansion}
  \varpi(u,J) = 
  \varpi^{(0)}(u) 
  + \sum_{a=1}^h \varpi^{(1)}_a(u)J_a 
  + \sum_{a=1}^h \varpi^{(2)}_a(u) \kappa_{abc} J_bJ_c
  - \varpi^{(3)}(u) \diff\Vol
  ,
\end{equation}
where $\diff\Vol$ is the volume form. The coefficients in
eq.~\eqref{eq:periodexpansion} are the fundamental period
$\varpi^{(0)}(u)$, that is, the unique solution to the Picard-Fuchs
equations holomorphic at $u_a=0$, and
\begin{equation}
  \begin{gathered}
    \varpi^{(1)}_a(u) = \partial_{J_a}\varpi(u,J)|_{J=0}, \qquad
    \varpi^{(2)}_a(u) = \frac{1}{2}
    \kappa_{abc} \partial_{J_b}\partial_{J_c}\varpi(u,J)|_{J=0}, \\
    \varpi^{(3)}(u) = -\frac{1}{6}
    \kappa_{abc} \partial_{J_a}\partial_{J_b}\partial_{J_c}\varpi(u,J)|_{J=0}.
  \end{gathered}
\end{equation}
These coefficients coincide with the basis of solutions of the Picard-
Fuchs equations obtained from
the Frobenius method in~\cite{Hosono:1994ax, Klemm:2004km}. The
B-model prepotential $\FprepotBY$ is
\begin{equation}
  \label{eq:BPreprot}
  \FprepotBY(u) = 
  \frac{1}{2\varpi^{(0)}(u)^2}\left(\varpi^{(0)}(u)\varpi^{(3)}(u) 
    + \sum_{a=1}^h \varpi^{(1)}_a(u)\varpi^{(2)}_a(u)  \right). 
\end{equation}
At the large complex structure point the mirror map defines natural
flat coordinates on the \Kahler{} moduli space of the original
manifold $Y$, which are
\begin{equation}
  \label{eq:mirrormap}
  t_i=\frac{\varpi_i^{(1)}(u)}{\varpi_0(u)}
  , \qquad 
  i=1,\ldots, h
  .
\end{equation}
We also define $q_j = e^{2\pi i t_j} = u_j + O(u^2)$.  One way to
obtain the prepotential is to compute its third derivatives
\begin{equation}
  \label{eq:threepoint}
  C^*_{abc} = 
  D_aD_bD_c \FprepotBY=
  \int_{Y^*} \Omega \wedge \partial_{a}\partial_{b}\partial_{c} \Omega   
  , 
\end{equation}
and apply the Picard-Fuchs operators. This leads to linear
differential equations, which determine $C^*_{abc}$ up to a common
constant, see again~\cite{Hosono:1994ax, Cox:1999ab} for details. The
quantum corrected three point function $C_{ijk}(q)$ on $Y$ follows
from $C^*_{abc}(u)$ using the inverse mirror map
eq.~\eqref{eq:mirrormap} $u=u(t)$, and one obtains
\begin{equation}
  \label{eq:Cabc}
  C_{ijk}(q)=
  \frac{1}{\varpi^{(0)}(u(q))^2}
  \frac{\partial u_a}{\partial t_i}
  \frac{\partial u_b}{\partial t_j} 
  \frac{\partial u_c}{\partial t_k}
  C^*_{abc}(u(q))
  .
\end{equation}
In practice, we use the formula
\begin{equation}
  \label{eq:Cijk2}
  C_{ijk}(q) = \partial_{t_i}\partial_{t_j} 
  \frac{\varpi^{(2)}_k(u(q))}{\varpi^{(0)}(u(q))}.
\end{equation}
Integrating three times with respect to $t_i$ yields the prepotential
$\FprepotBY(t)$ up to a polynomial of degree three in $t_i$ which can
be determined partially by the topological data of $Y$.

Mirror symmetry then ensures that the B-model prepotential,
eq.~\eqref{eq:BPreprot}, is equal to the A-model prepotential. That
is,
\begin{equation}
  \label{eq:mirror}
 \Fprepot{Y}(q)  =  \FprepotBY(u(q)) 
 .
\end{equation}
This allows us to compute the instanton numbers $n_d$. For the case of
interest,
\begin{equation}
  \Xt \in \mP_{\Delta^*} = \mP^2\times\mP^1\times\mP^2 
  ,
\end{equation}
we refer to~\cite{Hosono:1997hp} where this program been carried out
in detail. The same calculation can in principle be done on the mirror 
$X^*$, but the large number of toric moduli again makes it highly extensive.
Instead, we refer to the next section where a suitable
quotient of $\Xt^*$ will be treated in detail for which the computations 
are reasonably simple.


%% file: Bmodel-Quotient.tex
\section{Quotienting the B-Model}
\label{sec:Bquotient}

In this section we consider the quotient $X = \Xt/G$ in terms of toric
geometry and study the mirror of $X$ in this context. In order to
achieve this, we first analyze the partial quotient $\Xb=\Xt/G_1$.
Using the techniques introduced in \autoref{sec:BmodelIntro}, we
construct the mirror $\Xb^*$. Using their toric realization, we
perform the B-model computation for the non-perturbative prepotentials
$\FprepotNP{\Xb}$ and $\FprepotNP{\Xb^\ast}$, respectively. Finally,
we explain how one can implement the quotient by $G_2$ on both sides
in order to obtain $X$ and $X^*$.

\subsection
[The Quotient by $G_1$]
[The Quotient by G1]
{The Quotient by $\mathbf{G_1}$}
\label{sec:quotient-g_1}

We start with a review of the general discussion of free quotients of
complete intersections in toric geometry in~\cite{Klemm:2004km}.
Consider a fan $\Sigma \subset N_\R$ and pick a lattice refinement 
$\Bar{N}$ such that $\Gamma=\Bar{N}/N$ is a finite abelian group. Such a
lattice refinement consists of a finite sequence of lattice refinements 
of the form $N\to N+w_p\IZ$ which are described by a vector
$w_p=\tfrac{1}{k_p}\sum \alpha_{pi}\rho_i$ with $\alpha_{pi} \in \Z$.
The group $\Gamma$ is then isomorphic to $\prod_{p} \mZ_{k_p}$.
Let $\bar\Sigma$ be the fan obtained from $\Sigma$ by relating
everything to the lattice $\bar N$. In this context, we make some
additional identifications in the toric 
quotient eq.~(\ref{eq:toricvariety})~\cite{Cox:1993fz}.
One finds that $V_{\bar\Sigma}=V_{\Sigma}/\Gamma$ is the quotient of
$V_{\Sigma}$ by the finite abelian group $\Gamma$. Its action
on the homogeneous coordinates is by multiplication by phases
\begin{equation}
  \big[ z_1: \cdots :z_n\big]
  \mapsto
  \big[ \xi^{\alpha_1} z_1: \cdots : \xi^{\alpha_n} z_n\big]  
  , \qquad
  \xi = e^{\frac{2\pi\iunit}{k}},  
\end{equation}
for every cyclic subgroup of order $k$. We will denote such group 
actions by $\mZ_k: (\alpha_1,\dots,\alpha_n)$.
If $V_{\Sigma}$ is a compact toric variety, then the
quotient $V_{\bar\Sigma}$ is never free~\cite{Danilov:1978ab}.
However, a hypersurface or complete intersection in $V_{\Sigma}$ need
not intersect the set of fixed points, and in that case we get a
smooth quotient manifold with nontrivial fundamental group.

We now apply this to $\mP_{\Delta^*} = \mP^2\times\mP^1\times\mP^2$
defined in eq.~\eqref{eq:Delta}. The first step in performing the
quotient of $\mP_{\Delta^*}$ by $G_1$ thus amounts to a refinement
$\bar N=w\,\mZ+ N$ of the lattice $N$ with index $|G_1|=3$. From the
definition eq.~\eqref{eq:g1action} of the action of $G_1$ on
$\mP_{\Delta^*}$ and eq.~\eqref{eq:varchange} we read off that the
refinement is by a vector 
\begin{equation}
  w \in \frac13 \big(\rho_2+2\rho_3+\rho_7+2\rho_8 \big) + \mZ^5
  .
\end{equation}
The resulting
polytope $\bar{\Delta}^*$ admits the same nef partition as $\Delta^*$ in
eq.~\eqref{eq:nefcover},
\begin{align}
  \label{eq:nefpartial}
  \bar\nabla_1&=\langle \rhob_1,\dots,\rhob_4 , 0 \rangle, 
  & 
  \bar\nabla_2& = \langle \rhob_5,\dots,\rhob_8 , 0 \rangle
  .
\end{align}
where we express the generators $\rhob$ in terms of $\rho$ as
\begin{equation}
  \label{eq:Deltapartial}
  \begin{gathered}
    \rhob_i=\rho_i,
    \quad i=1,\dots,6\,,
    \\
    \rhob_7=\rho_7+e_1+2e_2+e_4+2e_5
    , \quad
    \rhob_8=\rho_8-e_1-2e_2-e_4-2e_5
    .
  \end{gathered}
\end{equation}
It is easy to check that the $\rhob_i$ satisfy the same linear 
relations eq.~\eqref{eq:linreldelta} as the $\rho_i$, and that
$w=\tfrac{1}{3}(\rhob_1-\rhob_2+\rhob_6-\rhob_7)=-e_2-e_5$. The
$\rhob_i$ together with $w$ therefore indeed generate the lattice 
$\bar N$. Note that, while all $8$ non-zero lattice points of 
$\bar\Delta^*$ are vertices, the dual polytope $\bar \Delta$ has $18$ 
vertices and $102$ points. Using
PALP~\cite{Kreuzer:2002uu} again, we compute the lattice points of the
polytope $\bar\nabla^*=\langle\bar\Delta_1,\bar\Delta_2\rangle\subset
M_\mR$, which will describe the ambient space of the mirror $\Xb^*$ of
$\Xb$. We find
\begin{equation}
  \label{eq:nefpartial2}
  \bar\Delta_1=\langle \nub_1,\dots,\nub_6,0\rangle
  , \quad  
  \bar\Delta_2=\langle \nub_7,\dots,\nub_{12},0\rangle
  ,
\end{equation}
where we express the vertices $\nub_i$ in terms of the vertices $\nu_i$ 
of $\nabla^*$ as
\begin{equation}
  \label{eq:Nablapartial}
  \nub_{3k+1}=\nu_{3k+1}
  ,\quad
  \nub_{3k+2}=\nu_{3k+2}-e_5
  ,\quad
  \nub_{3k+3}=\nu_{3k+3}+e_5
  ,\quad k=0,\dots,3
  .
\end{equation}
Again, it is easy to check that the $\nub_i$ satisfy the same linear
relations eq.~\eqref{eq:linrelnabla} as the $\nu_i$. It turns out that
the lattice points of $\bar\nabla^*$ generate a sublattice $\bar M$ of
index $3$ in $M$, and the lattice refinement is generated by
\begin{equation}
  w^*=\frac{1}{3}\big(\nub_1+2\nub_2+2\nub_7+\nub_8\big) 
  = e_2+e_4-e_5
  .   
\end{equation}
Among the points of $\nabla^*$ listed in eq.~\eqref{eq:points} only
$\nu_{13}$ and $\nu_{14}$ are also lattice points of the sublattice
$\bar M$. In fact, we have $\nub_{13}=\nu_{13}$ and
$\nub_{14}=\nu_{14}$. Hence, $\bar\nabla^*$ has 12 vertices and 15
lattice points; its dual $\bar\nabla=\bar\nabla_1+\bar\nabla_2$ has 42
lattice points among which 15 are vertices\footnote{\label{fn:HST}Note
  that all of our polytopes differ from the non-free
  $\mZ_3\times\mZ_3$ quotient of $\Delta^*$ defined
  in~\cite{Hosono:1997hp}, Proposition 7.1. In the notation 
  of~\cite{Klemm:2004km} their quotient is 
  \begin{equation}
    \nabla^*~\not=~
    \mP\begin{pmatrix}
      1&1&1&0&0&0&0&0\\
      0&0&0&1&1&1&0&0\\
      0&0&0&0&0&0&1&1
    \end{pmatrix}	
    \begin{bmatrix}3&0\\0&3\\1&1\end{bmatrix}   
    ~\Bigg/~
    \begin{matrix}
      \mZ_3:~0~1~2~0~0~0~0~0\\
      \mZ_3:~0~0~0~0~1~2~0~0
    \end{matrix}	
  \end{equation}
  and has $21$ points and $8$ vertices in the lattice $N$. 
}.

Once we have the polytopes $\bar\Delta^*$ and $\bar\nabla^*$, we can
construct $\Xb$ and $\Xb^*$ as complete intersections entirely analogous 
to $\Xt$ and $\Xt^*$, see \autoref{sec:BmodelIntro}. That is, using
eq.~\eqref{eq:CICY}, we define
\begin{equation}
  \label{eq:XbCICY}
  \Xb = \Db_{0,1} \cap \Db_{0,2}
  ,\quad
  \Xb^* = \Db^*_{0,1} \cap \Db^*_{0,2}
\end{equation}
in terms of the nef partitions eq.~\eqref{eq:nefpartial}
and~\eqref{eq:nefpartial2}, respectively.  Here, $\Db_i$ and $\Db^*_i$
denote the divisors associated to the generators $\rhob_i$ and
$\nub_i$, respectively.  The absence of fixed points of the $G_1$
action on the complete intersection $\Xt$ is guaranteed by the fact
that the resulting polytope $\bar\Delta^*\subset \bar N_\mR$ has no
additional lattice points~\cite{Klemm:2004km}. Hence, $\Xb = \Xt/G_1$
has a non-trivial fundamental group $\pi_1(\Xb) = \mZ_3$.
Surprisingly, it turns out that the mirror $\Xb^*$ is a free quotient
as well. To see this recall that, as noticed above, the lattice points
of $\bar\nabla^*$ generate a sublattice $\bar M$ of index $3$ in $M$.
Furthermore, $\bar\nabla^*$ also has no additional lattice points with
respect to $\nabla^*$. Therefore, there is a group $G_1^* \simeq
\mZ_3$ acting torically on $\mP_{\nabla^*}$. On the homogeneous
coordinates this action is
\begin{equation}
  \label{eq:g1*action}
  g_1^*: 
  \big[z_1:\dots:z_{12}\big] 
  \mapsto 
  \big[
  \zeta z_1: \zeta^2 z_2: z_3:\cdots:z_6:
  \zeta^2 z_7: \zeta z_8: z_9:\cdots:z_{12}
  \big]
  .
\end{equation}
Hence, $\Xb^* = \Xt^*/G_1^*$ also has a non-trivial fundamental group
$\pi_1(\Xb^*)=\mZ_3$. Note that this never happens for hypersurfaces
in toric varieties~\cite{Batyrev:2005jc}. Having the toric
representation of $\Xb$ and $\Xb^\ast$, we can now compute their Hodge
numbers. It turns out that
\begin{equation}
  \label{eq:HodgeXb}
  h^{1,1}\big(\Xb\big)=
  h^{1,2}\big(\Xb\big)=
  h^{1,1}\big(\Xb^*\big)=
  h^{1,2}\big(\Xb^*\big)=
  7
  ,
\end{equation}
in agreement with \partA~\cite{Braun:2007xh},
eq.~\xeqref{eq:HomCohXtZ3}.

\subsection
[The Quotient by $G_2$]
[The Quotient by G2]
{The Quotient by $\mathbf{G_2}$}
\label{sec:quotient-g_2}

We now turn to the $G_2$ action, which does not act torically. Hence,
we cannot, in principle, find a toric variety containing $X=\Xb/G_2$
as we did for the $G_1$ quotient above. However, at least we have to
ensure that $\Xb$ and $\Xb^\ast$ are $G_2$-symmetric. This can be
achieved via suitable symmetries in the toric data. 

The easy part of the toric data for $\Xb$ is the polytope
$\bar\Delta^\ast$. The $G_2$ action on the ambient space permutes the
homogeneous coordinates, see eq.~\eqref{eq:g2action}. In terms of
toric geometry, this means that it permutes the corresponding points
of the polytope. That is\footnote{We define the modulus operation such
  that $(i\tmod3) \in \{0,1,2\}$.},
\begin{equation}
  \label{eq:permZ3}
  \begin{split}
    g_2:&~ \rhob_{i} \mapsto \rhob_{1+(i\tmod3)}
    \quad 
    \forall 
    i\in\{1,2,3\}
    ,\\
    g_2:&~ \rhob_4 \mapsto \rhob_4
    ,\quad
    \rhob_5 \mapsto \rhob_5
    ,\\
    g_2:&~ \rhob_{5+i} \mapsto \rhob_{6+(i\tmod3)}
    \quad 
    \forall 
    i\in\{1,2,3\}
    .
  \end{split}
\end{equation}
It induces a mirror group action $G_2^*$ on $\Xb^*$ which is
geometrical, rather than a quantum symmetry as discussed
in~\cite{Kreuzer:1994uc}. The action of $G_2^*$ is obviously the dual
group action on the dual lattice $M$, which again must be a symmetry
of the relevant polytope $\bar\nabla^*$. We find that
\begin{equation}
  \label{eq:mirrorZ3}
  g_2^* : \nub_{3k+i} \mapsto \nub_{3k+1+(i\tmod 3)}
  \quad 
  \forall k=0,\dots,3
  ,\;
  i\in\{1,2,3\}
  .
\end{equation}
As a check on the mirror group action, note that the matrix of
scalar products, see eq.~\eqref{eq:exponents} below, is
invariant. That is,
\begin{equation}
  \big< g_2(\rhob_l) ,~ g_2^*(\nub_{l'}) \big>
  = 
  \big< \rhob_l ,~ \nub_{l'} \big>
  \quad
  \forall\; l,\, l'
  .
\end{equation}
By abuse of notation, we denote the corresponding cyclic permutation
of homogeneous coordinates by $g_2^\ast$ as well. Using this action,
we define the mirror of $X$ to be $X^*=\Xb^*/G_2^*$. This idea has
already been used for the construction of mirrors of orbifolds of the
quintic~\cite{Aspinwall:1990xe} soon after the discovery of the first
mirror construction by Greene and Plesser.

Following eq.~\eqref{eq:CIeqs}, the equations for the Calabi-Yau
complete intersections $\Xb$ and $\Xb^*$ are defined by evaluating the
matrix of scalar products $\langle \rhob_i,\nub_j \rangle +
\delta_{l\,l'}$, which are
\begin{equation}
  \label{eq:exponents}
  \begin{array}{c|ccccccc|ccccccccc}
    \langle\,,\rangle+\delta_{l\,l'}     & 
    \nub_1&\nub_2&\nub_3&\nub_4&\nub_5&\nub_6&\nub_{13}&
    \nub_7&\nub_8&\nub_9&\nub_{10}&\nub_{11}&\nub_{12}&\nub_{14}\\[2pt]
    \hline
    \rhob_1 & 3 & 0 & 0 & 3 & 0 & 0 & 1 &0&0&0&0&0&0&0\\
    \rhob_2 & 0 & 3 & 0 & 0 & 3 & 0 & 1 &0&0&0&0&0&0&0\\
    \rhob_3 & 0 & 0 & 3 & 0 & 0 & 3 & 1 &0&0&0&0&0&0&0\\
    \rhob_4 & 1 & 1 & 1 & 0 & 0 & 0 & 0 &0&0&0&1&1&1&1\\
    \hline
    \rhob_5 &0&0&0&1&1&1&1&  1 & 1 & 1 & 0 & 0 & 0 & 0\\
    \rhob_6 &0&0&0&0&0&0&0&  3 & 0 & 0 & 3 & 0 & 0 & 1\\
    \rhob_7 &0&0&0&0&0&0&0&  0 & 3 & 0 & 0 & 3 & 0 & 1\\
    \rhob_8 &0&0&0&0&0&0&0&  0 & 0 & 3 & 0 & 0 & 3 & 1
  \end{array}
\end{equation}
The equations of $X$ can now be read off from the columns of
eq.~\eqref{eq:exponents}, and one finds
\begin{subequations}
  \label{eq:F12}
  \begin{align}
    F_1=&\;
    (\lambda_5 t_0+\lambda_6 t_1)(x_0^3+x_1^3+x_2^3)+
    (\lambda_7 t_0+\lambda_8 t_1)x_0x_1x_2
    ,
    \\
    F_2=&\;
    (\lambda_1 t_0+\lambda_4 t_1)(y_0^3+y_1^3+y_2^3)+
    (\lambda_2 t_0+\lambda_3 t_1)y_0y_1y_2
    ,
  \end{align}
\end{subequations}
where the $G_2$-symmetry has been imposed. Note that the last monomial
in each equation corresponds to the vector $0\in\bar\Delta_l,\,
l=1,2$.  Two of the eight coefficients $\lambda_m$ can be fixed by
normalizing the equations, say $\lambda_4=\lambda_5=1$, and three
correspond to the symmetries of $\mP^1$, that is, $SL(2)$
transformations of $[t_0:t_1]$. Hence, we can, for example, set
$\lambda_6=\lambda_7=\lambda_8 = 0$.  This leaves us with $3$ complex
structure deformations $\lambda_1$, $\lambda_2$, and $\lambda_3$, see
eqns.~\eqref{eq:B1} and~\eqref{eq:B2}.

The equations defining $X^*$ correspond to the rows of
eq.~\eqref{eq:exponents}, that is,
\begin{subequations}
  \label{eq:F12mirror}
  \begin{align}
    F_1^*&=a_1 (z_1^3z_4^3+z_2^3z_5^3+z_3^3z_6^3)z_{13} 			
    + (a_2 z_{10}z_{11}z_{12}z_{14}+ a_3
    z_4z_5z_6z_{13})z_1z_2z_3,
    \\
    F_2^*&=a_4 (z_7^3z_{10}^3+z_8^3z_{11}^3+z_9^3z_{12}^3)z_{14}+
    (a_5 z_4z_5z_6z_{13}+a_6 z_{10}z_{11}z_{12}z_{14})z_7z_8z_9,
  \end{align}
\end{subequations}
where, again, invariance under $G_2^\ast$ has been imposed and the
last monomial of each equation comes from the lattice point
$0\in\bar\nabla_l,\,, l=1,2$. Both equations are homogeneous with
respect to all seven scaling degrees that follow from the linear
relations eq.~\eqref{eq:linrelnabla}. Among the twelve scalings of the
coordinates $z_i$, six are compatible with the cyclic permutations
$g_2^\ast$, see eq.~\eqref{eq:mirrorZ3}. Subtracting the three $G_2$
symmetric independent scalings among the relations
eq.~\eqref{eq:linrelnabla}, there remains one torus action that acts
effectively on the parameters plus two normalizations of the
equations. As expected, the six parameters $a_m$ of the equations of
$X^*$ thus become the $3$ complex structure moduli.

So far, we only considered the polytopes $\bar\Delta^\ast$ and
$\bar\nabla^\ast$. However, this is only part of the toric data
defining the manifolds $\Xb$ and $\Xb^\ast$, respectively.  In
addition, we need the triangulations and the corresponding exceptional
sets. A change in the triangulation corresponds to a flop of the toric
variety. The very real danger is that not all, and perhaps none, of
the flopped Calabi-Yau manifolds are $G_2$-symmetric. For $\Xb\subset
\mP_{\bar\Delta^*}$ this turns out to be unproblematic, but for
$\Xb^\ast\subset \mP_{\bar\nabla^*}$ we will find a condition for the
choice of a triangulation.

\subsection
[B-Model on $\Xb$]
[B-Model on Xbar]
{B-Model on $\mathbf{\Xb}$}
\label{sec:BmodelXb}

We now return to the discussion of the triangulations and the
intersection ring of $\Xb$. The analogous, but technically much more
involved discussion of $\Xb^*$ will be presented in
\autoref{sec:BmodelXbmirror}.

For $\Xb$ everything is straightforward since the $G_1$-quotient did
not introduce additional lattice points in the associated polytope
$\bar\Delta^\ast$. Therefore, just like for the polytope $\Delta^\ast$ of
the covering space $\Xt$, there exists a unique triangulation. In
particular the primitive collections, the Stanley-Reisner ideal, and
the ideal $I_{\Xb}$ are identical to the ones in
eqns.~\eqref{eq:primitive}, \eqref{eq:SRI1}, and~\eqref{eq:idealXt}
since they are derived from the same triangulation. Moreover, one can
easily see that this triangulation is $G_2$-invariant and, hence,
$\Xb$ is $G_2$ symmetric.

The only change is in the normalization of the intersection ring in
eq.~\eqref{eq:normalization}, since the total volume has to be divided
by $3=|G_1|$. This can also be seen in eq.~\eqref{eq:Deltapartial},
where the volume of the cone is now $3$ instead of $1$. Hence, on
$\Xb$ the intersection ring and the second Chern class are
\begin{equation}
  \label{eq:ringXb}
  \begin{gathered} 
    \Jb_2^2\Jb_3 = 1    , \quad 
    \Jb_1\Jb_2\Jb_3 = 3   , \quad
    \Jb_2\Jb_3^2  =1    ,\\
    \ch_2\big(\Xb\big) \cdot \Jb_1 = 0  ,\quad
    \ch_2\big(\Xb\big) \cdot \Jb_2 = 12   ,\quad
    \ch_2\big(\Xb\big) \cdot \Jb_3 = 12
    .
  \end{gathered}
\end{equation}
Comparing these intersection numbers with eq.~\eqref{eq:t1t2fcubic},
it is clear that the toric divisors should be identified with the
$G_1$-invariant divisors on $\Xb$ as
\begin{equation}
  \Jb_1 = \phi
  ,\quad
  \Jb_2 = \tau_1
  ,\quad
  \Jb_3 = \tau_2
  .
\end{equation}
The curves spanning the Mori cone on the cover turn out to be
$G_1$-invariant as well. Therefore, the Mori cones
$\NE(\mP_{\bar\Delta^*})$ and $\NE(\Xb)_{\text{toric}}$ are identical
to those in eqns.~\eqref{eq:MoriDelta} and~\eqref{eq:MoriXt},
respectively.

Following the steps given in~\autoref{sec:BmodelIntro} we now want to
compute the B-model prepotential $\FprepotentialB_{\Xb^*,0}$, plug in
the mirror map, and obtain the prepotential on $\Xb$
\begin{equation}
  \FprepotNP{\Xb} (P,Q_1,Q_2,Q_3,R_1,R_2,R_3,b_1)
  . 
\end{equation}
We immediately realize the following two caveats:
\begin{itemize}
\item We do not know how to incorporate the torsion curves
  $H_2(\Xb,\Z)_\tors = \Z_3$ into the toric mirror symmetry
  calculation.
\item Of the $7$ \Kahler{} classes on $\Xb$, only $3$ are toric.
\end{itemize}
This means that only $3$ out of the $7+1$ variables in the
prepotential are accessible, and the remaining ones are set to one.
Looking at the intersection numbers eq.~\eqref{eq:ringXb}, it is clear
that the $3$ divisors are precisely the $G_2$-invariant divisors on
$\Xb$, see eq.~\eqref{eq:t1t2fcubic}. Therefore, these $3$ variables
must be those that map to the variables $p$, $q$, and $r$ on $X$.  By
comparing with eq.~\eqref{eq:Xb2X}, we see that the corresponding
variables on $\Xb$ are $P$, $Q_1$, and $R_1$. Hence, we actually only
compute
\begin{equation}
  \label{eq:FXbpart}
  \FprepotNP{\Xb} (P,Q_1,1,1,R_1,1,1,1)
  = 
  \sum_{n_1,n_2,n_3}
  n^{\Xb}_{(n_1,n_2,n_3)} 
  \Li_3\big( P^{n_1} Q_1^{n_2} R_1^{n_3} \big)
  . 
\end{equation}
In effect, this means that the resulting instanton numbers are not
just the instantons in a single integral homology class, but the
instanton numbers in a whole set of integral homology classes. The
instanton numbers sum over all curve classes that cannot be
distinguished by $P,Q_1,R_1\in \Hom\big(H_2(\Xb,\Z),\Cunits\big)$. Up
to total degree $4$ and the symmetry
\begin{equation}
  n^\Xb_{(n_1,n_2,n_3)} = n^\Xb_{(n_1,n_3,n_2)}
  ,
\end{equation}
the resulting instanton numbers are
\begin{equation}
  \label{eq:instantonsXb}
  \begin{aligned}
     n^\Xb_{(1,0,0)}
    & = 27
    &n^\Xb_{(1,0,1)}
    & = 108
    &n^\Xb_{(1,0,2)}
    & =378
    &n^\Xb_{(1,0,3)}
    & =1080
    \\
     n^\Xb_{(1,1,1)}
    & =432
    &n^\Xb_{(1,1,2)}
    & =1512
    &n^\Xb_{(2,0,1)}
    & =-54
    &n^\Xb_{(2,0,2)}
    & =-756
    \\
     n^\Xb_{(2,1,1)}
    & =864
    &n^\Xb_{(3,0,1)}
    & =9
    .
  \end{aligned}
\end{equation}


\subsection
[Instanton Numbers of $X$]
[Instanton Numbers of X]
{Instanton Numbers of $\mathbf{X}$}
\label{sec:XInst}

Knowing the prepotential on $\Xb$, we now want to divide out the free
$G_2$ action and arrive at the prepotential on $X$. Since we do not
know the complete expansion but only eq.~\eqref{eq:FXbpart}, we have
to set $b_1=b_2=1$ in the descent equation~\eqref{eq:Xb2X}. This
yields
\begin{equation}
  \begin{split}
    \FprepotXNP\big(p,q,r,1,1) 
    =&\;
    \frac{1}{3}    
    \FprepotNP{\Xb}\big(
    \begin{array}[t]{l}
      p ,\,
      q ,\,
      1 ,\,
      1 ,\,
      r ,\,
      1 ,\,
      1 ,\,
      1
      \big)
    \end{array}
    \\ 
    =&\;
    \sum_{n_1,n_2,n_3}
    n^X_{(n_1,n_2,n_3)} 
    \Li_3\big( p^{n_1} q^{n_2} r^{n_3} \big)
    . 
  \end{split}
\end{equation}
Up to the symmetry $n^X_{(n_1,n_2,n_3)}=n^X_{(n_1,n_3,n_2)}$, the
non-vanishing instanton numbers for $X$ up to total degree $5$ are
\begin{equation}
  \begin{aligned}
    \label{eq:instantonsX}
    n^X_{(1,0,0)}
    & =9
    &n^X_{(1,0,1)}
    & =36
    &n^X_{(1,0,2)}
    & =126
    &n^X_{(1,0,3)}
    & =360
    \\
    n^X_{(1,0,4)}
    & =945
    &n^X_{(1,1,1)}
    & =144
    &n^X_{(1,1,2)}
    & =504
    &n^X_{(1,1,3)}
    & =1440
    \\
    n^X_{(1,2,2)}
    & =1764
    &n^X_{(2,0,1)}
    & =-18
    &n^X_{(2,0,2)}
    & =-252
    &n^X_{(2,0,3)}
    & =-1728
    \\
    n^X_{(2,1,1)}
    & =288
    &n^X_{(2,1,2)}
    & =3960
    &n^X_{(3,0,1)}
    & =3
    &n^X_{(3,0,2)}
    & =252
    \\
    n^X_{(3,1,1)}
    & =756
    ,
\end{aligned}  
\end{equation}
Unfortunately, this direct calculation misses the torsion information
and only yields the expansion $\FprepotXNP(p,q,r,1,1)$. The $b_1$
dependence was lost because the toric methods do not yield this part,
and the $b_2$ dependence was lost because the relevant divisor on
$\Xb$ was not toric. Comparing with the full expansion of the
prepotential
\begin{equation}
  \label{eq:FXexpansion}
  \FprepotXNP\big(p,q,r,b_1,b_2) 
  =
  \sum_{\substack{ n_1,n_2,n_3 \\ m_1,m_2 }}
  n^X_{(n_1,n_2,n_3,m_1,m_2)} 
  \Li_3\big( p^{n_1} q^{n_2} r^{n_3} b_1^{m_1} b_2^{m_2} \big)
  ,
\end{equation}
see \partA{} eq.~\xeqref{eq:XTorsPrepotExpansion}, this means we only
obtain the sum of the instanton numbers over all torsion classes
\begin{equation}
  \label{eq:nXsumtorsion}
  n^X_{(n_1,n_2,n_3)} 
  =
  \sum_{m_1,m_2=0}^2  
  n^X_{(n_1,n_2,n_3,m_1,m_2)}
  .
\end{equation}
Clearly, this destroys the torsion information, that is, the instanton
numbers $n^X_{(n_1,n_2,n_3)}$ do not depend on the torsion part of the
integral homology. For comparison purposes, we list the instanton
numbers $n^X_{(n_1,n_2,n_3)}$ for $0\leq n_1,n_2,n_3\leq 5$ in
\autoref{tab:n1n2n3Inst}.
\begin{sidewaystable}[htpb]
  \centering
  \renewcommand{\arraystretch}{1.1}
  \newcommand{\s}{\scriptscriptstyle}
  \newcommand{\st}{\scriptstyle}
  \newcommand{\sss}{\hspace{3mm}}
  \begin{tabular}{|@{\sss}c@{\sss}|@{\sss}c@{\sss}|}
    \hline
    & \\[-1em]
    $n^X_{(0,n_2,n_3)}$ &
    $n^X_{(3,n_2,n_3)}$
    \\
    \begin{tabular}{c|cccccc}
      \backslashbox{$\mathrlap{n_2}$}{$\mathclap{n_3~}$}
      &
      $0$ & $1$ & $2$ & $3$ & $4$ & $5$
      \\ \hline
      $0$ &
      $0$&$0$&$0$&$0$&$0$&$0$
      \\
      $1$ &
      $0$&$0$&$0$&$0$&$0$&$0$
      \\
      $2$ &
      $0$&$0$&$0$&$0$&$0$&$0$
      \\
      $3$ &
      $0$&$0$&$0$&$0$&$0$&$0$
      \\
      $4$ &
      $0$&$0$&$0$&$0$&$0$&$0$
      \\
      $5$ &
      $0$&$0$&$0$&$0$&$0$&$0$
    \end{tabular}
    &
    \begin{tabular}{c|cccccc}
      \backslashbox{$\mathrlap{n_2}$}{$\mathclap{n_3~}$}
      &
      $0$ & $1$ & $2$ & $3$ & $4$ & $5$
      \\ \hline
      $0$ &
      $0$&$3$&$252$&$4158$&$\st40173$&$\st287415$
      \\
      $1$ &
      $3$&$756$&$\st15390$&$\st164280$&$\s1259685$&$\s7763364$
      \\
      $2$ &
      $252$&$\st15390$&$\st426708$&$\s5427684$&$\s46537092$&$\s310465062$
      \\
      $3$ &
      $4158$&$\st164280$&$\s5427684$&$\s73971360$&$\s657552966$&$\s4487097816$
      \\
      $4$ &
      $\st40173$&$\s1259685$&$\s46537092$&$\s657552966$&$\s5948103483$&$\s41016575313$
      \\
      $5$ &
      $\st287415$&$\s7763364$&$\s310465062$&$\s4487097816$&$\s41016575313$&$\s284581389204$
    \end{tabular}
    \\
    & \\[-1em] \hline
    & \\[-1em] 
    $n^X_{(1,n_2,n_3)}$ &
    $n^X_{(4,n_2,n_3)}$
    \\
    \begin{tabular}{c|cccccc}
      \backslashbox{$\mathrlap{n_2}$}{$\mathclap{n_3~}$}
      &
      $0$ & $1$ & $2$ & $3$ & $4$ & $5$
      \\ \hline
      $0$ &
      $9$&$36$&$126$&$360$&$945$&$2268$
      \\
      $1$ &
      $36$&$144$&$504$&$1440$&$3780$&$9072$
      \\
      $2$ &
      $126$&$504$&$1764$&$5040$&$\st13230$&$\st31752$
      \\
      $3$ &
      $360$&$1440$&$5040$&$\st14400$&$\st37800$&$\st90720$
      \\
      $4$ &
      $945$&$3780$&$\st13230$&$\st37800$&$\st99225$&$\s238140$
      \\
      $5$ &
      $2268$&$9072$&$\st31752$&$\st90720$&$\s238140$&$\s571536$
      \\
    \end{tabular}
    &
    \begin{tabular}{c|c@{~~}c@{~~}c@{~~}c@{~~}c@{~~}c}
      \backslashbox{$\mathrlap{n_2}$}{$\mathclap{n_3~}$}
      &
      $0$ & $1$ & $2$ & $3$ & $4$ & $5$
      \\ \hline
      $0$ &
      $0$&$0$&$-144$&$\st-6048$&$\s-107280$&$\s-1235520$
      \\
      $1$ &
      $0$&$-306$&$\st-12348$&$\s-207000$&$\s-2273400$&$\s-19066500$
      \\
      $2$ &
      $-144$&$\st-12348$&$\st348480$&$\s14609520$&$\s235219680$&$\s2505155400$
      \\
      $3$ &
      $\st-6048$&$\s-207000$&$\s14609520$&$\s520226784$&$\s8245864800$&$\s87989812560$
      \\
      $4$ &
      $\s-107280$&$\s-2273400$&$\s235219680$&$\s8245864800$&$\s131759049600$&$\s1417949658000$
      \\
      $5$ &
      $\s-1235520$&$\s-19066500$&$\s2505155400$&$\s87989812560$&$\s1417949658000$&$\s15365394415800$
    \end{tabular}
    \\
    & \\[-1em] \hline
    & \\[-1em] 
    $n^X_{(2,n_2,n_3)}$ &
    $n^X_{(5,n_2,n_3)}$
    \\
    \begin{tabular}{c|c@{~~}c@{~~}c@{~~}c@{~~}c@{~~}c}
      \backslashbox{$\mathrlap{n_2}$}{$\mathclap{n_3~}$}
      &
      $0$ & $1$ & $2$ & $3$ & $4$ & $5$
      \\ \hline
      $0$ &
      $0$&$-18$&$\st-252$&$\s-1728$&$\s-9000$&$\s-38808$
      \\
      $1$ &
      $-18$&$288$&$\st3960$&$\s27648$&$\s143748$&$\s620928$
      \\
      $2$ &
      $\st-252$&$\st3960$&$\s54432$&$\s380160$&$\s1976472$&$\s8537760$
      \\
      $3$ &
      $\s-1728$&$\s27648$&$\s380160$&$\s2654208$&$\s13799808$&$\s59609088$
      \\
      $4$ &
      $\s-9000$&$\s143748$&$\s1976472$&$\s13799808$&$\s71748000$&$\s309920688$
      \\
      $5$ &
      $\s-38808$&$\s620928$&$\s8537760$&$\s59609088$&$\s309920688$&$\s1338720768$
    \end{tabular}
    &
    \begin{tabular}{c|c@{~~}c@{~~}c@{~~}c@{~~}c@{~~}c}
      \backslashbox{$\mathrlap{n_2}$}{$\mathclap{n_3~}$}
      &
      $0$ & $1$ & $2$ & $3$ & $4$ & $5$
      \\ \hline
      $0$ &
      $0$&$0$&$45$&$\st5670$&$\st189990$&$\s3508920$
      \\
      $1$ &
      $0$&$36$&$\st13140$&$\st474840$&$\s8793648$&$\s111499020$
      \\
      $2$ &
      $45$&$\st13140$&$\s1112886$&$\s38961252$&$\s777759975$&$\s10723515300$
      \\
      $3$ &
      $\st5670$&$\st474840$&$\s38961252$&$\s1952428464$&$\s47357606430$&$\s732897531720$
      \\
      $4$ &
      $\st189990$&$\s8793648$&$\s777759975$&$\s47357606430$&$\s1237373786439$&$\s19911043749420$
      \\
      $5$ &
      $\s3508920$&$\s111499020$&$\s10723515300$&$\s732897531720$&$\s19911043749420$&$\s327006066948660$
    \end{tabular}
    \\[-1em] &
    \\\hline    
  \end{tabular}
  \caption{Summed instanton numbers
    $n^X_{(n_1,n_2,n_3)}=\sum_{m_1,m_2} n^X_{(n_1,n_2,n_3,m_1,m_2)}$
    (hence not distinguishing torsion) computed by mirror symmetry.
    The table contains all non-vanishing instanton numbers for $0\leq
    n_1,n_2,n_3\leq 6$.}
  \label{tab:n1n2n3Inst}
\end{sidewaystable}

\subsection
[B-Model on $\Xb^*$]
[B-Model on the Mirror of Xbar]
{B-Model on $\mathbf{\Xb^*}$}
\label{sec:BmodelXbmirror}

We now study the mirror $\Xb^\ast$, which sits in a more
complicated ambient toric variety. 
Consequently, the analysis
is more involved. 
The big advantage, however, will turn out to be that all
$h^{11}(\Xb^\ast)=7$ \Kahler{} moduli are toric, which will enable us
to obtain the full instanton expansion.

Since the polytope $\bar\nabla^*$ in eq.~\eqref{eq:Nablapartial} is
not simplicial, we have to specify a resolution of the singularities,
that is, a triangulation $T(\bar\nabla^*)$. Moreover, not any
triangulation will do, but we have to make sure that it is compatible
with the action of the permutation group $G_2^*$. While a tedious
technicality, the existence of such a resolution has to be shown in
order to establish the existence of a geometrical mirror family of
$X$. In particular, we show in \autoref{sec:nablatriangs} that there
is no projective resolution of the ambient space among the $720$
coherent star triangulations of $\bar\nabla^*$ that respects the
permutation symmetry eq.~\eqref{eq:mirrorZ3}. In other words, if one
demands $G_2^\ast$ symmetry then the ambient toric variety cannot be
chosen to be \Kahler{}, but only a complex manifold. Clearly, in that
case there is no \Kahler{} cone and the usual toric mirror symmetry
algorithm does not work. What comes to the rescue is that there are
two classes of non-symmetric projective resolutions for which the
symmetry-violating exceptional sets do not intersect $\Xb^*$. Hence
the complete intersection is $G_2$-symmetric, even though the ambient
space is not.

We conclude that the extended \Kahler{} moduli space of $\Xb^*$
contains two symmetric phases. We will denote these two classes of
triangulations by $T_{\pm}=T_{\pm}(\bar\nabla^*)$,
see~\autoref{sec:nablatriangs}. In fact, the two phases are
topologically distinct, and only the triangulation $T_+$ describes the
threefold $\Xb^\ast$ that we are interested in. In
\autoref{sec:X*flop}, we will investigate the other triangulation
$T_-$ which describes a flop of $\Xb^\ast$.

Following~\autoref{sec:intersection-ring}, given the triangulation 
$T_+$, we can determine the primitive collections. This immediately 
yields the Stanley-Reisner ideal
\begin{multline}
  \IdealSR = 
  \smash{\Big\langle} 
  \Db_{{1}}\Db_{{13}},\Db_{{2}}\Db_{{4}},\Db_{{2}}\Db_{{13}},\Db_{{3}}\Db_{{4}},
  \Db_{{3}}\Db_{{5}},\Db_{{3}}\Db_{{13}},\Db_{{4}}\Db_{{10}},\Db_{{4}}\Db_{{11}},
  \Db_{{4}}\Db_{{12}},
  \\
  D_{{4}}\Db_{{14}},
  \Db_{{5}}\Db_{{10}},\Db_{{5}}\Db_{{11}},\Db_{{5}}\Db_{{12}},\Db_{{5}}\Db_{{14}}, 
  \Db_{{6}}\Db_{{10}},\Db_{{6}}\Db_{{11}},\Db_{{6}}\Db_{{12}},
  \\
  \Db_{{6}}\Db_{{14}},
  \Db_{{13}}\Db_{{10}},
  \Db_{{13}}\Db_{{11}},\Db_{{13}}\Db_{{12}},\Db_{{13}}\Db_{{14}},
  \Db_{{7}}\Db_{{14}},\Db_{{8}}\Db_{{10}},\Db_{{8}}\Db_{{12}},
  \\
  \Db_{{8}}\Db_{{14}},
  \Db_{{9}}D_{{10}}, \Db_{{9}}\Db_{{14}},
  \Db_{{1}}\Db_{{2}}\Db_{{3}},
  \Db_{{1}}\Db_{{2}}\Db_{{6}},
  D_{{1}}\Db_{{5}}\Db_{{6}},\Db_{{4}}\Db_{{5}}\Db_{{6}},
  \\
  \Db_{{7}}\Db_{{8}}\Db_{{9}},
  \Db_{{7}}\Db_{{9}}\Db_{{11}},
  \Db_{{7}}\Db_{{11}}\Db_{{12}},\Db_{{10}}\Db_{{11}}\Db_{{12}} 
  \smash{\Big\rangle}
\end{multline}
where we dropped the superscript $*$ on $\Db$ for ease of notation. 
From this, in turn, we obtain the generators $\lb_{+}^{(a)}$ of the 
Mori cone $\NE(\mP_{\bar\nabla^*})$:
\begin{equation}
  \renewcommand{\arraystretch}{1.3}
  \begin{array}{r@{(}r@{,}r@{,}r@{,}r@{,}r@{,}r@{,}
      r@{,}r@{,}r@{,}r@{,}r@{,}r@{,}r@{,}r@{)}l} 
    \lb_{+}^{(1)}=&0&0&0&0&0&0&1&0&0&-1&0&0&0&1\\
    \lb_{+}^{(2)}=&1&0&0&-1&0&0&0&0&0&0&0&0&1&0\\
    \lb_{+}^{(3)}=&-1&1&0&1&-1&0&0&0&0&0&0&0&0&0\\
    \lb_{+}^{(4)}=&0&-1&1&0&1&-1&0&0&0&0&0&0&0&0\\
    \lb_{+}^{(5)}=&0&0&-1&0&0&1&0&-1&0&0&1&0&0&0\\
    \lb_{+}^{(6)}=&0&0&0&0&0&0&-1&0&1&1&0&-1&0&0\\
    \lb_{+}^{(7)}=&0&0&0&0&0&0&0&1&-1&0&-1&1&0&0\\
    \lb_{+}^{(8)}=&0&0&0&1&1&1&0&0&0&0&0&0&-3&0\\
    \lb_{+}^{(9)}=&0&0&0&0&0&0&0&0&0&1&1&1&0&-3&.
  \end{array}
\end{equation}
A dual basis for the generators of the K\"ahler cone
$\Kcone(\mP_{\bar\nabla^*})$ is
\begin{equation}
  \begin{split}
    \Kb_{{1}} \;&=\Db_{{13}}+2\,\Db_{{1}}-\Db_{{2}}
    -\Db_{{3}}+\Db_{{9}}+\Db_{{7}}+\Db_{{8}}+3\,\Db_{{4}}
    ,\\
    \Kb_{{2}} \;&=3\,\Db_{{1}}+\Db_{{13}}+3\,\Db_{{4}}
    ,\\
    \Kb_{{3}} \;&=\Db_{{13}}+2\,\Db_{{1}}+3\,\Db_{{4}}
    ,\\
    \Kb_{{4}} \;&=\Db_{{13}}+2\,\Db_{{1}}-\Db_{{2}}+3\,\Db_{{4}}
    ,\\
    \Kb_{{5}} \;&=\Db_{{13}}+2\,\Db_{{1}}-\Db_{{2}}-\Db_{{3}}+3\,\Db_{{4}}
    ,\\
    \Kb_{{6}} \;&=\Db_{{13}}+2\,\Db_{{1}}-\Db_{{2}}-\Db_{{3}}
    +\Db_{{9}}+\Db_{{8}}+3\,\Db_{{4}}
    ,\\
    \Kb_{{7}} \;&=\Db_{{8}}+\Db_{{13}}+2\,\Db_{{1}}-\Db_{{2}}-\Db_{{3}}+3\,\Db_{{4}}
    ,\\
    \Kb_{{8}} \;&=\Db_{{4}}+\Db_{{1}}
    ,\\
    \Kb_{{9}} \;&=\Db_{{10}}+\Db_{{7}}
    .
  \end{split}
\end{equation}
The Calabi-Yau complete intersection $\Xb^*$ is then defined by
$\Xb^*=\Kb_{{1}}\Kb_{{2}}$.  It turns out that the divisors
$\Db_{{13}}$, $\Db_{{14}}$ do not intersect $\Xb^*$. Therefore, all
\begin{equation}
  h^{1,1}_{\text{toric}}\big(\Xb^*\big) = h^{1,1}\big(\Xb^*\big)=7  
\end{equation}
\Kahler{} moduli are realized torically. Since there are two divisors
that do not intersect, finding the Mori cone is somewhat subtle.
First, we have to restrict the lattice of linear relations to the
sublattice orthogonal to these two directions. For the generators of
the toric Mori cone $\NE(\Xb^*)_{\text{toric}}$, this means that
$\lb_{+}^{(1)} \to 3\lb_{+}^{(1)}+\lb_{+}^{(9)},\,\lb_{+}^{(2)} \to
3\lb_{+}^{(2)}+\lb_{+}^{(8)}$ and that we drop
$\lb_{+}^{(8)},\,\lb_{+}^{(9)}$ as well as the entries corresponding
to intersections with $\Db_{{13}}$, $\Db_{{14}}$. In addition, we
prepend the intersection numbers with $\Db_{0,1}$ and $\Db_{0,2}$.
This yields
\begin{equation}
  \label{eq:MoriCY}
  \renewcommand{\arraystretch}{1.3}
  \begin{array}{r@{(}r@{,}r@{;}r@{,}r@{,}r@{,}r@{,}
      r@{,}r@{,}r@{,}r@{,}r@{,}r@{,}r@{,}r@{)}l} 
    \lb_{+}^{(1)}=&-3&0&0&0&0&0&0&0&3&0&0&-2&1&1\\
    \lb_{+}^{(2)}=&0&-3&3&0&0&-2&1&1&0&0&0&0&0&0\\
    \lb_{+}^{(3)}=&0&0&-1&1&0&1&-1&0&0&0&0&0&0&0\\
    \lb_{+}^{(4)}=&0&0&0&-1&1&0&1&-1&0&0&0&0&0&0\\
    \lb_{+}^{(5)}=&0&0&0&0&-1&0&0&1&0&-1&0&0&1&0\\
    \lb_{+}^{(6)}=&0&0&0&0&0&0&0&0&-1&0&1&1&0&-1\\
    \lb_{+}^{(7)}=&0&0&0&0&0&0&0&0&0&1&-1&0&-1&1&.
  \end{array}
\end{equation}
The dual basis of divisors is
\begin{equation}
  \label{eq:KaehlerCY}
  \begin{gathered}
    \Jbstar_{{1}}=\tfrac{1}{3}\,\Kb_{{1}}^2\Kb_{{2}}, \quad
    \Jbstar_{{2}}=\tfrac{1}{3}\,\Kb_{{1}}\Kb_{{2}}^2, \quad
    \Jbstar_{{5}}=\Kb_{{1}}\Kb_{{2}}\Kb_{{5}}, \\ 
    \Jbstar_{{3}}=\Kb_{{1}}\Kb_{{2}}\Kb_{{3}}, \quad 
    \Jbstar_{{4}}=\Kb_{{1}}\Kb_{{2}}\Kb_{{4}}, \\
    \Jbstar_{{6}}=\Kb_{{1}}\Kb_{{2}}\Kb_{{6}}, \quad
    \Jbstar_{{7}}=\Kb_{{1}}\Kb_{{2}}\Kb_{{7}}.
  \end{gathered}
\end{equation}
We now try to identify this basis $\Jbstar_1,\dots,\Jbstar_7$ of
divisors on $\Xb^\ast$ with the basis
$\{\phi,\tau_1,\upsilon_1,\psi_1,\tau_2,\upsilon_2, \psi_2\}$ of
divisors on $\Xb$ in eq.~\eqref{eq:H2Xb}. It turns out that there is
more than one way to identify the bases if one only wants to preserve
the triple intersection numbers. To obtain a unique answer, we also
need to identify the actions by $G_2^\ast$ and $G_2$ as well. First, the
$G_2^\ast$ action on $H^2(\CP_{\bar\nabla^\ast},\Z)$ is defined by
eq.~\eqref{eq:mirrorZ3}. Using the linear equivalence relations
\begin{equation}
  \label{eq:lineq_nablab}
  \begin{split}
    2\Db_1 - \Db_2 - \Db_3 + 2\Db_4 - \Db_5 - \Db_6 &= 0\\
    -\Db_1 + 2\Db_2 - \Db_3 - \Db_4 + 2\Db_5 - \Db_6 &= 0\\
    2\Db_7 - \Db_8 - \Db_9 + 2\Db_{10} - \Db_{11} - \Db_{12} &= 0\\
    -\Db_2 + \Db_3 - \Db_5 + \Db_6 -\Db_8 + \Db_9 -\Db_{11} + \Db_{12} &= 0\\
    -\Db_4 - \Db_5 - \Db_6 + \Db_{10} + \Db_{11} + \Db_{12} - \Db_{13} + \Db_{14} &= 0  
  \end{split}
\end{equation}
and the definition eq.~\eqref{eq:KaehlerCY}, one can compute the induced
group action on $H^2(\Xb^\ast,\Z)$. We find
\begin{equation}
  \label{eq:g2*action}
  g_2^\ast
  \begin{pmatrix}
    \Jbstar_1 \\ 
    \Jbstar_2 \\ 
    \Jbstar_3 \\ 
    \Jbstar_4 \\ 
    \Jbstar_5 \\ 
    \Jbstar_6 \\ 
    \Jbstar_7 \\ 
  \end{pmatrix}
  = 
  \begin{pmatrix}
    1 & 0 &  0 & 0 & 0 & 0 &  0 \\
    0 & 1 &  0 & 0 & 0 & 0 &  0 \\
    0 & 3 & -1 & 1 & 0 & 0 &  0 \\
    0 & 3 & -1 & 0 & 1 & 0 &  0 \\
    0 & 0 &  0 & 0 & 1 & 0 &  0 \\
    3 & 0 &  0 & 0 & 1 & 0 & -1 \\
    0 & 0 &  0 & 0 & 1 & 1 & -1 \\
  \end{pmatrix}
  \begin{pmatrix}
    \Jbstar_1 \\ 
    \Jbstar_2 \\ 
    \Jbstar_3 \\ 
    \Jbstar_4 \\ 
    \Jbstar_5 \\ 
    \Jbstar_6 \\ 
    \Jbstar_7 \\ 
  \end{pmatrix}
  .
\end{equation}
Second, recall that the $G_2$ action on the divisors of $\Xb^\ast$ is
\begin{equation}
  g_2
  \begin{pmatrix}
    \phi \\ 
    \tau_1 \\
    \upsilon_1 \\
    \psi_1 \\
    \tau_2 \\
    \upsilon_2 \\
    \psi_2    
  \end{pmatrix}
  = 
  \begin{pmatrix}
    1  &  0 &  0 &  0 &  0 &  0 &  0 \\
    0  &  1 &  0 &  0 &  0 &  0 &  0 \\
    1  &  3 &  0 & -1 &  0 &  0 &  0 \\
    0  &  3 &  1 & -1 &  0 &  0 &  0 \\
    0  &  0 &  0 &  0 &  1 &  0 &  0 \\
    1  &  0 &  0 &  0 &  3 &  0 & -1  \\
    0  &  0 &  0 &  0 &  3 &  1 & -1 
  \end{pmatrix}
  \begin{pmatrix}
    \phi \\ 
    \tau_1 \\
    \upsilon_1 \\
    \psi_1 \\
    \tau_2 \\
    \upsilon_2 \\
    \psi_2    
  \end{pmatrix}
  ,
\end{equation}
see \partA{} eq.~\xeqref{eq:Bg1invg2act}. 

The essentially unique\footnote{Up to the $\Hat{g}_2$ and
  $g_2^{-1}\Hat{g}_2$ symmetry.} identification of divisors on
$\Xb^\ast$ and $\Xb$ then turns out to be
\begin{equation}
  \label{eq:XbDivisorIdent}
  \begin{gathered}
    \Jbstar_1=\tau_1, \quad
    \Jbstar_2=\tau_2, \quad
    \Jbstar_3=\psi_2, \quad 
    \Jbstar_4=\upsilon_2,
    \\
    \Jbstar_5=\phi, \quad
    \Jbstar_6=3\tau_1+\upsilon_1-\psi_1= g_2(\psi_1), \quad
    \Jbstar_7=\upsilon_1
    .    
  \end{gathered}
\end{equation}
Note that we are identifying divisors on $\Xb^\ast$ with divisors on
$\Xb$ in eq.\eqref{eq:XbDivisorIdent}, something that one would
usually not do. However, in view of the anticipated self-mirror
property, $\Xb^* \cong \Xb$, this is a sensible thing to try to attempt.
And, indeed, the identification above is an isomorphism of the
intersection rings.

Regardless of this identification, we now continue to apply mirror
symmetry. First, the second Chern class is
\begin{equation}
  \label{eq:c2Xbar*}
  \begin{gathered}
    \ch_2\big(\Xb^\ast\big) \cdot \Jbstar_1 = 12 ,\quad    
    \ch_2\big(\Xb^\ast\big) \cdot \Jbstar_5 = 0  ,\quad    
    \ch_2\big(\Xb^\ast\big) \cdot \Jbstar_2 = 12 ,
    \\
    \ch_2\big(\Xb^\ast\big) \cdot \Jbstar_3 =
    \ch_2\big(\Xb^\ast\big) \cdot \Jbstar_6 = 24 ,\quad    
    \ch_2\big(\Xb^\ast\big) \cdot \Jbstar_4 =
    \ch_2\big(\Xb^\ast\big) \cdot \Jbstar_7 = 12
    .
  \end{gathered}
\end{equation}
Using this information, we now compute the B-model prepotential
\begin{equation}
  \label{eq:FBXb}
  \begin{split}
    \FprepotentialB_{\Xb,0}(q_1,\dots,q_7) 
    =&\; 
    3q_5+3q_4q_5
    +\frac{3}{8}q_5^2
    +3q_5q_7
    +3q_7q_5q_6
    +3q_4q_5q_7
    +\frac{1}{9}q_5^3
    \\ &\;
    +3q_3q_4q_5
    +\frac{3}{8}q_5^2 q_7^2
    +3q_3q_4q_5q_7
    +\frac{3}{8} q_4^2 q_5^2
    \\ &\;
    +3q_4q_3q_2q_5
    +3q_7q_5q_4q_6
    +3q_1q_5q_6q_7
    +\frac{3}{64}q_5^4
    +O(q^5).
  \end{split}
\end{equation}
Finally, we insert the mirror map and obtain the A-model prepotential
on $\Xb^\ast$. Since we already identified the bases
$\Jbstar_1,\dots,\Jbstar_7$ with the divisors on $\Xb$, we will use
the same names (but with an added $\ast$ superscript) for the
Fourier-transformed variables to expand the prepotential. With this
notation, we obtain
\begin{equation}
  \begin{split}
    \FprepotNP{\Xb^\ast}&(
    P^\ast,
    Q_1^\ast, Q_2^\ast, Q_3^\ast,
    R_1^\ast, R_2^\ast, R_3^\ast,
    1
    ) 
    =
    3 P^{\ast} + 
    \tfrac{3}{8} P^{\ast 2} +
    \tfrac{1}{9} P^{\ast 3} +
    \tfrac{3}{64} P^{\ast 4} +
    \tfrac{3}{125} P^{\ast 5} 
    \\ & +
    3 P^{\ast} Q_2^{\ast} +
    \tfrac{3}{8} P^{\ast 2} Q_2^{\ast 2} +
    3 P^{\ast} Q_2^{\ast} Q_3^{\ast} +
    3 P^{\ast} R_2^{\ast} +
    \tfrac{3}{8} P^{\ast 2} R_2^{\ast 2} +
    3 P^{\ast} R_2^{\ast} Q_2^{\ast} 
    \\ & +
    3 P^{\ast} R_2^{\ast} Q_2^{\ast} Q_3^{\ast} +
    3 P^{\ast} R_2^{\ast} R_3^{\ast} +
    3 P^{\ast} R_2^{\ast} R_3^{\ast} Q_2^{\ast} +
    3 P^{\ast} R_2^{\ast} R_3^{\ast} Q_2^{\ast} Q_3^{\ast}
    \\ & +
    3 P^{\ast} Q_1^{\ast} R_2^{\ast} R_3^{\ast}  + 
    3 P^{\ast} Q_1^{\ast} R_2^{\ast} R_3^{\ast} Q_2^{\ast} + 
    9 P^{\ast} Q_1^{\ast} R_2^{\ast} R_3^{\ast} R_3^{\ast} +
    3 P^{\ast} Q_2^{\ast} Q_3^{\ast} R_1^{\ast} 
    \\ & +
    3 P^{\ast} Q_2^{\ast} Q_3^{\ast} R_1^{\ast} R_2^{\ast} + 
    9 P^{\ast} Q_2^{\ast} Q_3^{\ast} R_1^{\ast} Q_3^{\ast}
    + \big(\text{total degree}\geq6 \big)
    ,
  \end{split}
\end{equation}
see also \partA{} eq.~\xeqref{eq:PrepotXG1fromXt}. The instanton
numbers on $\Xb^\ast$ are the expansion coefficients
\begin{multline}
  \label{eq:FXb*}
  \FprepotNP{\Xb^\ast}(
  P^\ast,
  Q_1^\ast, Q_2^\ast, Q_3^\ast,
  R_1^\ast, R_2^\ast, R_3^\ast,
  1
  )
  \\
  =
  \sum_{n_1,\dots,n_7}
  n^{\Xb^\ast}_{(n_1,n_2,n_3,n_4,n_5,n_6,n_7)} 
  \Li_3\big( 
  P^{\ast n_1} 
  Q_1^{\ast n_2} Q_2^{\ast n_3} Q_3^{\ast n_4} 
  R_1^{\ast n_5} R_2^{\ast n_6} R_3^{\ast n_7} 
  \big)
  . 
\end{multline}
We see that we almost get the complete instanton expansion
eq.~\eqref{eq:Xt2Xb}, we only miss the expansion in the $b_1^\ast$
variable which is not computed by the toric mirror symmetry algorithm.
Up to total degree $5$, the instanton numbers are
\begin{equation}
  \label{eq:instantonsXb*+}
  \begin{aligned}
    n^{\Xb^\ast}_{(1,0,0,0,0,0,0)}=&3 &\quad
    n^{\Xb^\ast}_{(1,0,0,0,0,1,0)}=&3 &\quad
    n^{\Xb^\ast}_{(1,0,0,0,0,1,1)}=&3 &\quad
    n^{\Xb^\ast}_{(1,0,1,0,0,0,0)}=&3 
    \\
    n^{\Xb^\ast}_{(1,0,1,0,0,1,0)}=&3 &
    n^{\Xb^\ast}_{(1,0,1,0,0,1,1)}=&3 &
    n^{\Xb^\ast}_{(1,0,1,1,0,0,0)}=&3 &
    n^{\Xb^\ast}_{(1,0,1,1,0,1,0)}=&3 
    \\
    n^{\Xb^\ast}_{(1,0,1,1,0,1,1)}=&3 &
    n^{\Xb^\ast}_{(1,1,0,0,0,1,2)}=&9 &
    n^{\Xb^\ast}_{(1,0,1,2,1,0,0)}=&9 &
    n^{\Xb^\ast}_{(1,0,1,1,1,1,0)}=&3 
    \\
    n^{\Xb^\ast}_{(1,1,1,0,0,1,1)}=&3 &
    n^{\Xb^\ast}_{(1,1,0,0,0,1,1)}=&3 &
    n^{\Xb^\ast}_{(1,0,1,1,1,0,0)}=&3 
    .
  \end{aligned}
\end{equation}

Finally, let us take a look at the $G_2^\ast$ action, see
eq.~\eqref{eq:mirrorZ3}. Of the $7$ generators of the toric Mori cone,
eq.~\eqref{eq:MoriCY}, only the $3$ generators $\lb_{+}^{(1)}$,
$\lb_{+}^{(2)}$ and $\lb_{+}^{(5)}$ are invariant. Not
surprisingly, the dual $G_2^\ast$-invariant divisors 
\begin{equation}
  \label{eq:JbG2*inv}
  \Jb^\ast_5 = \phi ,\quad
  \Jb^\ast_1 = \tau_1 ,\quad 
  \Jb^\ast_2 = \tau_2
\end{equation}
were identified with the $G_2$-invariant divisors on $\Xb$ in
eq.~\eqref{eq:XbDivisorIdent}. Therefore, only $3$ \Kahler{}
parameters survive to the quotient $X^\ast=\Xb^\ast/G_2^\ast$, and we
have
\begin{equation}
  \label{eq:HodgeX}
  h^{1,1}\big(X\big)=
  h^{1,2}\big(X\big)=
  h^{1,1}\big(X^*\big)=
  h^{1,2}\big(X^*\big)=3
  .
\end{equation}

\subsection
[Instanton Numbers of  $X^*$]
[Instanton Numbers of the Mirror of X]
{Instanton Numbers of $\mathbf{X^*}$}
\label{sec:InstXmirror}

Now that we have the expression eq.~\eqref{eq:FXb*} for the
prepotential on $\Xb^\ast$, we can again apply a suitable variable
substitution 
\begin{equation}
  \Big\{
  P^\ast,
  Q_1^\ast, Q_2^\ast, Q_3^\ast,
  R_1^\ast, R_2^\ast, R_3^\ast,
  b_1^\ast, b_2^\ast
  \Big\}
  \longrightarrow
  \Big\{
  p^\ast, q^\ast, r^\ast,
  b_1^\ast, b_2^\ast
  \Big\}
\end{equation}
and obtain the prepotential on the quotient
$X^\ast=\Xb^\ast/G_2^\ast$. The correct way to replace the variables
is determined by the group action on the homology and cohomology as we
explained in \partA. Having computed the $G_2^\ast$-action in
eq.~\eqref{eq:g2*action}, we determine the descent equation for the
prepotential to be\footnote{Interestingly, eq.~\eqref{eq:Xb*2X*} turns
  out to be exactly analogous to eq.~\eqref{eq:Xb2X}, even though the
  identification of divisors on $\Xb^\ast$ and $\Xb$ is not just a
  relabeling of divisors.}
\begin{equation}
  \label{eq:Xb*2X*}
  \FprepotNP{X^\ast}\big(p^\ast,q^\ast,r^\ast,b_1^\ast,b_2^\ast) 
  =
  \frac{1}{|G_2^\ast|}    
  \FprepotNP{\Xb^\ast}\big(
    p^\ast ,
    q^\ast ,
    b_2^\ast ,
    b_2^\ast ,
    r^\ast ,
    b_2^{\ast 2} ,
    b_2^{\ast 2} ,
    b_1^\ast
  \big)
  .
\end{equation}
Using the series expansion of the prepotential for $b_1^\ast=1$ on
$\Xb^\ast$ from \autoref{sec:BmodelXbmirror}, we now find that
\begin{equation}
  \label{eq:PrepotX*nob1}
  \begin{split}
    \FprepotNP{X^\ast}&(p^\ast,q^\ast,r^\ast,1,b_2^\ast)
    \\ =&\;
    \sum_{j=0}^2
    3 \times
    \Big(
    \begin{array}[t]{ll}
        \Li_3(p^\ast b_2^{\ast j})
      + 4  \Li_3(p^\ast q^\ast b_2^{\ast j})
      + 4  \Li_3(p^\ast r^\ast b_2^{\ast j})
      \\~
      + 14  \Li_3(p^\ast q^{\ast 2} b_2^{\ast j})
      + 16  \Li_3(p^\ast q^\ast r^\ast b_2^{\ast j})
      + 14  \Li_3(p^\ast r^{\ast 2} b_2^{\ast j})
      \\~
      + 40  \Li_3(p^\ast q^{\ast 3} b_2^{\ast j})
      + 56  \Li_3(p^\ast q^{\ast 2} r b_2^{\ast j})
      + 56  \Li_3(p^\ast q^\ast r^2 b_2^{\ast j})
      \\~
      + 40  \Li_3(p^\ast r^{\ast 3} b_2^{\ast j})
      + 105 \Li_3(p^\ast q^{\ast 4} b_2^{\ast j})
      + 160 \Li_3(p^\ast q^{\ast 3} r^\ast b_2^{\ast j})
      \\~
      -2  \Li_3(p^{\ast 2} q^\ast b_2^{\ast j})
      -2  \Li_3(p^{\ast 2} r^\ast b_2^{\ast j})
      \\~
      -28 \Li_3(p^{\ast 2} q^{\ast 2} b_2^{\ast j})
      +32 \Li_3(p^{\ast 2} q^\ast r^\ast b_2^{\ast j})
      -28 \Li_3(p^{\ast 2} r^{\ast 2} b_2^{\ast j})
      \hspace{1ex}\smash{\Big)}
    \end{array}
    \\ &+
    3 \Li_3(p^{\ast 3} q^\ast ) + 3 \Li_3(p^{\ast 3} r^\ast ) 
    \\ &+
    \big(\text{total }p^\ast,q^\ast,r^\ast\text{-degree }\geq 5\big)
    .
  \end{split}  
\end{equation}
The corresponding instanton numbers
\begin{equation}
  \FprepotNP{X\ast}(p^\ast,q^\ast,r^\ast,1,b_2^\ast)
  = 
  \sum_{n_1,n_2,n_3,m_2}
  n^{X^\ast}_{(n_1,n_2,n_3,m_2)}
  \Li_3\big( p^{\ast n_1} q^{\ast n_2} r^{\ast n_3} b_2^{\ast m_2} \big)
\end{equation}
are listed in \autoref{tab:n1n2n3m2mirror}.
\begin{table}[htbp]
  \centering
  \renewcommand{\arraystretch}{1.2}
  \begin{tabular}{c|ccc|c}
    $(n_1,n_2,n_3)$  & 
    $n^{X^\ast}_{(n_1,n_2,n_3,0)}$ & 
    $n^{X^\ast}_{(n_1,n_2,n_3,1)}$ & 
    $n^{X^\ast}_{(n_1,n_2,n_3,2)}$ & 
    $n^X_{(n_1,n_2,n_3)}$ \\
    \hline
    $(1,0,0)$    & $     3$ & $     3$ & $    3$ & $     9$ \\
    $(1,0,1)$    & $    12$ & $    12$ & $   12$ & $    36$ \\
    $(1,0,2)$    & $    42$ & $    42$ & $   42$ & $   126$ \\
    $(1,0,3)$    & $   120$ & $   120$ & $  120$ & $   360$ \\
    $(1,1,1)$    & $    48$ & $    48$ & $   48$ & $   144$ \\
    $(1,1,2)$    & $   168$ & $   168$ & $  168$ & $   504$ \\
    $(2,0,1)$    & $    -6$ & $    -6$ & $   -6$ & $   -18$ \\
    $(2,0,2)$    & $   -84$ & $   -84$ & $  -84$ & $  -252$ \\
    $(2,1,1)$    & $    96$ & $    96$ & $   96$ & $   288$ \\
    $(3,0,1)$    & $     3$ & $     0$ & $    0$ & $     3$ \\    
  \end{tabular}
  \caption{Instanton numbers $n^{X^\ast}_{(n_1,n_2,n_3,m_2)}$ computed
    by toric mirror symmetry. They are invariant under the exchange
    $n_2\leftrightarrow n_3$, so we only display 
    them for $n_2\leq n_3$.}
  \label{tab:n1n2n3m2mirror}
\end{table}
For comparison purposes, we list the summed instanton numbers on $X$
as well, see eq.~\eqref{eq:nXsumtorsion}. One observes that the sum
over the more refined instanton numbers on $X^*$ equals the summed
instanton number on $X$, another clue towards $X$ being self-mirror.

\subsection{Instanton Numbers Assuming The Self-Mirror Property}
\label{sec:SelfMirrorNumbers}

So far, we have alluded to $X$ being possibly self-mirror, but not
actually made use of this property. Now we are going to assume the
self-mirror property and, hence, obtain the prepotential on $X$ as
\begin{equation}
  \FprepotNP{X}(p,q,r,b_1,b_2) = 
  \FprepotNP{X^\ast}(p,q,r,b_1,b_2)
  .
\end{equation}
Note that at linear and quadratic order in $p$ we can actually recover
the $b_1$, $b_2$ expansion from the summed instanton numbers in
\autoref{sec:XInst} and the factorization which we will prove in
\autoref{sec:p3check}. 

In contrast, for the prepotential terms at order $p^3$ we have to use
the $X^\ast$ prepotential to obtain the $b_2$ expansion from
eq.~\eqref{eq:PrepotX*nob1}. Since this is based on a toric
computation on $\Xb^\ast$, we do not directly obtain the $b_1$
expansion. However, note that the fact that $g_1$ acted torically,
eq.~\eqref{eq:g1action}, and $g_2$ non-torically,
eq.~\eqref{eq:g2action}, is just a consequence of the choice of
coordinate system on $\CPambient$. By a suitable coordinate choice, we
could have made any one of the four $\Z_3$ subgroups of $G=\ZZZ$ act
torically. Therefore, any combination of $b_1, b_2$ other than
$1=b_1^0b_2^0$ has to occur in the same way in the complete series
expansion of the prepotential. We conclude that the prepotential can
only depend on $b_1$ and $b_2$ through the combinations
\begin{equation}
  1
  ,\qquad
  \sum_{i,j=0}^2 b_1^i b_2^j
  .
\end{equation}
This observation lets us recover the full $b_1$, $b_2$ expansion of
the prepotential. To summarize, we obtain
\begin{equation}
  \label{eq:PrepotX*}
  \begin{split}
    \FprepotNP{X\ast}&(p,q,r,b_1,b_2)
    \\ =&\;
    \sum_{i,j=0}^2
    \Big(
    \begin{array}[t]{ll}
        \Li_3(p b_1^i b_2^j)
      + 4  \Li_3(p q b_1^i b_2^j)
      + 4  \Li_3(p r b_1^i b_2^j)
      \\~
      + 14  \Li_3(p q^2 b_1^i b_2^j)
      + 16  \Li_3(p q r b_1^i b_2^j)
      + 14  \Li_3(p r^2 b_1^i b_2^j)
      \\~
      + 40  \Li_3(p q^3 b_1^i b_2^j)
      + 56  \Li_3(p q^2 r b_1^i b_2^j)
      + 56  \Li_3(p q r^2 b_1^i b_2^j)
      \\~
      + 40  \Li_3(p r^3 b_1^i b_2^j)
      + 105 \Li_3(p q^4 b_1^i b_2^j)
      + 160 \Li_3(p q^3 r b_1^i b_2^j)
      \\~
      + 196 \Li_3(p q^2 r^2 b_1^i b_2^j)
      + 160 \Li_3(p q r^3 b_1^i b_2^j)
      + 105 \Li_3(p r^4 b_1^i b_2^j)
      \\~
      -2 \Li_3(p^2 q b_1^i b_2^j)
      -2 \Li_3(p^2 r b_1^i b_2^j)
      -28 \Li_3(p^2 q^2 b_1^i b_2^j)
      \\~
      +32 \Li_3(p^2 q r b_1^i b_2^j)
      -28 \Li_3(p^2 r^2 b_1^i b_2^j)
      -192 \Li_3(p^2 q^3 b_1^i b_2^j)
      \\~
      +440 \Li_3(p^2 q^2 r b_1^i b_2^j)
      +440 \Li_3(p^2 q r^2 b_1^i b_2^j)
      -192 \Li_3(p^2 r^3 b_1^i b_2^j)
      \mathrlap{\hspace{1ex}\smash{\Big)}}
    \end{array}
    \\ &+
    3 \Li_3(p^3 q ) + 3 \Li_3(p^3 r) 
    \\ &+
    9 \Li_3(p^3 q^2 ) + 
    27 \sum_{i,j=0}^2 \Li_3(p^3 q^2 b_1^i b_2^j)
    \\ &+
    9 \Li_3(p^3 q^2 ) + 
    27 \sum_{i,j=0}^2 \Li_3(p^3 q^2 b_1^i b_2^j)
    \\ &+
    27 \Li_3(p^3 q r ) + 
    81 \sum_{i,j=0}^2 \Li_3(p^3 q r b_1^i b_2^j)
    \\ &+ 
    \big(\text{total }p,q,r\text{-degree }\geq 6\big)
    .
  \end{split}  
\end{equation}
Obtaining all of these terms required a computation of
$\FprepotentialB_{\Xb,0}$ in eq.~\eqref{eq:FBXb} up to total
degree $23$ in the $7$ variables, which is close to the limit of what
can be done with current desktop computers.
\begin{table}[htpb]
  \centering
  \renewcommand{\arraystretch}{1.3}
  \newcommand{\s}{\scriptstyle}
  \newcommand{\sss}{\hspace{5mm}}
  \begin{tabular}{|@{\sss}c@{\sss}|@{\sss}c@{\sss}|}
    \hline
    & \\[-1em]
    $n^X_{(1,n_2,n_3,0,0)}$ &
    $n^X_{(1,n_2,n_3,m_1,m_2)},~(m_1,m_2)\not=(0,0)$
    \\
    \begin{tabular}{c|cccccc}
      \backslashbox{$\mathrlap{n_2}$}{$\mathclap{n_3~}$}
      &
      $0$ & $1$ & $2$ & $3$ & $4$
      \\ \hline
      $0$ &
      $1$&$4$&$14$&$40$&$105$
      \\
      $1$ &
      $4$&$16$&$56$&$160$
      \\
      $2$ &
      $14$&$56$&$196$
      \\
      $3$ &
      $40$&$160$
      \\
      $4$ &
      $105$
    \end{tabular}
    &
    \begin{tabular}{c|ccccc}
      \backslashbox{$\mathrlap{n_2}$}{$\mathclap{n_3~}$}
      &
      $0$ & $1$ & $2$ & $3$ & $4$
      \\ \hline
      $0$ &
      $1$&$4$&$14$&$40$&$105$
      \\
      $1$ &
      $4$&$16$&$56$&$160$
      \\
      $2$ &
      $14$&$56$&$196$
      \\
      $3$ &
      $40$&$160$
      \\
      $4$ &
      $105$
    \end{tabular}
    \\
    & \\[-1em] \hline
    & \\[-1em] 
    $n^X_{(2,n_2,n_3,0,0)}$ &
    $n^X_{(2,n_2,n_3,m_1,m_2)},~(m_1,m_2)\not=(0,0)$
    \\
    \begin{tabular}{c|cccc}
      \backslashbox{$\mathrlap{n_2}$}{$\mathclap{n_3~}$}
      &
      $0$ & $1$ & $2$ & $3$
      \\ \hline
      $0$ &
      $0$&$-2$&$-28$&$-192$
      \\
      $1$ &
      $-2$&$32$&$440$
      \\
      $2$ &
      $-28$&$440$
      \\
      $3$ &
      $-192$
    \end{tabular}
    &
    \begin{tabular}{c|cccc}
      \backslashbox{$\mathrlap{n_2}$}{$\mathclap{n_3~}$}
      &
      $0$ & $1$ & $2$ & $3$
      \\ \hline
      $0$ &
      $0$&$-2$&$-28$&$-192$
      \\
      $1$ &
      $-2$&$32$&$440$
      \\
      $2$ &
      $-28$&$440$
      \\
      $3$ &
      $-192$
    \end{tabular}
    \\
    & \\[-1em] \hline
    & \\[-1em] 
    $n^X_{(3,n_2,n_3,0,0)}$ &
    $n^X_{(3,n_2,n_3,m_1,m_2)},~(m_1,m_2)\not=(0,0)$
    \\
    \begin{tabular}{c|ccc}
      \backslashbox{$\mathrlap{n_2}$}{$\mathclap{n_3~}$}
      &
      $0$ & $1$ & $2$ 
      \\ \hline
      $0$ &
      $0$&$\mathemph{3}$&$\mathemph{36}$
      \\
      $1$ &
      $\mathemph{3}$&$\mathemph{108}$
      \\
      $2$ &
      $\mathemph{36}$
    \end{tabular}
    &
    \begin{tabular}{c|ccc}
      \backslashbox{$\mathrlap{n_2}$}{$\mathclap{n_3~}$}
      &
      $0$ & $1$ & $2$ 
      \\ \hline
      $0$ &
      $0$&$\mathemph{0}$&$\mathemph{27}$
      \\
      $1$ &
      $\mathemph{0}$&$\mathemph{81}$
      \\
      $2$ &
      $\mathemph{27}$
    \end{tabular}
    \\[-1em] &
    \\\hline    
  \end{tabular}
  \caption{Instanton numbers $n^X_{(n_1,n_2,n_3,m_1,m_2)}$ computed 
    by mirror symmetry. The table contains all non-vanishing
    instanton numbers for $n_1+n_2+n_3\leq 5$. The entries marked in
    \textcolor{red}{\textbf{bold}} 
    depend non-trivially on the torsion part of their 
    respective homology class.}
  \label{tab:n1n2n3b1b2Inst}
\end{table}
We list the instanton numbers in \autoref{tab:n1n2n3b1b2Inst}. Observe
that the instanton numbers sometimes do depend on the torsion part of
their homology class.

\section{The Self-Mirror Property}
\label{sec:self-mirror}

When one speaks of a Calabi-Yau manifold $Y$ being self-mirror, one
has to indicate which level of invariants one is referring to. In
particular, one might think of four types of invariants that are
natural from the point of view of string theory. The weakest level is
just the Euler number. In general, exchanging complex structure and
\Kahler{} moduli changes the sign of $\chi(Y)=2 h^{11}(Y)-2h^{21}(Y)$.
Therefore, a necessary condition for $Y$ and its mirror $Y^*$ to be
equal is obviously that
\begin{equation}
  \label{eq:Euler}
  \chi(Y) = -\chi(Y^*) = 0.  
\end{equation}
This level of invariants, however, is much too crude and therefore
insufficient. A much stronger level is based on the fact that the
cohomology groups of even degree come with an integral lattice
structure and form a ring, and therefore have a product. Because of
Poincar\'e duality, that is, $H^2(Y) = H^4(Y)\spcheck$, it is
sufficient to look at $H^2(Y)$. There is a product $H^2(Y) \times
H^2(Y) \to H^2(Y)$ whose structure constants $\kappa_{ijk}$ are the
triple intersection numbers. These intersection numbers are finer
invariants than just the dimensions of the cohomology groups, and a
self-mirror Calabi-Yau threefold should satisfy 
\begin{equation}
  \label{eq:kappaijk}
  \kappa_{ijk}(Y) = \kappa_{ijk}(Y^*)
  .
\end{equation}
For simply connected threefolds with torsion-free homology a theorem
of Wall~\cite{Wall:1966ab} states that the cohomology groups with the
intersection product $\kappa_{ijk}(Y)$ together with the second Chern
class $\ch_2(Y)$ determine the diffeomorphism type of $Y$.

If, however, $Y$ and $Y^*$ have non-trivial fundamental groups then we
cannot conclude that easily that they are diffeomorphic. But the
non-trivial fundamental group is often reflected in torsion in
homology (for example if $\pi_1(Y)$ is Abelian). In that case, the
conjecture of~\cite{Batyrev:2005jc} says that for any Calabi-Yau
threefold $Z$
\begin{align}
  \label{eq:BaytrevKreuzer}
  H^3\big(Z,\mZ\big)_\tors &\simeq H^2\big(Z^*,\mZ\big)_\tors
  ,
  &
  H^2\big(Z,\mZ\big)_\tors &\simeq H^3\big(Z^*,\mZ\big)_\tors
  . 
\end{align}
Therefore, a self-mirror manifold $Y=Y^\ast$ is expected to satisfy
\begin{equation}
  \label{eq:H2H3tors}
  H^2\big(Y,\mZ\big)_\tors \simeq H^3\big(Y,\mZ\big)_\tors
  .
\end{equation}
Of the many spaces $Y$ satisfying eq.~\eqref{eq:Euler} there are only
a few which also satisfy eq.~\eqref{eq:kappaijk}.

So far we only considered classical topology, but we know that the
ring $H^2(Y)$ experiences quantum corrections when going far away from
the large volume limit.  At small volume the intersection numbers are
replaced by the three-point functions $C_{ijk}(q)$ of (topological)
conformal field theory in eq.~\eqref{eq:Cabc}. In the large volume
limit $q$ goes to zero and the $C_{ijk}(q)$ go to $\kappa_{ijk}$, as
expected.  The $C_{ijk}(q)$ are characterized by the genus zero
instanton numbers $n^{(0)}_d=n_d$. In mathematical terms, these are
resummations of the Gromow-Witten invariants of $Y$ and characterize the
symplectic structure of $Y$. This level of invariants is even stronger
than the cohomology ring, since there are examples of diffeomorphic
manifolds which have different Calabi-Yau structures, i.e. different
$n^{(0)}_d$~\cite{MR1416344, MR1390655, Klemm:2004km}. Therefore, a
self-mirror Calabi-Yau threefold $Y$ must satisfy
\begin{equation}
  \label{eq:genus0}
  n^{(0)}_d(Y) = n^{(0)}_d(Y^*)
  .
\end{equation}
One can go even further and couple the topological conformal field
theory to topological gravity and define higher genus instanton
numbers $n^{(g)}_d$, where now
\begin{equation}
  \label{eq:genusg}
  n^{(g)}_d(Y) = n^{(g)}_d(Y^*), \qquad g > 0
\end{equation}
has to hold. These invariants are very difficult to compute,
however see~\cite{Huang:2006hq, Grimm:2007tm} for recent progress. We
do not know whether they contain more information about the symplectic
structure than the genus zero invariants. In other words, there are
presently no examples known whose $n^{(g)}_d$ agree for $g=0$ but
differ for $g>0$.

Now, one can start with any $Y$ and use some method to construct the
mirror $Y^*$. Among these are the Greene-Plesser construction in
conformal field theory, or its geometric generalizations by Batyrev
and Borisov for complete intersections in toric varieties. Then, to
show that $Y$ is self-mirror one proceeds to compute the various
invariants. The simplest condition, eq.~\eqref{eq:Euler}, can directly
be checked in terms of the toric data.  This concretely means that one
starts with a mirror pair $Y$ and $Y^*$ satisfying
eq.~\eqref{eq:Euler} and checks whether eqns.~\eqref{eq:kappaijk},
\eqref{eq:H2H3tors}, \eqref{eq:genus0}, and~\eqref{eq:genusg} are
satisfied. In fact, in \autoref{sec:Bquotient} we collected a large
amount of evidence in favor of the claim that $X$ and its
Batyrev-Borisov mirror threefold $X^*$ are the same. Indeed,
eqns.~\eqref{eq:HodgeXt}, \eqref{eq:HodgeXb} and~\eqref{eq:HodgeX}
show that $\Xt$, $\Xb$, and $X$ satisfy by construction the constraint
eq.~\eqref{eq:Euler} on the Euler number.  More interestingly, by the
identifications found in eqns.~\eqref{eq:XbDivisorIdent}
and~\eqref{eq:JbG2*inv} we observed that the condition on the
intersection ring, eq.~\eqref{eq:kappaijk}, is satisfied for $\Xb$ and
$X$, respectively.  Next, eq.~\eqref{eq:instantonsX} and
\autoref{tab:n1n2n3m2mirror} show that $X$ also fulfils the
requirement eq.~\eqref{eq:genus0} on the genus zero instanton numbers.
It would be very interesting to see whether also the condition
eq.~\eqref{eq:genusg} for higher genus curves can be met.

Finally, we consider the torsion in cohomology. In \partA{}
\xautoref{sec:CoHomology} we have shown that 
\begin{equation}
  H^3\big(X,\Z\big)_\tors
  \simeq 
  H^2\big(X,\Z\big)_\tors
  \simeq 
  \Z_3\oplus\Z_3  
  ,
\end{equation}
as we expect from a self-mirror threefold. Moreover, we can actually
compute the fundamental group of the Batyrev-Borisov mirror
independently. For that, first notice that the quotient
$\Xb^*=\Xt^*/G_1^*$ is fixed-point free, see
\autoref{sec:quotient-g_2}. The mirror permutation $G_2^*$ on $\Xb^*$
acts freely as well. Therefore, both $X$ and $X^*$ are free quotients
by a group isomorphic to $\Z_3\oplus\Z_3$, thus their fundamental
groups are 
\begin{equation}
  \pi_1\big(X\big) 
  \simeq 
  \pi_1\big(X^*\big) 
  \simeq 
  \mZ_3\oplus\mZ_3
  .
\end{equation}
Moreover, on can easily show that on a proper\footnote{A proper
  Calabi-Yau threefold has holonomy group the full $SU(3)$. In
  particular, this implies that the fundamental group is finite.}
Calabi-Yau threefold $Z$ one has $H^2(Z,\Z)_\tors =
\pi_1(Z)_\text{ab}$, the Abelianization of the fundamental group.
Hence, we see that
\begin{equation}
  H^3(X,\Z)_\tors
  \simeq
  \Z_3\oplus\Z_3
  \simeq
  H^2(X^\ast,\Z)_\tors
\end{equation}
and the first of eq.~\eqref{eq:BaytrevKreuzer} is true. This provides
the first evidence for the conjecture of~\cite{Batyrev:2005jc} in a
context other than toric hypersurfaces.

Another point of view is that there is a geometrical or rather
combinatorial reason for the self-mirror property in this case.  From
eqns.~\eqref{eq:Nabla} and~\eqref{eq:points} one can easily see that
the lattice points $\nu_i,\nu_{6+i},\nu_{13},\nu_{14}$, $i=1,\dots,3$,
span a sub-polytope of $\nabla^*$ satisfying the same linear relations
as all the lattice points $\rho_i$ of $\Delta^*$ in
eq.~\eqref{eq:linreldelta}. Hence, this sub-polytope is isomorphic to
$\Delta^*$. The same is true for the polytopes $\bar\nabla^*$ and
$\bar\Delta^*$. The toric variety $\mP_{\bar\nabla^*}$ which is the
ambient space of $\Xb^*$ can therefore be regarded as a blow-up of a
quotient of $\mP_{\bar\Delta^*}$, the ambient space of $\Xb$.
Actually, this blow-up makes all 7 divisors of $\Xb^*$ toric.
Similarly, $\mP_{\nabla^*}$ can be regarded as a blow-up of a quotient
of $\mP_{\Delta^*}$. As shown in \autoref{sec:intersection-ring} this
entails that all $19$ K\"ahler moduli of $\Xt^*$ are realized
torically.  Note that it is possible that the mirror polytopes
$\Delta^*$ and $\nabla^*$ are actually isomorphic. In fact, for toric
hypersurfaces there are $41,710$ self-dual
polytopes~\cite{Kreuzer:2000xy}. The novel feature in our case is that
non-isomorphic polytopes lead to self-mirror complete intersections,
consistent with the nef partitions.


\section
[Factorization vs. The (3,1,0,0,0) Curve]
{Factorization vs. The $\mathbf{(3,1,0,0,0)}$ Curve}
\label{sec:p3check}

One interesting observation is that the prepotential $\FprepotXNP$ at
order $p$, see eq.~\eqref{eq:PrepotX*} in this paper and
eq.~\xeqref{eq:PrepotX} in \partA~\cite{Braun:2007xh}, factors into
$\sum_{i,j=0}^2 b_1^i b_2^j$ times a function of $p$, $q$, $r$ only.
This means that the instanton number for any pseudo-section (curve
contributing at order $p$) does not depend on the torsion part of its
homology class. In other words, for any pseudo-section there are $8$
other pseudo-sections with the same class in $H_2(X,\Z)_\free$ and
together filling up all of $H_2(X,\Z)_\tors=\Z_3\oplus\Z_3$. In
contrast, this factorization does not hold at order $p^3$. For
example,
\begin{equation}
  \label{eq:FXp3q}
  \begin{split}
    \FprepotXNP(p,q,r,b_1,b_2) =
    \cdots +&\; 
    3 p^3 q 
    \\
    +&\;
    0 
    \big(
    b_1+b_1^2+ 
    b_2+b_1 b_2+b_1^2 b_2+ 
    b_2^2+b_1 b_2^2+b_1^2 b_2^2
    \big) 
    p^3 q
    \\
    +&\;
    \cdots
    .    
  \end{split}
\end{equation}
The purpose of this subsection is to understand this behavior.

First, the factorization of the prepotential at any order of $p$ not
divisible by $3$ follows from an extra symmetry that we have not
utilized so far. The covering space $\Xt$ is, in addition to
eqns.~\eqref{eq:g1action} and~\eqref{eq:g2action}, also invariant
under another $\Hat{G}= \ZZZ$ action generated by ($\zeta\eqdef
e^{\frac{2\pi i}{3}}$)
\begin{subequations}
\begin{equation}
  \label{eq:g1hataction}
  \Hat{g}_1:  
  \begin{cases}
    [x_0:x_1:x_2] \mapsto
    [x_0:\zeta x_1:\zeta^2 x_2]
    \\
    [t_0:t_1] \mapsto
    [t_0:t_1] 
    ~\text{(no action)}
    \\
    [y_0:y_1:y_2] \mapsto
    [y_0:y_1:y_2]
    ~\text{(no action)}
  \end{cases}
\end{equation}
and
\begin{equation}
  \label{eq:g2hataction}
  \Hat{g}_2:  
  \begin{cases}
    [x_0:x_1:x_2] \mapsto
    [x_1:x_2:x_0]
    \\
    [t_0:t_1] \mapsto
    [t_0:t_1] 
    ~\text{(no action)}
    \\
    [y_0:y_1:y_2] \mapsto
    [y_0:y_1:y_2]
    ~\text{(no action)}
  \end{cases}
\end{equation}
\end{subequations}
This symmetry has fixed points and, therefore, cannot be used if one
is looking for a smooth quotient of $\Xt$. However, it commutes with
$G$ and hence descends to a $\Hat{G}=\ZZZ$ symmetry of $X$ (with fixed
points). Clearly, the instanton sum must observe this additional
geometric symmetry. To make use of this symmetry, we have to express
its action on the variables in $\FprepotXNP(p,q,r,b_1,b_2)$. We can do
so by first noting that the basic $81$ curves
\begin{equation}
  s_1\FPtimes s_2
  \subset
  \Xt
  , \qquad
  s_1 \in MW(B_1)
  ,~
  s_2 \in MW(B_2)
\end{equation}
are really one orbit under $G\times\Hat{G}$. Recall that, after
dividing out $G$, these curves became the $9$ sections in
$MW(X)=\Z_3\oplus\Z_3$, see \partA{} \xautoref{sec:Aquotient}. We now
observe that $MW(X)=\{s_{ij}\}$ is one $\Hat{G}$-orbit; since each of
these sections contributes $p b_1^i b_2^j$, $i,j=0,\dots,2$ the
induced $\Hat{G}$ action on the prepotential must be
\begin{equation}
  \begin{split}
    \Hat{g}_1
    :& \quad
    \FprepotXNP(p,q,r,b_1,b_2)
    \mapsto
    \FprepotXNP(b_1 p,q,r,b_1,b_2)  
    , \\
    \Hat{g}_2
    :& \quad
    \FprepotXNP(p,q,r,b_1,b_2)
    \mapsto
    \FprepotXNP(b_2 p,q,r,b_1,b_2)  
    .
  \end{split}
\end{equation}
Clearly, the prepotential must be invariant under the $\Hat{g}_1$,
$\Hat{g}_2$ action. While imposing no constraint on the $p^{3n}$ terms
in the prepotential, all other powers of $p$ must appear in the
combination
\begin{equation}
  p^n \Big( \sum_{i,j=0}^2 b_1^i b_2^j \Big)
  ,\qquad n\not\equiv 0\mod 3
  .
\end{equation}
This proves the factorization observed at the beginning of this
subsection. 

Second, we would like to understand the $p^3 q$ terms in
eq.~\eqref{eq:FXp3q}. These are the curves in the homology
classes\footnote{Recall that the exponent of $p$ is the degree along
  the base $\CP^1$. This is why we pick a basis in $H_2(X,\Z)_\free$
  such that a curve in $(n_1,n_2,n_3,m_1,m_2)$ contributes at order
  $p^{n_1} q^{n_2} r^{n_3} b_1^{m_1} b_2^{m_2}$ in the prepotential.}
\begin{equation}
  (3,1,0,\ast,\ast) \in \Z^3\oplus\Z_3\oplus\Z_3 = 
  H_2\big(X,\Z\big)
  .
\end{equation}
We will show that the rational curves in this class come in a single
family, that is, the moduli space of genus $0$ curves on $X$ in these
homology classes
\begin{equation}
  \Moduli_0\Big(X, (3,1,0,\ast,\ast) \Big)
\end{equation}
is connected. In particular, all such curves have the same homology
class $(3,1,0,0,0)$ and only contribute to $p^3q$ in the prepotential
eq.~\eqref{eq:FXp3q}. As discussed in \partA{}
\xautoref{sec:CoHomology}, any such map $C_X: \CP^1\to X$ factors
\begin{equation}
  \vcenter{\xymatrix{
      \CP^1 
      \ar[rr]^{C_X}
      \ar[dr]_{C_\Xt}
      & 
      & 
      X
      \\
      &
      \Xt
      \ar[ur]_{q}
    }}      
  .
\end{equation}
The map $C_\Xt$ can be written in terms of homogeneous coordinates as
a function
\begin{equation}
  \label{eq:CXtdef}
  C_\Xt
  :~
  \CP^1_{[z_0:z_1]} \mapsto 
  \CP^2_{[x_0:x_1:x_2]} \times
  \CP^1_{[t_0:t_1]} \times
  \CP^2_{[y_0:y_1:y_2]} 
\end{equation}
satisfying the equations~\eqref{eq:B1} and~\eqref{eq:B2} defining
$\Xt$,
\begin{equation}
  \label{eq:F12CXtconstraint}
  F_1 \circ C_\Xt \big([z_0:z_1]\big)
  = 0 = 
  F_2 \circ C_\Xt \big([z_0:z_1]\big)
  \qquad
  \forall [z_0:z_1]\in\CP^1
  .
\end{equation}
The curve $C_X$ ends up in the homology class $(3,1,0,\ast,\ast)$ if
and only if the defining equation~\eqref{eq:CXtdef} is of degree
$(3,1,0)$ in $\CPambient$. Hence, eq.~\eqref{eq:CXtdef} is defined by
complex constants $\alpha_{ij}$, $\beta_{ij}$, $\gamma_i$ (up to
rescaling) such that
\begin{equation}
  \label{eq:CXtcoord}
  \begin{aligned}
    x_i =&\; 
    \alpha_{i0}\, z_0 + 
    \alpha_{i1}\, z_1
    &
    i=&\; 0,1,2
    \\
    t_i =&\; 
    \beta_{i0}\, z_0^3 + 
    \beta_{i1}\, z_0^2 z_1 + 
    \beta_{i2}\, z_0 z_1^2 + 
    \beta_{i3}\, z_1^3 
    &
    i=&\; 0,1
    \\
    y_i =&\; \gamma_i
    &
    i=&\; 0,1,2
    .
  \end{aligned}
\end{equation}
These constants have to be picked such that the resulting curve lies
on the complete intersection $\Xt$, that is, they have to satisfy
eq.~\eqref{eq:F12CXtconstraint}. Inserting eq.~\eqref{eq:CXtcoord}, we
find that $F_1\circ C_\Xt\big([z_0:z_1]\big)$ is a homogeneous degree
$6$ polynomial in $[z_0:z_1]$. Since the coefficients of $z_0^k
z_1^{6-k}$ must vanish individually, this yields $7$ constraints for
the parameters $\alpha_{ij}$, $\beta_{ij}$. What makes this system of
constraint equations tractable is the fact that they are all linear in
$\beta_{ij}$,
\begin{equation}
  F_1\circ C_\Xt = 0
  \quad\Leftrightarrow\quad
  \begin{pmatrix}
    A_1 &   0 &   0 &   0 & A_5 &   0 &   0 &   0 \\
    A_2 & A_1 &   0 &   0 & A_6 & A_5 &   0 &   0 \\
    A_3 & A_2 & A_1 &   0 & A_7 & A_6 & A_5 &   0 \\
    A_4 & A_3 & A_2 & A_1 & A_8 & A_7 & A_6 & A_5 \\
    0 & A_4 & A_3 & A_2 &   0 & A_8 & A_7 & A_6 \\
    0 &   0 & A_4 & A_3 &   0 &   0 & A_8 & A_7 \\
    0 &   0 &   0 & A_4 &   0 &   0 &   0 & A_8 \\
  \end{pmatrix}
  \begin{pmatrix}
    \beta_{00} \\ \beta_{01} \\ \beta_{02} \\ \beta_{03} \\
    \beta_{10} \\ \beta_{11} \\ \beta_{12} \\ \beta_{13}
  \end{pmatrix}
  = 0
\end{equation}
where
\begin{equation}
  \begin{aligned}
    A_1 \eqdef&\; 
    \alpha_{00}^3+\alpha_{10}^3+\alpha_{20}^3 
    &
    A_5 \eqdef&\; 
    \alpha_{00} \alpha_{10} \alpha_{20}
    \\
    A_2 \eqdef&\; 
    3 \alpha_{01} \alpha_{00}^2+3 \alpha_{11} \alpha_{10}^2
    +3 \alpha_{21} \alpha_{20}^2 +\alpha_{20}^3
    &
    A_6 \eqdef&\; 
    ( \alpha_{01} \alpha_{10}+\alpha_{00} \alpha_{11} ) 
    \alpha_{20}+\alpha_{00} \alpha_{10} \alpha_{21}
    \\
    A_3 \eqdef&\; 
    3 \alpha_{01}^2\alpha_{00}+3 \alpha_{11}^2\alpha_{10}+3
    \alpha_{21}^2\alpha_{20} 
    &
    A_7 \eqdef&\; 
    \alpha_{01} \alpha_{11} \alpha_{20}+ 
    ( \alpha_{01} \alpha_{10}+\alpha_{00} 
    \alpha_{11} ) \alpha_{21}
    \\
    A_4 \eqdef&\; 
    \alpha_{01}^3+\alpha_{11}^3+\alpha_{21}^3 
    &
    A_8 \eqdef&\; 
    \alpha_{01} \alpha_{11} \alpha_{21}
    .
  \end{aligned}
\end{equation}
Thinking of this as $7$ linear equations for the $8$ parameters
$\beta_{ij}$, there is always a non-zero solution. The solution is
generically unique up to an overall factor, and turns into an $\CP^n$
for special values of the $\alpha_{ij}$. Moreover, the parameter space
of the $\alpha_{ij}$ is connected (essentially, the moduli space of
lines in $\CP^2$). Since we just identified the parameter space of the
$(\alpha_{ij},\beta_{ij})$ as a blow-up thereof, it is therefore
connected as well.

It remains to satisfy $F_2\circ C_\Xt=0$. One can easily see that the
only way is to pick the $\gamma_i$ to be simultaneous solutions of
\begin{equation}
  \gamma_0^3+\gamma_1^3+\gamma_2^3 = 0 = \gamma_1 \gamma_2 \gamma_3
  .
\end{equation}
Since two cubics intersect in $9$ points, there are $9$ such
solutions, permuted by $G$. Therefore, the parameter space of
$(\alpha_{ij},\beta_{ij},\gamma_i)$ has $9$ connected components,
permuted by the $G$-action. The moduli space of curves $C_X$ on $X$ is
the $G$-quotient of the moduli space of curves $C_\Xt$ on $\Xt$, and
therefore has only a single connected component. By continuity, every
curve $C_X$ in this connected family has the same homology class,
explaining the piece of the prepotential given in
eq.~\eqref{eq:FXp3q}.


%% file: Bmodel-Fin.tex
\section{Towards a Closed Formula}
\label{sec:conjecture}

Putting all the information together we found out about the
prepotential on $X$, one can try to divine a closed form for the
prepotential. We guess that the order $p^n$ terms have the closed form
\begin{flalign}
  \label{eq:guess12}
  \hspace{7mm}
  \FprepotXNP(p,q,r, b_1,b_2)\Big|_{p^n} 
  =&\;
  \frac{p^n}{8^{n-1}} 
  \left(\sum_{i,j\in \Z_3} b_1^{i} b_2^{j} \right)
  \Big(P(q)^4 P(r)^4 \Big)^n 
  M_{2n-2}(q,r)
  &
\end{flalign}
if $n$ is not a multiple of $3$ and, slightly weaker, that
\begin{flalign}
  \label{eq:guess3}
  \hspace{7mm}
  \FprepotXNP(p,q,r,1,1)\Big|_{p^n} 
  =&\;
  \frac{9 p^n}{8^{n-1}} 
  \Big(P(q)^4 P(r)^4 \Big)^n 
  M_{2n-2}(q,r)
  &
\end{flalign}
if $n$ is a multiple of $3$. Here,
\begin{itemize}
\item $P(q)$ is the usual generating function of partitions
  eq.~\eqref{eq:GenfnPartitions}. 
\item The $M_{2n-2}$ are polynomials in the Eisenstein series
  $E_2(q)$, $E_4(q)$, $E_6(q)$ and $E_2(r)$, $E_4(r)$, $E_6(r)$,
  starting with
  \begin{equation}
    \label{eq:Mpolynomials}
    \begin{split}
      M_{-2}(q,r) =&\;
      0
      \\
      M_0(q,r) =&\;
      1
      \\
      M_2(q,r) =&\;
      E_2(q) E_2(r)
      \\
      M_4(q,r) =&\;
       \tfrac{13}{108} E_4(q)E_4(r)
      +\tfrac{1}{4} \Big( E_4(q) E_2(r)^2 + E_2(q)^2 E_4(r) \Big)
      +\tfrac{7}{4} E_2(q)^2 E_2(r)^2 
      \\
      M_6(q,r) =&\;
       \tfrac{1}{27} E_6(q) E_6(r) 
      +\tfrac{13}{54} 
        \Big( E_6(q) E_4(r) E_2(r) + E_4(q) E_2(q) E_6(r) \Big)
      \\ &\;
      +\tfrac{1}{6} \Big( E_6(q) E_2(r)^3 + E_2(q)^3 E_6(r) \Big)
      +\tfrac{79}{108} E_4(q) E_2(q) E_4(r) E_2(r)
      \\ &\;
      +\tfrac{5}{4} 
        \Big( E_2(q)^3 E_4(r) E_2(r) + E_4(q) E_2(q) E_2(r)^3  \Big)
      +\tfrac{47}{12} E_2(q)^3 E_2(r)^{3}
      \\
      M_8(q,r) =&\;
       \tfrac{2}{3} E_6(q)E_2(q) E_6(r)E_2(r) 
      +\tfrac{1309}{6750} E_4(q)E_4(q)E_4(r)E_4(r) 
      \\ &\;
      +\tfrac{25}{108} 
        \Big(E_6(q)E_2(q)E_4(r)E_4(r) + E_4(q)E_4(q)E_6(r)E_2(r)\Big) 
      \\ &\;
      +\tfrac{85}{54} 
        \Big(E_6(q)E_2(q)E_4(r)E_2(r)^2 +  E_4(q)E_2(q)^2E_6(r)E_2(r)\Big) 
      \\ &\;
      +\tfrac{13}{12} 
        \Big(E_6(q)E_2(q)E_2(r)^4 + E_2(q)^4E_6(r)E_2(r)\Big) 
      \\ &\;
      +\tfrac{137}{216} 
        \Big(E_4(q)E_4(q)E_4(r)E_2(r)^2 + E_4(q)E_2(q)^2E_4(r)E_4(r)\Big) 
      \\ &\;
      +\tfrac{3}{8} \Big(E_4(q)E_4(q)E_2(r)^4 + E_2(q)^4E_4(r)E_4(r)\Big) 
      \\ &\;
      +\tfrac{34}{9} E_4(q)E_2(q)^2E_4(r)E_2(r)^2 
      +\tfrac{121}{12} E_2(q)^4E_2(r)^4
      \\ &\;
      +\tfrac{41}{8} 
        \Big(E_4(q)E_2(q)^2E_2(r)^4 + E_2(q)^4E_4(r)E_2(r)^2\Big) 
      .
    \end{split}
  \end{equation}
  They are symmetric under the exchange $q\leftrightarrow r$ and of
  weight $2n$ in $q$ and $r$ separately.  But, for example, $M_4$
  above does not factor into a function of $q$ and a function of $r$.
  So the $M_{2n-2}$ are not the products of the polynomials
  appearing in the \dP9 prepotential.  However, by setting $q=0$ or
  $r=0$ one recovers the corresponding polynomials in the \dP9
  prepotential~\cite{Hosono:1999qc}.
\item The $E_{2i}$ are the usual Eisenstein series
  \begin{equation}
    \label{eq:Eisenstein}
    \begin{split}
      E_2(q) =&\;
      1-24 q-72 q^2-96 q^3-168 q^4-144 q^5 -288 q^6 + O(q^7)
      \\
      E_4(q) =&\;
      1+240 q+2160 q^2+6720 q^3+17520 q^4+30240 q^5+ O(q^6)      
      \\
      E_6(q) =&\;
      1-504 q-16632 q^2-122976 q^3-532728 q^4 + O(q^5)
      .
    \end{split}  
  \end{equation}
\end{itemize}
Note that the naive Taylor series coefficients of the prepotential are
fractional, but when expanding in terms of $\Li_3$'s (which account
for the multicover contributions) one finds integral instanton
numbers.

These expressions for the prepotential agree with all instanton
numbers computed in this paper. Unfortunately, we have not been able
to guess a closed formula that includes the $b_1$ and $b_2$ dependence
of the prepotential $\FprepotXNP(p,q,r,b_1,b_2)|_{p^n}$ if $n$ is
divisible by $3$. We expect that these involve extra functions beyond
the Eisenstein series.

\section{Conclusion}
\label{sec:conclusion}

In the initial paper \partA{}~\cite{Braun:2007xh}, we analyzed the
topology of the Calabi-Yau manifold of interest and found that
\begin{equation}
  H_2\big( X,\Z\big) = 
  \Z^3 \oplus 
  \Z_3 \oplus \Z_3
  .
\end{equation}
Although the presence of torsion curve classes complicates the
counting of rational curves, we managed to derive the A-model
prepotential to linear order in $p$.

The goal of this paper is to go beyond the results of
\partA{} using mirror symmetry. By carefully adapting methods designed
for complete intersections in toric varieties, we can apply mirror
symmetry to compute the instanton numbers on $X$, even though $X$ is
not toric. Using that $X$ is self-mirror, we completely solve this
problem and are able to calculate the complete A-model prepotential to
any desired precision (and for arbitrary degrees in $p$), limited only
by computer power. Carrying out this computation, we find the first
examples of instanton numbers that do depend on the torsion part of
their integral homology class, see \autoref{tab:n1n2n3b1b2Inst} on
Page~\pageref{tab:n1n2n3b1b2Inst}.

Since the self-mirror property of $X$ is important, we investigate it
in detail. In doing so, we go far beyond just checking that the Hodge
numbers are self-mirror. In particular, we find that the intersection
rings are identical and that torsion in homology obeys the conjectured
mirror relation~\cite{Batyrev:2005jc}. Finally, going beyond classical
geometry, we independently calculate certain instanton numbers on $X$
and its Batyrev-Borisov mirror $X^\ast$. Again, we find that $X$ and
$X^\ast$ are indistinguishable, providing strong evidence for $X$
being self-mirror. Both of these results extend those found in
\partA~\cite{Braun:2007xh}.

Using these results, we are able to guess certain closed expressions
for the prepotential of $X$ in terms of modular forms. In certain
limits it specializes to the $dP_9$ prepotential
of~\cite{Hosono:1999qc}.  There it is known that the coefficients in
$p$ of the $dP_9$ prepotential satisfy a recursion relation. Moreover,
there is a gap condition, that is, a certain number of subsequent
terms in a series expansion is absent. This condition provides
sufficient data to determine the integration constants for the
recursion and allows to determine the prepotential completely, even at
higher genus. We expect a similar story to be valid for the
prepotential of $X$.

\section*{Acknowledgments}

The authors would like to thank Albrecht Klemm, Tony Pantev, and
Masa-Hiko Saito for valuable discussions. We also thank Johanna Knapp
for providing a Singular~\cite{GPS05} code to compute the intersection
ring of Calabi-Yau manifolds in toric varieties.
\GrantsAcknowledgements
E.~S. thanks the Math/Physics Research group at the University of
Pennsylvania for kind hospitality.

\appendix

\makeatletter
\def\Hy@chapterstring{section}
\makeatother

\section
[Triangulation of $\bar\nabla^*$ and $\nabla^*$]
[Triangulations]
{Triangulation of $\mathbf{\bar\nabla^*}$ and $\mathbf{\nabla^*}$}
\label{sec:nablatriangs}

In principle the coherent triangulations of the fan over
$\bar\nabla^*$ can be computed with TOPCOM by finding the $720$ star
triangulations in the total of $230,832$ coherent triangulations of
$\bar\nabla^*$. The discussion of the symmetry properties is greatly
facilitated, however, by an explicit understanding of their structure.
We will work out the triangulations by first triangulating the facets
and then checking the compatibility of their maximal intersections and
the coherence of the resulting star triangulations.

We start with a couple of useful definitions. A circuit is a minimal
collection of $n$ affinely dependent points $p_1,\ldots,p_n$,
\begin{equation}
  \lambda_1p_1+\ldots \lambda_np_n=0 
  \qquad 
  \text{with}
  \qquad 
  \lambda_1+\ldots+\lambda_n=0,\;
  \lambda_i \not = 0,
\end{equation}
any proper subset of which is affinely independent. The coefficient
vector $\lambda_n$ hence has nonzero entries and is unique up to a
prefactor.  We indicate the unique separation into points with
positive and negative coefficients with the notation $\langle
p_{i_1}\ldots p_{i_s}| p_{i_{s+1}}\ldots p_{i_n}\rangle$. Each circuit
admits two different triangulations, which are obtained by dropping
one of the points with positive coefficients and one of the remaining
points, respectively. We indicate this with a hat over the relevant 
subset. The two resulting triangulations
\begin{equation}
  \big\langle 
  \widehat{p_{i_1}\ldots p_{i_s}}| p_{i_{s+1}}\ldots p_{i_n}
  \big\rangle
  ,\qquad
  \big\langle 
  p_{i_1}\ldots p_{i_s}| \widehat{p_{i_{s+1}}\ldots p_{i_n}}
  \big\rangle
\end{equation}
hence consist of $s$ and $n-s$ simplices, respectively. If the first
point is in the convex hull of the others, that is, $s=1$, then only
one of the triangulations is maximal (all points are vertices of at
least one simplex).

Furthermore, we introduce the notation:
\begin{equation}
  \label{eq:abcdef}
  \begin{gathered}
    a_i = \nub_i, \quad
    b_i = \nub_{3+i}, \quad
    c_i = \nub_{6+i}, \quad 
    d_i = \nub_{12+i},
    \qquad
    i=1,2,3
    ,
    \\
    e = \nub_{13},     \quad
    f = \nub_{14}
    . 
  \end{gathered}
\end{equation}
Among these $14$ vectors in eq.~\eqref{eq:abcdef} there are $9$
independent linear relations, see eq.~\eqref{eq:linrelnabla},
\begin{equation}
  \label{eq:scale}
  \begin{gathered}
    a_1+a_2+a_3=0, \qquad 
    c_1+c_2+c_3=0, 
    \\
    e+f=0, \qquad
    b_i=a_i+e, \qquad
    d_l=c_l+f
    ,
  \end{gathered}
\end{equation}
which imply others like $a_i+b_j=a_j+b_i$ and $a_i+c_l=b_i+d_l$ or
$e=\frac13(b_1+b_2+b_3)$ and $f=\frac13(d_1+d_2+d_3)$.

\begin{lemma}
  $\bar\nabla^*$ has 15 facets, 6 of which are simplicial:
  \begin{equation}
    \big[a_ia_jb_ib_jc_lc_md_ld_m]_{\substack{i<j \\ l<m}}
    ,\quad
    \big[a_ia_jd_1d_2d_3\big]_{i<j}
    ,\quad
    \big[b_1b_2b_3c_lc_m\big]_{l<m}
    .
  \end{equation}
\end{lemma}
The nine non-simplicial facets form an orbit under the permutation
symmetries $\mZ_3^{ab}\times\mZ_3^{cd}$ generated by $g_{ab}:
\left(\begin{smallmatrix} a_i \\ b_i \end{smallmatrix}\right) \to
\left(\begin{smallmatrix} a_{i+1} \\ b_{i+1} \end{smallmatrix}\right)$
and $g_{cd}: \left(\begin{smallmatrix} c_l \\ d_l
  \end{smallmatrix}\right) \to \left(\begin{smallmatrix} c_{l+1} \\
    d_{l+1} \end{smallmatrix}\right)$. According to the linear
relations eq.~\eqref{eq:scale} the eight points on each non-simplicial
facet form quadratic circuits $a_i+b_j=b_i+a_j$, $a_i+c_l=b_i+d_l$,
and $c_l+d_m=c_m+d_l$, which we call \emph{mixed} if they contain
vertices of both elements of the nef partition $\langle
a_ic_l|b_id_l\rangle$, and \emph{pure circuits} $\langle
a_ib_j|b_ia_j\rangle$, $\langle c_ld_m|c_md_l\rangle$ otherwise.

The coherent triangulations of the facets 
$[a_ia_jb_ib_jc_lc_md_ld_m]$ are most easily obtained
from their Gale transform 
\begin{equation}
  \begin{pmatrix}
    1&-1&-1&1&0&0&0&0\\
    1&0&-1&0&1&0&-1&0\\
    0&0&0&0&1&-1&-1&1
  \end{pmatrix}
  ,  
\end{equation}
which is the coefficient matrix of the basis $a_i-a_j-b_i+b_j=0$,
$a_i-b_i+c_l-d_l=0$ and $c_l-c_m-d_l+d_m=0$ of linear relations.  The
coherent triangulations are in one-to-one correspondence to chambers
that are seperated by the facets of the cones generated by all linear
bases $\mu=\{v_1,v_2,v_3\}$ with $v_i$ selected among the 8 column
vectors of the Gale transform \cite{MR1074022, MR1264417}.
\begin{figure}  
  \begin{center}\unitlength=3pt\thicklines
    \begin{picture}(40,40)(-20,-20)
      \psline(12,-12)(20,20)(0,20)(-12,12)(-20,-20)(0,-20)(12,-12)
      \psline(8,12)(20,20)\psline(8,12)(-12,12)\psline(8,12)(0,-20)
      \psline[linestyle=dotted](-8,-12)(-20,-20)
      \psline[linestyle=dotted](-8,-12)(0,20)
      \psline[linestyle=dotted](-8,-12)(12,-12)       
      \psset{linecolor=red,linewidth=0.4pt}
      \psline{->}(0,0)(15,-15)        \psline{->}(0,0)(-15,15)
      \psline{->}(0,0)(10,15)    \psline{->}(0,0)(-10,-15)
      \psline{->}(0,0)(23,23) \psline{->}(0,0)(0,24)
      \psline{->}(0,0)(-23,-23)       \psline{->}(0,0)(0,-24)
      \psset{linecolor=blue,linewidth=1.2pt,linestyle=dashed,dash=4pt 2pt}
      \psline(-12,12)(20,20)\psline(8,12)(0,20)
      \psline(-12,12)(0,-20)\psline(8,12)(-20,-20)
      \psline(0,-20)(20,20)\psline(8,12)(12,-12)
      \psset{linecolor=black}
      \putcx(12,-12)  \putlab(12,-13.5)t{$ a_i$}      
      \putcx(8,12)    \putlab(7.6,13.5)b{$ a_j$}      
      \putcx(-12,12)  \putlab(-12,14)b{$   b_i$}      
      \putcx(-8,-12)  \putlab(-6,-13.5)t{$ b_j$}            
      \putcx(20,20)   \putlab(20,22)b{$  c_l$}        
      \putcx(0,-20)   \putlab(3,-21)t{$  c_m$}        
      \putcx(-20,-20)\putlab(-19,-21)t{$ d_l$}        
      \putcx(0,20)    \putlab(-1,22)r{$  d_m$}
    \end{picture}
  \end{center}
  \caption{
    Secondary fan of the non-simplicial facets. 
    Chambers are indicated by dashed lines.}
  \label{fig:chambers}
\end{figure}
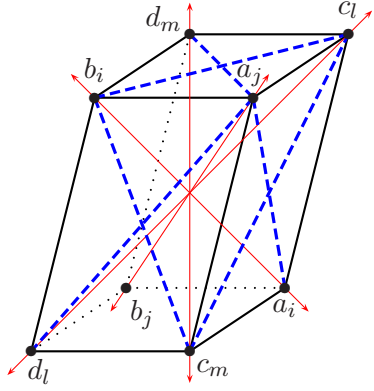
In the present case the cones over the faces of the parallel-epiped in
\autoref{fig:chambers} are subdivided into $24$ chambers, which are
indicated by dashed lines. The triangulations, which we can label by
the facet containing and the edge adjoining the chamber, are obtained
as the sets of complements of those bases $\mu$ that span a cone
containing the respective chamber.

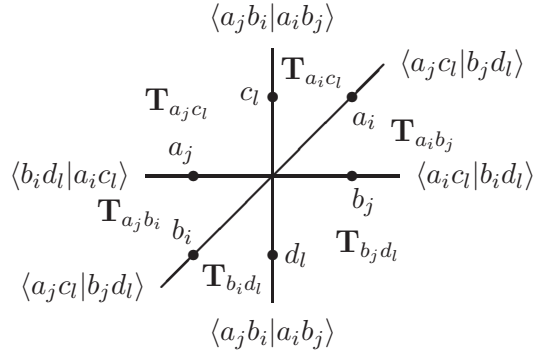
\begin{figure}[htbp]
  \begin{center}
    \unitlength=3pt\thicklines
    \begin{picture}(40,40)(-20,-20)

      \putlab(19,5)c{\normalsize                    $\mathbf{T}_{a_ib_j}$}
      \putlin(0,0,1,0,16)\putlab(18,0)l{$\langle a_ic_l|b_id_l\rangle$}
      \putcx(10,0) \putlab(11.5,-1.5)t{$b_j$}

      \putlab(5,13)c{\normalsize                    $\mathbf{T}_{a_ic_l}$}
      \putlin(0,0,1,1,14)\putlab(16,14)l{$\langle a_jc_l|b_jd_l\rangle$}
      \putcx(10,10) \putlab(10,7)l{$a_i$}

      \putlab(-12,9)c{\normalsize                    $\mathbf{T}_{a_jc_l}$}
      \putlin(0,0,0,1,16)\putlab(0,18)b{$\langle a_jb_i|a_ib_j \rangle$}
      \putcx(0,10) \putlab(-1.5,10)r{$c_l$}

      \putlab(-18,-5)c{\normalsize                    $\mathbf{T}_{a_jb_i}$}
      \putlin(0,0,-1,0,16)\putlab(-18,0)r{$\langle b_id_l|a_ic_l\rangle$}
      \putcx(-10,0) \putlab(-11.5,1.5)b{$a_j$}

      \putlab(-5,-13)c{\normalsize                    $\mathbf{T}_{b_id_l}$}
      \putlin(0,0,-1,-1,14)\putlab(-16,-14)r{$\langle a_jc_l|b_jd_l\rangle$}
      \putcx(-10,-10) \putlab(-10,-7)r{$b_i$}

      \putlab(12,-9)c{\normalsize                    $\mathbf{T}_{b_jd_l}$}
      \putlin(0,0,0,-1,16)\putlab(0,-18)t{$\langle a_jb_i|a_ib_j\rangle$}
      \putcx(0,-10) \putlab(1.5,-10)l{$d_l$}
    \end{picture}
  \end{center}
  \caption{
    Secondary fan of the codimension two face $[a_ia_jb_ib_jc_ld_l]$.}
  \label{fig:cd2face} 
\end{figure}

Hence, each non-simplicial facet has 24 coherent triangulations, which
can be characterized by the triangulations of its 2 pure and of its 4
mixed circuits: Calling the triangulation $\langle
\widehat{a_ic_l}|b_id_l\rangle$ positive and the triangulation
$\langle a_ic_l|\widehat{b_id_l}\rangle$ negative, and arranging the
cyclic permutations $g_{ab}$ and $g_{cd}$ in the horizontal and
vertical direction, respectively, we can assign one of 16 different
types \PMdos\pm\pm\pm\pm{} to each triangulation, where the signs
indicate the induced triangulations of the mixed circuits.  The
constraints that reduce the a priori $32=2^6$ combinations to 24 all
derive from the following rules:
\begin{equation}
  \label{eq:prism}
  \begin{picture}(20,0)(10,4)     
    \psline[linestyle=dashed](14,8)(0,0)
    \psline(14,8)(14,0)(0,0)(0,8)(14,8)(8,10)(0,8)
    \psline[linestyle=dotted](0,0)(8,2)(14,0)
    \psline[linestyle=dotted](0,0)(8,10)(8,2)(14,8)
    \putlab(-.4,-1)r{$a_i$}\putlab(14.4,-1)l{$a_j$}
    \putlab(-.6,9)r{$b_i$}\putlab(14.6,9)l{$b_j$}\putlab(8,11)b{$c_l$}
  \end{picture}
  \begin{array}{c}
    \big\langle a_ic_l|\widehat{b_id_l}\big\rangle 
    ~\wedge~
    \big\langle \widehat{a_jc_l}|b_jd_l\big\rangle
    \quad\Rightarrow\quad
    \big\langle a_ib_j|\widehat{a_jb_i}\big\rangle
    \\
    \big\langle \widehat{a_ic_l}|b_id_l\big\rangle 
    ~\wedge~
    \big\langle a_jc_l|\widehat{b_jd_l}\big\rangle 
    \quad\Rightarrow\quad
    \big\langle \widehat{a_ib_j}|a_jb_i\big\rangle 
  \end{array}
\end{equation}
i.e. a triangular prism can be triangulated in 6 different ways, which
correlates the a priori 8 combinations of the triangulations of the 3
squares (with analogous constraints for the two ``horizontal'' prisms
$[a_ib_ic_lc_md_ld_m]$ contained in the facet
$[a_ia_jb_ib_jc_lc_md_ld_m]$).  Putting the pieces together we obtain
\begin{lemma}
  The $24$ triangulations of the non-simplicial facets can be
  assorted as follows: 
  \begin{itemize}
  \item For \PMdos++++{},~\PMdos----{} the pure circuits are
    unconstrained, yielding $2\cdot 2^2=8$ triangulations.
  \item For \PMdos++--{},~\PMdos--++{} the pure $ab$-circuit is
    unconstrained; with the transposed types
    \PMdos+-+-{},~\PMdos-+-+{} this accounts for another $8$
    triangulations.
  \item The final $8$ triangulations come from the $8$ types with an
    odd number of positive signs, for which the triangulation of the
    pure circuits is unique.
  \item The two types \PMdos+--+{} and~\PMdos-++-{} cannot occur
    because of contradictory implications for the triangulations of
    the pure circuits.
  \end{itemize}
\end{lemma}
The secondary fan and the induced triangulations for the
codimension-two faces at which the non-simplical facets intersect can
be obtained from \autoref{fig:chambers} by projection along the
dropped vertices. The secondary fan of the prism of
eq.~\eqref{eq:prism}, for example, which is shown in
\autoref{fig:cd2face}, is obtained from \autoref{fig:chambers} by
projection along the diagonal $\langle c_md_m\rangle$. The wall
crossings between the six cones in \autoref{fig:cd2face} are labeled
by the circuits whose flops relate the adjoining
triangulations~\cite{MR1074022}.

For the construction of the complete star triangulation we now observe
that the non-simplicial intersections of the $9$ non-simplicial facets
$[a_ia_jb_ib_jc_lc_md_ld_m]$ are given by the 18 triangular prisms
$[a_ia_jb_ib_jc_ld_l]$ and $[a_ib_ic_lc_md_ld_m]$.  If we interpret
the former as vertices and the latter as links then the resulting
compatibility conditions correspond to a graph with the topology of a
torus. The vertices of this graph are decorated by signs as shown in
\autoref{tab:PMtorus} and connected by horizontal and vertical links. 
\begin{table}[bhtp]
  \begin{center}
    \begin{tabular}{||c|c|c|c|c|c|c||}\hline\hline
      \PMtres+++++++++&\PMtres-++++++++&\PMtres--+++++++&\PMtres---++++++ 
      &\PMtres--+-+++++&\PMtres----+++++&\PMtres--+--++++\\\hline
      \PMtres---------&\PMtres+--------&\PMtres++-------&\PMtres+++------  
      &\PMtres++-+-----&\PMtres++++-----&\PMtres++-++----\\\hline
      \vrule height 14pt width 0pt 
      $1\cdot 2^6$ & 
      $9\cdot 2^2$ & 
      $18\cdot 2^2$ & $6\cdot 2^4$ & $36\cdot 2^0$ 
      & $36\cdot 2^1$ & $9\cdot 2^2$\\\hline\hline
    \end{tabular}
  \end{center}
  \caption{The $824 = 2\,(64+36+72+96+36+72+36)$ star  
        triangulations of $\nabla^*$, including
	the $720 = 2\,(36+36+72+72+36+72+36)$ 
	coherent triangulations.}
  \label{tab:PMtorus}
\end{table}
The restriction on the compatible signs is due to the absence of the
inconsistent types \PMdos+--+{} and \PMdos-++-{} as subgraphs on the
torus. The multiplicities $\mu\cdot 2^n$ come from the number $n$ of
unconstrained pure circuits and from the order $\mu$ of the effective
part of the symmetry group generated by transposition and permutations
of lines and columns. We thus find a total of $824$ triangulations.
The cyclic permutation symmetry that we want to keep on the Calabi-Yau
manifold $\Xb^*$ amounts to a diagonal shift, i.e. its induced action
on the graph is generated by $g_{ab}g_{cd}$. We are hence left with
the types \PMplus\, and \PMminus\,, and the shift symmetry furthermore
aligns the triangulations of the pure circuits and thus reduced the
multiplicities from $2^6$ to $2^2$, yielding a total of 8
triangulations for which $\mP_{\bar\nabla^*}$ is $G_2^*$ symmetric.

The resulting triangulations of the facet $[a_2a_3b_2b_3c_2c_3d_2d_3]$
are
\begin{gather}	
  \label{eq:triP}
  \begin{tabular}{c|c}
    \vrule width 0pt depth 9pt
    {\tiny
      \begin{picture}(5,5)(0,1)\put(0,0)+\put(0,2)+\put(0,4)+
        \put(2,0)+\put(2,2)+\put(2,4)+\put(4,0)+\put(4,2)+\put(4,4)+
      \end{picture}}& 
    \hbox{triangulation of $[a_2a_3b_2b_3c_2c_3d_2d_3]$} \\
    \hline	
    \vrule width 0pt height 16pt
    $\langle \widehat{a_2b_3}|a_3b_2 \rangle$, 
    $\langle \widehat{c_2d_3}|c_3d_2 \rangle$ 
    & 
    $\{[a_3b_2b_3d_2d_3],[b_2b_3c_3d_2d_3],
    [a_2a_3b_2d_2d_3],[b_2b_3c_2c_3d_2]\}$		\\[4pt]
    $\langle \widehat{a_2b_3}|a_3b_2 \rangle$, 
    $\langle c_2d_3|\widehat{c_3d_2} \rangle$
    & 
    $\{[a_3b_2b_3d_2d_3],[b_2b_3c_2d_2d_3],
    [a_2a_3b_2d_2d_3],[b_2b_3c_2c_3d_3]\}$		\\[4pt]
    $\langle a_2b_3|\widehat{a_3b_2} \rangle$, 
    $\langle \widehat{c_2d_3}|c_3d_2 \rangle$
    & 
    $\{[a_2b_2b_3d_2d_3],[b_2b_3c_3d_2d_3],
    [a_2a_3b_3d_2d_3],[b_2b_3c_2c_3d_2]\}$ 		\\[4pt]
    $\langle a_2b_3|\widehat{a_3b_2} \rangle$, 
    $\langle c_2d_3|\widehat{c_3d_2} \rangle$
    & 
    $\{[a_2b_2b_3d_2d_3],[b_2b_3c_2d_2d_3],
    [a_2a_3b_3d_2d_3],[b_2b_3c_2c_3d_3]\}$ 	
  \end{tabular}
  \\
  \label{eq:triM}
  \begin{tabular}{c|c}
    \vrule width 0pt depth 9pt
    {\tiny
      \begin{picture}(5,5)(0,1)\put(0,0){--}\put(0,2){--}\put(0,4){--}
        \put(2,0){--}\put(2,2){--}\put(2,4){--}\put(4,0){--}
        \put(4,2){--}\put(4,4){--}				
      \end{picture}}& 
    \hbox{triangulation of $[a_2a_3b_2b_3c_2c_3d_2d_3]$} \\
    \hline
    \vrule width 0pt height 16pt
    $\langle a_2b_3|\widehat{a_3b_2} \rangle$, 
    $\langle c_2d_3|\widehat{c_3d_2} \rangle$
    & 
    $\{[a_2a_3b_3c_2c_3],[a_2a_3c_2c_3d_3],
    [a_2b_2b_3c_2c_3],[a_2a_3c_2d_2d_3]\}$       	\\[4pt]
    $\langle a_2b_3|\widehat{a_3b_2} \rangle$, 
    $\langle \widehat{c_2d_3}|c_3d_2 \rangle$       
    & 
    $\{[a_2a_3b_3c_2c_3],[a_2a_3c_2c_3d_2],
    [a_2b_3b_2c_2c_3],[a_2a_3c_3d_2d_3]\}$		\\[4pt]
    $\langle \widehat{a_2b_3}|a_3b_2 \rangle$, 
    $\langle c_2d_3|\widehat{c_3d_2} \rangle$       
    & 
    $\{[a_2a_3b_2c_2c_3],[a_2a_3c_2c_3d_3],
    [a_3b_2b_3c_2c_3],[a_2a_3c_2d_2d_3]\}$		\\[4pt]
    $\langle \widehat{a_2b_3}|a_3b_2 \rangle$, 
    $\langle \widehat{c_2d_3}|c_3d_2 \rangle$
    & 
    $\{[a_2a_3b_2c_2c_3],[a_2a_3c_2c_3d_2],
    [a_3b_2b_3c_2c_3],[a_2a_3c_3d_2d_3]\}$ 	
  \end{tabular}
\end{gather}
It can be checked that the triangulations listed in
eqns.~\eqref{eq:triP} and~\eqref{eq:triM} come from the chambers
contained in the cones over $[a_2a_3c_2c_3]$ and $[b_2b_3d_2d_3]$,
respectively. For the first of these triangulations we consider the
chamber adjoining the edge $[a_2c_2]$, which is contained in the span
of the four bases $\mu_1=\{a_2c_2c_3\}$, $\mu_2=\{a_2a_3c_2\}$,
$\mu_3=\{b_3c_2c_3\}$ and $\mu_4=\{a_2a_3d_3\}$, whose complements are
$[a_3b_2b_3d_2d_3]$, $[b_2b_3c_3d_2d_3]$, $[a_2a_3b_2d_2d_3]$ and
$[b_2b_3c_2c_3d_2]$ in agreement with the first triangulation in
eq.~\eqref{eq:triP}.

Unfortunately, coherent triangulations of the facets that induce the
same triangulations on their common (maximal) intersections do not
automatically combine to \emph{coherent} star triangulations of the
polytope, and indeed only 720 of the 824 triangulations 
in~\autoref{tab:PMtorus} turn out to be coherent. The non-coherent ones
are easily isolated by observing that coherent triangulations (via
their height functions) induce coherent triangulations of the prisms
$[a_1a_2a_3b_1b_2b_2]$ and $[c_1c_2c_3d_1d_2d_3]$, which eliminates
the triangulations for which $\mZ^{ab}_3$ or $\mZ^{cd}_3$ is not
broken by the triangulation of the pure circuits. For the
triangulation types \PMplus\, and \PMminus\, this reduces the
multiplicity from $8^2$ to $6^2$. The only other affected types are
the ones in the middle column of~\autoref{tab:PMtorus}, which have
unbroken horizontal symmetry and for which the multiplicity is reduced
from $12\cdot 8$ to $12\cdot6$. This poses a problem for the eight
$\mZ_3$-symmetric triangulations, which are all non-coherent.
Coherence of the remaining 720 triangulations can be established by 
checking that their Mori cones are all strictly 
convex~\cite{Wisniewski:2002ab}.

What comes to our rescue is that, even if all projective ambient
spaces break the diagonal $\mZ_3$ permutation symmetry, it may be
preserved on $\Xb^*$ if the obstructing exceptional sets do not
overlap with the complete intersection.  In the present case these are
the blow-ups of the singularities coming from the pure circuits, i.e.
codimension two sets of the form $a_i\cdot b_j$ or $c_l\cdot d_m$,
where we use, for simplicity, the symbol of the vertex $\nub_j$ for
the corresponding divisor $D_j$. Recall from
eq.~\eqref{eq:nefpartial2} that $\Xb^*$ is given by the product $\bar
D^*_{0,1}\cdot \bar D^*_{0,2}$ of the divisors
\begin{equation}
  \bar D^*_{0,1}=a_1+a_2+a_3+b_1+b_2+b_3+e,
  \qquad
  \bar D^*_{0,2}=c_1+c_2+c_3+c_1+d_2+d_3+f
\end{equation}
defined by the nef partition. Taking into account the five linear
equivalences, we observe that
\begin{equation}
  \begin{gathered}
    a_1+b_1= a_2+b_2= a_3+b_3,
    \qquad
    c_1+d_1= c_2+d_2= c_3+d_3,
    \\
    b_1+b_2+b_3+e= d_1+d_2+d_3+f,
  \end{gathered}
\end{equation}
for divisor classes in the intersection ring. We first show that $e$ 
and $f$ do not intersect $\Xb^*$: In any maximal triangulation $e$ and 
$f$ belong only to the simplices 
\begin{equation}
  \big[\widehat{b_1b_2b_3}c_mc_le\big]
  ,
  \qquad
  \big[\widehat{d_1d_2d_3}a_ma_lf\big],	
  \label{eq:triEF}
\end{equation}
respectively, so that
\begin{equation}
  e\cdot a_i=e\cdot d_l=e\cdot f=0 
  \quad\Rightarrow\quad
  e\cdot \bar D^*_{0,1}=e\cdot(b_1+b_2+b_3+e)= 
  e\cdot(d_1+d_2+d_3+f)=0
\end{equation}
and similarly $f\cdot \bar D^*_{0,2}=0$. Putting everything together, we
conclude that
\begin{equation}
  a_1\cdot b_2\cdot \bar D^*_{0,1}= a_1\cdot b_2\cdot 3(a_3+b_3)=0
\end{equation}
because none of the facets, and hence no triangle in any of the
triangulations contains $\{a_1,b_2,a_3\}$ or $\{a_1,b_2,b_3\}$ as a
subset. Similarly $c_l\cdot d_m \cdot \bar D^*_{0,2}=0$ in the intersection
ring for $l\neq m$. Consequently, all exceptional sets arising from
triangulations of pure circuits do not intersect $\Xb^*$ and
hence do not obstruct the cyclic permutation symmetry $G_2^*$.
We will denote any of the remaining 36 coherent triangulations of 
type \PMplus\, and \PMminus\, by $T_{+}=T_{+}(\bar\nabla^*)$ and
$T_{-}=T_{-}(\bar\nabla^*)$, respectively.

The polytope $\nabla^*$ of the mirror $\Xt^*$ of the universal cover
has $39$ lattice points, with the same $12$ vertices as $\bar\nabla^*$
but living on the finer lattice $\bar M$. The $24$ additional lattice
points, see eq.~\eqref{eq:points}, are
\begin{align}
  a_{ij}&\;=\frac13(a_i+2a_j),&	
  b_{ij}&\;=\frac13(b_i+2b_j),
  \\
  c_{ij}&\;=\frac13(c_i+2c_j),&	
  d_{ij}&\;=\frac13(d_i+2d_j)
  ,
  \label{eq:abcdij}
\end{align}
where $i\neq j$. These additional points are all located on edges of
$\nabla^*$. It is natural to consider triangulations that are
refinements of the ones that we just discussed. Observing that the
additional points turn all simplices in eqns.~\eqref{eq:triP},
\eqref{eq:triM} and~\eqref{eq:triEF} into pyramids over a tetrahedron
with interior points on opposite edges it is easy to see that the
maximal triangulations are unique and multiply the number
$54=9\cdot4+6\cdot3$ of triangles in the original triangulations by a
factor of $9$. The resulting triangulations have been used to show that 
the divisors corresponding to the vertices $a_{ij}$ and $c_{ij}$ do not 
intersect $\tilde X^*$.

\section
[The Flop of $\mathbf{X^\ast}$]
[The Flop of X*]
{The Flop of $\mathbf{X^\ast}$}
\label{sec:X*flop}

In~\autoref{sec:BmodelXbmirror} we have taken into account only one of the 
triangulations $T_{+}(\bar\nabla^*)$. We can repeat the same calculation 
with one of the triangulations $T_{-}$. We denote the resulting Calabi-Yau 
manifold by $\Xb_{-}^*$. Skipping the details, we find that the generators 
of the Mori cone $\NE(\Xb_{-}^*)$ can be expressed in terms of those of 
$\NE(\Xb^*)$ in eq.~\eqref{eq:MoriCY} as
\begin{equation}
  \label{eq:MoriCY-}
  \begin{split}
    \lb_{-}^{(1)} =&\; \lb_{+}^{(1)} + 
    3\big(\lb_{+}^{(4)}+\lb_{+}^{(5)}+\lb_{+}^{(6)}+\lb_{+}^{(7)}\big)
    ,\\
    \lb_{-}^{(2)} =&\; \lb_{+}^{(2)} +
    3\big(\lb_{+}^{(3)}+\lb_{+}^{(4)}+\lb_{+}^{(5)}\big)
    ,\\ 
    \lb_{-}^{(3)} =&\; \lb_{+}^{(3)}+\lb_{+}^{(4)}
    ,\\ 
    \lb_{-}^{(4)} =&\; -\lb_{+}^{(4)}
    ,\\ 
    \lb_{-}^{(5)} =&\; -\lb_{+}^{(5)} -\lb_{+}^{(3)}-\lb_{+}^{(4)}
    -\lb_{+}^{(6)}-\lb_{+}^{(7)}
    , \\ 
    \lb_{-}^{(6)} =&\; \lb_{+}^{(6)}
    , \\
    \lb_{-}^{(7)} =&\; \lb_{+}^{(7)}
    . 
  \end{split}
\end{equation}
One can also express the dual basis of divisors $\Jb'_i$ on the flop
in terms of the dual basis $\Jb^*_i$ on $X$, see
eq.~\eqref{eq:KaehlerCY}. We find
\begin{equation}
  \label{eq:JXb-}
  \begin{gathered}
    \Jb'_1 = \Jb^*_1, \qquad 
    \Jb'_2 = \Jb^*_2, \qquad 
    \Jb'_5 = 3\Jb^*_1 + 3\Jb^*_2 - \Jb^*_5\\
    \Jb'_3 = 3\Jb^*_1 + \Jb^*_3 - \Jb^*_5, \qquad 
    \Jb'_4 = 3\Jb^*_1 + \Jb^*_3 - \Jb^*_4,\\
    \Jb'_6 = 3\Jb^*_2 - \Jb^*_5 + \Jb^*_6, \qquad 
    \Jb'_7 = 3\Jb^*_2 - \Jb^*_5 + \Jb^*_7.
  \end{gathered}
\end{equation}
The intersection ring is
\begin{equation}
  \label{eq:cubicformXb*-}
  \begin{aligned}
    \Jb_1^{\prime 2} \Jb_2'   &=1,  & 
    \Jb_1^{\prime 2} \Jb_3'   &=2,   & 
    \Jb_1^{\prime 2} \Jb_4'   &=1,  & 
    \Jb_1^{\prime 2} \Jb_5'   &=3, & 
    \Jb_1^{\prime 2} \Jb_6'   &=3,  \\
    \Jb_1^{\prime 2} \Jb_7'   &=3,  & 
    \Jb_1' \Jb_2^{\prime 2}   &=1, & 
    \Jb_1' \Jb_2' \Jb_3'      &=3, & 
    \Jb_1' \Jb_2' \Jb_4'      &=3, & 
    \Jb_1' \Jb_2' \Jb_5'      &=3, \\
    \Jb_1' \Jb_2' \Jb_6'      &=3, & 
    \Jb_1' \Jb_2' \Jb_7'      &=3, & 
    \Jb_1' \Jb_3^{\prime 2}   &=6, & 
    \Jb_1' \Jb_3' \Jb_4'      &=6, & 
    \Jb_1' \Jb_3' \Jb_5'      &=9, \\
    \Jb_1' \Jb_3' \Jb_6'      &=9, & 
    \Jb_1' \Jb_3' \Jb_7'      &=9, & 
    \Jb_1' \Jb_4^{\prime 2}   &=3, & 
    \Jb_1' \Jb_4' \Jb_5'      &=9, & 
    \Jb_1' \Jb_4' \Jb_6'      &=9, \\ 
    \Jb_1' \Jb_4' \Jb_7'      &=9, &
    \Jb_1' \Jb_5^{\prime 2}   &=9, & 
    \Jb_1' \Jb_5' \Jb_6'      &=9, & 
    \Jb_1' \Jb_5' \Jb_7'      &=9, & 
    \Jb_1' \Jb_6^{\prime 2}   &=9, \\
    \Jb_1' \Jb_6' \Jb_7'      &=9, & 
    \Jb_1' \Jb_7^{\prime 2}   &=9, & 
    \Jb_2^{\prime 2}\Jb_3'    &=3, & 
    \Jb_2^{\prime 2}\Jb_4'    &=3, & 
    \Jb_2^{\prime 2}\Jb_5'    &=3, \\
    \Jb_2^{\prime 2}\Jb_6'    &=2, & 
    \Jb_2^{\prime 2}\Jb_7'    &=1, & 
    \Jb_2' \Jb_3^{\prime 2}   &=9, & 
    \Jb_2' \Jb_3' \Jb_4'      &=9, & 
    \Jb_2' \Jb_3' \Jb_5'      &=9, \\ 
    \Jb_2' \Jb_3' \Jb_6'      &=9, & 
    \Jb_2' \Jb_3' \Jb_7'      &=9, & 
    \Jb_2' \Jb_4^{\prime 2}   &=9, & 
    \Jb_2' \Jb_4' \Jb_5'      &=9, & 
    \Jb_2' \Jb_4' \Jb_6'      &=9, \\
    \Jb_2' \Jb_4' \Jb_7'      &=9, & 
    \Jb_2' \Jb_5^{\prime 2}   &=9, & 
    \Jb_2' \Jb_5' \Jb_6'      &=9, & 
    \Jb_2' \Jb_5' \Jb_7'      &=9, & 
    \Jb_2' \Jb_6^{\prime 2}   &=6, \\
    \Jb_2' \Jb_6' \Jb_7'      &=6, & 
    \Jb_2' \Jb_7^{\prime 2}   &=3, & 
    \Jb_3^{\prime 3}          &=18, & 
    \Jb_3^{\prime 2} \Jb_4'   &=18, & 
    \Jb_3^{\prime 2} \Jb_5'   &=27, \\
    \Jb_3^{\prime 2} \Jb_6'   &=27, & 
    \Jb_3^{\prime 2} \Jb_7'   &=27, & 
    \Jb_3' \Jb_4^{\prime 2}   &=18, & 
    \Jb_3' \Jb_4' \Jb_5'      &=27, & 
    \Jb_3' \Jb_4' \Jb_6'      &=27, \\
    \Jb_3' \Jb_4' \Jb_7'      &=27, & 
    \Jb_3' \Jb_5^{\prime 2}   &=27, & 
    \Jb_3' \Jb_5' \Jb_6'      &=27, & 
    \Jb_3' \Jb_5' \Jb_7'      &=27, & 
    \Jb_3' \Jb_6^{\prime 2}   &=27, \\
    \Jb_3' \Jb_6' \Jb_7'      &=27, & 
    \Jb_3' \Jb_7^{\prime 2}   &=27, & 
    \Jb_4^{\prime 3}          &=9, & 
    \Jb_4^{\prime 2} \Jb_5'   &=27, & 
    \Jb_4^{\prime 2} \Jb_6'   &=27, \\ 
    \Jb_4^{\prime 2} \Jb_7'   &=27, & 
    \Jb_4' \Jb_5^{\prime 2}   &=27, & 
    \Jb_4' \Jb_5' \Jb_6'      &=27, &
    \Jb_4' \Jb_5' \Jb_7'      &=27, & 
    \Jb_4' \Jb_6^{\prime 2}   &=27, \\
    \Jb_4' \Jb_6' \Jb_7'      &=27, & 
    \Jb_4' \Jb_7^{\prime 2}   &=27, & 
    \Jb_5^{\prime 3}          &=27, & 
    \Jb_5^{\prime 2}\Jb_6'    &=27, & 
    \Jb_5^{\prime 2}\Jb_7'    &=27, \\
    \Jb_5 \Jb_6^{\prime 2}    &=27, & 
    \Jb_5' \Jb_6' \Jb_7'      &=27, & 
    \Jb_5 \Jb_7^{\prime 2}    &=27, & 
    \Jb_6^{\prime 3}          &=18, & 
    \Jb_6^{\prime 2} \Jb_7'   &=18, \\
    \Jb_6' \Jb_7^{\prime 2}   &=18, & 
    \Jb_7^{\prime 3}          &=9.
\end{aligned}
\end{equation}
The second Chern class is
\begin{equation}
  \label{eq:c2Xb*-}
  \begin{gathered}
    \ch_2\big(\Xb^\ast_{-}\big) \cdot \Jb'_1 = 12 ,\qquad    
    \ch_2\big(\Xb^\ast_{-}\big) \cdot \Jb'_5 = 18 ,\qquad    
    \ch_2\big(\Xb^\ast_{-}\big) \cdot \Jb'_2 = 12 ,
    \\
    \ch_2\big(\Xb^\ast_{-}\big) \cdot \Jb'_3 =
    \ch_2\big(\Xb^\ast_{-}\big) \cdot \Jb'_6 = 24 ,\qquad    
    \ch_2\big(\Xb^\ast_{-}\big) \cdot \Jb'_4 =
    \ch_2\big(\Xb^\ast_{-}\big) \cdot \Jb'_7 = 30
    .
  \end{gathered}
\end{equation}
We observe that both the intersection ring and the second Chern class 
cannot be brought into~(\ref{eq:cubicformXb}) and~(\ref{eq:c2Xbar*}) by
a linear transformation with integer coefficients, respectively. 
Hence, the second phase really is topologically distinct.

We denote the Fourier-transformed variables in the B-model 
prepotential~(\ref{eq:mirror}) by $q'_i,\,i=1,\dots,7$. With this 
notation, we obtain
\begin{equation}
  \label{eq:FBXb-}
  \begin{split}
    \FprepotentialB_{\Xb_{-},0}(q'_1,\dots,q'_7) 
    =&\;
    3\,q'_{{1}}+3\,q'_{{2}}+3\,q'_{{5}}  
    -{\frac {45}{8}}\,{q'_{{1}}}^{2}-{\frac {45}{8}}\,{q'_{{2}}}^{2}
    +\frac{3}{8}\,{q'_{{5}}}^{2}+3\,q'_{{3}}q'_{{5}}+3\,q'_{{5}}q'_{{6}} 
    \\ &\;
    + {\frac {244}{9}}\,{q'_{{1}}}^{3}+{\frac {244}{9}}\,{q'_{{2}}}^{3}
    +\frac{1}{9}\,{q'_{{5}}}^{3}+3\,q'_{{3}}q'_{{4}}q'_{{5}}
    +3\,q'_{{3}}q'_{{5}}q'_{{6}}+3\,q'_{{5}}q'_{{6}}q'_{{7}}
    \\ &\;
    -{\frac {12333}{64}}\,{q'_{{1}}}^{4}-{\frac {12333}{64}}\,{q'_{{2}}}^{4}
    +{\frac {3}{64}}\,{q'_{{5}}}^{4}+3\,q'_{{1}}{q'_{{4}}}^{3}
    +3\,q'_{{2}}{q'_{{7}}}^{3}
    \\ &\;
    +\frac{3}{8}\,{q'_{{5}}}^{2}{q'_{{6}}}^{2}
    +\frac{3}{8}\,{q'_{{3}}}^{2}{q'_{{5}}}^{2}
    -6\,q'_{{1}}q'_{{3}}q'_{{4}}q'_{{5}}-6\,q'_{{2}}q'_{{5}}q'_{{6}}q'_{{7}}
    \\ &\;
    +3\,q'_{{3}}q'_{{4}}q'_{{5}}q'_{{6}}+3\,q'_{{3}}q'_{{5}}q'_{{6}}q'_{{7}}
    +O(q'^5).
  \end{split}
\end{equation}
The instanton numbers on $\Xb^*_{-}$ are the expansion coefficients in
\begin{equation}
  \label{eq:FXb*-}
  \FprepotNP{\Xb^\ast_{-}}(q'_1,\dots,q'_7,1)
  =
  \sum_{n_1,\dots,n_7}  n^{\Xb^\ast_-}_{(n_1,n_2,n_3,n_4,n_5,n_6,n_7)} 
  \Li_3\Big( \prod_{i=1}^7 q'_i{}^{n_i}  \Big)
  . 
\end{equation}
Up to degree 4, they read
\begin{equation}
  \begin{aligned}
    n^{\Xb^*_-}_{{1,0,0,0,0,0,0}}
    & =3
    &n^{\Xb^*_-}_{{0,1,0,0,0,0,0}}
    & =3
    &n^{\Xb^*_-}_{{0,0,0,0,1,0,0}}
    & =3
    &n^{\Xb^*_-}_{{2,0,0,0,0,0,0}}
    & =-6
    \\ 
    n^{\Xb^*_-}_{{0,2,0,0,0,0,0}}
    & =-6
    &n^{\Xb^*_-}_{{3,0,0,0,0,0,0}}
    & =27
    &n^{\Xb^*_-}_{{0,3,0,0,0,0,0}}
    & =27
    &n^{\Xb^*_-}_{{4,0,0,0,0,0,0}}
    & =-192
    \\ 
    n^{\Xb^*_-}_{{0,4,0,0,0,0,0}}
    & =-192
    &n^{\Xb^*_-}_{{0,0,1,0,1,0,0}}
    & =3
    &n^{\Xb^*_-}_{{0,0,0,0,1,1,0}}
    & =3
    &n^{\Xb^*_-}_{{0,0,1,1,1,0,0}}
    & =3
    \\ 
    n^{\Xb^*_-}_{{0,0,1,0,1,1,0}}
    & =3
    &n^{\Xb^*_-}_{{0,0,0,0,1,1,1}}
    & =3
    &n^{\Xb^*_-}_{{1,0,0,3,0,0,0}}
    & =3
    &n^{\Xb^*_-}_{{0,1,0,0,0,0,3}}
    & =3
    \\
    n^{\Xb^*_-}_{{1,0,1,1,1,0,0}}
    & =-6
    &n^{\Xb^*_-}_{{0,1,0,0,1,1,1}}
    & =-6
    &n^{\Xb^*_-}_{{0,0,1,1,1,1,0}}
    & =3
    &n^{\Xb^*_-}_{{0,0,1,0,1,1,1}}
    & =3
  \end{aligned}  
\end{equation}
It is easy to check that the symmetry $G_2^*$ acts without fixed
points on $\Xb_{-}^*$ so that there are two phases of the
quotient $X^*$, too, with $h^{1,1}(X^*)=h^{1,2}(X^*)=3$ and
fundamental group $\pi_1(X^*) = \mZ_3\times\mZ_3$. They correspond
to the two classes of triangulations $T_{+}$ and $T_{-}$. The first
phase was studied in detail in~\autoref{sec:BmodelXbmirror} 
and~\ref{sec:InstXmirror}. 

We denote the Calabi-Yau manifold in the second phase by 
$X_{-}^* = \Xb^*_{-} / G_2^*$. From the linear equivalence relations 
eq.~(\ref{eq:lineq_nablab}) and the definition eq.~(\ref{eq:JXb-}) we
can compute the induced group action on $H^2(\Xb^\ast_{-},\Z)$ and find
\begin{equation}
  \label{eq:g2*action-}
  g_2^\ast
  \begin{pmatrix}
    \Jb'_1 \\
    \Jb'_2 \\
    \Jb'_3 \\
    \Jb'_4 \\
    \Jb'_5 \\
    \Jb'_6 \\
    \Jb'_7 \\
  \end{pmatrix}
  = 
  \begin{pmatrix}
    1 & 0 &  0 &  0 & 0 & 0 &  0 \\
    0 & 1 &  0 &  0 & 0 & 0 &  0 \\
    3 & 0 &  0 & -1 & 1 & 0 &  0 \\
    3 & 0 &  1 & -1 & 0 & 0 &  0 \\
    0 & 0 &  0 &  0 & 1 & 0 &  0 \\
    0 & 3 &  0 &  0 & 1 & 0 & -1 \\
    0 & 3 &  0 &  0 & 0 & 1 & -1 \\
  \end{pmatrix}
  \begin{pmatrix}
    \Jb'_1 \\
    \Jb'_2 \\
    \Jb'_3 \\
    \Jb'_4 \\
    \Jb'_5 \\
    \Jb'_6 \\
    \Jb'_7 \\
  \end{pmatrix}
  .
\end{equation}
In terms of the three invariant divisors
$J_1'=\Jb_5',J_2'=\Jb_1',J_3'=\Jb_2'$ the intersection ring and the
second Chern class of $X_{-}^*$ then are
\begin{equation}
  \begin{aligned}
    J_2^{\prime 2} J'_3 &= 1, & 
    J_1'   J_2^{\prime 2} &= 3, & 
    J_2'   J_3^{\prime 2} &= 1, &  
    J_1'   J'_2J'_3 &= 3, \\
    J_1^{\prime 2} J'_2 &= 9, & 
    J_1'   J_3^{\prime 2} &= 3, & 
    J_1^{\prime 2} J'_3 &= 9, &  
    J_1^{\prime 3} &= 27,\\
    \ch_2 J_1' & = 18, & 
    \ch_2 J_2' &= 12, & 
    \ch_2 J_3' &= 12.
  \end{aligned}
  \label{eq:ringX*-}
\end{equation}
Again, we observe that there is no linear basis transformation with 
integer coefficients that brings both the intersection ring and the 
second Chern class into~(\ref{eq:ringXb}). Hence, also the phase $X^*_{-}$ 
is topologically distinct from $X^*$.

To give a geometrical interpretation of what happens, we look at the 
induced action of $G_2^*$ on the toric Mori cone $\NE(\Xb^*_{-})$. 
Only the generators $\lb_{-}^{(1)}$, $\lb_{-}^{(2)}$ and $\lb_{-}^{(5)}$ in 
eq.~\eqref{eq:MoriCY-} are invariant. This is exactly as in the first
phase. Denoting the invariant generators 
$\lb_{\pm}^{(5)}, \lb_{\pm}^{(1)}, \lb_{\pm}^{(2)}$ by 
$l_{\pm}^{(1)}, l_{\pm}^{(2)}, l_{\pm}^{(3)}$, respectively,
we observe that phase $T_{-}$ is obtained from the phase $T_{+}$ as a flop 
by the curve corresponding to the generator $l_{+}^{(1)}$:
\begin{align}
  \label{eq:flop}
  l_{-}^{(1)} &=  -l_{+}^{(1)}, & 
  l_{-}^{(2)} &= l_{+}^{(2)} + 3l_{+}^{(1)}, & 
  l_{-}^{(3)} &= l_{+}^{(3)} + 3l_{+}^{(1)}
  .
\end{align}
If we use the realization of $X$ in terms of the fiber product of two $dP_9$
surfaces, the above result means that the base $\mP^1$ of $X$ has been 
flopped. 

Furthermore, having computed the $G_2^\ast$-action in 
eq.~(\ref{eq:g2*action-}), we determine the descent equation 
for the prepotential to be
\begin{equation}
  \label{eq:Xb*-2X*-}
  \FprepotNP{X^\ast_{-}}\big(p',q',r',b_1',b_2') 
  =
  \frac{1}{|G_2^\ast|}    
  \FprepotNP{\Xb^\ast_{-}}\big(
    p' ,
    q' ,
    b'_2 ,
    b'_2 ,
    r' ,
    b'_2{}^{2} ,
    b'_2{}^{2} ,
    b'_1
  \big)
  .
\end{equation}
The corresponding instanton numbers
\begin{equation}
  \FprepotNP{X_{-}^\ast}(p',q',r',1,b_2')
  = 
  \sum_{n_1,n_2,n_3,m_2}
  n^{X^\ast_{-}}_{(n_1,n_2,n_3,m_2)}
  \Li_3\big( p'{}^{n_1} q'{}^{n_2} r'{}^{n_3} b_2'{}^{m_2} \big)
\end{equation}
are listed in \autoref{tab:n1n2n3m2mirror-}.
\begin{table}[tbp]
  \centering
  \renewcommand{\arraystretch}{1.2}
  \begin{tabular}{c|ccc|c}
    $(n_1,n_2,n_3)$  & 
    $n^{X_{-}^\ast}_{(n_1,n_2,n_3,0)}$ & 
    $n^{X_{-}^\ast}_{(n_1,n_2,n_3,1)}$ & 
    $n^{X_{-}^\ast}_{(n_1,n_2,n_3,2)}$ & 
    $\sum_{m_2=0}^2 n^{X_{-}^\ast}_{(n_1,n_2,n_3)}$ \\
    \hline
    $(1,0,0)$    & $     3$ & $     0$ & $    0$ & $     3$ \\
    $(2,0,0)$    & $    -6$ & $     0$ & $    0$ & $    -6$ \\
    $(3,0,0)$    & $    18$ & $     0$ & $    0$ & $    18$ \\
    $(0,1,0)$    & $     3$ & $     3$ & $    3$ & $     9$ \\
    $(1,1,0)$    & $    -6$ & $    -6$ & $   -6$ & $   -18$ \\
    $(1,1,1)$    & $    12$ & $    12$ & $   12$ & $    36$ \\
    $(2,1,0)$    & $    15$ & $    15$ & $   15$ & $    45$ \\
    $(1,2,0)$    & $    12$ & $    12$ & $   12$ & $    36$ \\
  \end{tabular}
  \caption{Instanton numbers $n^{X_{-}^\ast}_{(n_1,n_2,n_3,m_2)}$ computed
    by toric mirror symmetry. They are invariant under the exchange
    $n_2\leftrightarrow n_3$, so we only display 
    them for $n_2\leq n_3$.}
  \label{tab:n1n2n3m2mirror-}
\end{table}


%% file: Bmodel.bbl
\providecommand{\href}[2]{#2}\begingroup\raggedright\begin{thebibliography}{10}

\bibitem{Braun:2007xh}
V.~Braun, M.~Kreuzer, B.~A. Ovrut, and E.~Scheidegger, ``Worldsheet instantons
  and torsion curves. Part A: Direct computation,''
\href{http://arXiv.org/abs/hep-th/0703182}{{\tt hep-th/0703182}}.

\bibitem{Candelas:1990rm}
P.~Candelas, X.~C. De~La~Ossa, P.~S. Green, and L.~Parkes, ``A pair of
  Calabi-Yau manifolds as an exactly soluble superconformal theory,'' {\em
  Nucl. Phys.} {\bf B359} (1991)
21--74.

\bibitem{MR923487}
C.~Schoen, ``On fiber products of rational elliptic surfaces with section,''
  {\em Math. Z.} {\bf 197} (1988), no.~2, 177--199.

\bibitem{Braun:2004xv}
V.~Braun, B.~A. Ovrut, T.~Pantev, and R.~Reinbacher, ``Elliptic Calabi-Yau
  threefolds with Z(3) x Z(3) Wilson lines,'' {\em JHEP} {\bf 12} (2004) 062,
\href{http://arXiv.org/abs/hep-th/0410055}{{\tt hep-th/0410055}}.

\bibitem{Aspinwall:1995rb}
P.~S. Aspinwall, D.~R. Morrison, and M.~Gross, ``Stable singularities in string
  theory,'' {\em Commun. Math. Phys.} {\bf 178} (1996) 115--134,
\href{http://arXiv.org/abs/hep-th/9503208}{{\tt hep-th/9503208}}.

\bibitem{Batyrev:2005jc}
V.~Batyrev and M.~Kreuzer, ``Integral Cohomology and Mirror Symmetry for
  Calabi-Yau 3-folds,''
\href{http://arXiv.org/abs/math.AG/0505432}{{\tt math.AG/0505432}}.

\bibitem{gross-2005}
M.~Gross and S.~Pavanelli, ``A Calabi-Yau threefold with Brauer group
  (Z/8Z)${}^2$,'' \href{http://arXiv.org/abs/math.AG/0512182}{{\tt
  math.AG/0512182}}.

\bibitem{Ferrara:1995yx}
S.~Ferrara, J.~A. Harvey, A.~Strominger, and C.~Vafa, ``Second quantized mirror
  symmetry,'' {\em Phys. Lett.} {\bf B361} (1995) 59--65,
\href{http://arXiv.org/abs/hep-th/9505162}{{\tt hep-th/9505162}}.

\bibitem{Aspinwall:1995mh}
P.~S. Aspinwall, ``An N=2 Dual Pair and a Phase Transition,'' {\em Nucl. Phys.}
  {\bf B460} (1996) 57--76,
\href{http://arXiv.org/abs/hep-th/9510142}{{\tt hep-th/9510142}}.

\bibitem{Lima:2001nh}
E.~Lima, B.~A. Ovrut, and J.~Park, ``Five-brane superpotentials in heterotic
  M-theory,'' {\em Nucl. Phys.} {\bf B626} (2002) 113--164,
\href{http://arXiv.org/abs/hep-th/0102046}{{\tt hep-th/0102046}}.

\bibitem{Lima:2001jc}
E.~Lima, B.~A. Ovrut, J.~Park, and R.~Reinbacher, ``Non-perturbative
  superpotential from membrane instantons in heterotic M-theory,'' {\em Nucl.
  Phys.} {\bf B614} (2001) 117--170,
\href{http://arXiv.org/abs/hep-th/0101049}{{\tt hep-th/0101049}}.

\bibitem{Buchbinder:2002ic}
E.~I. Buchbinder, R.~Donagi, and B.~A. Ovrut, ``Superpotentials for vector
  bundle moduli,'' {\em Nucl. Phys.} {\bf B653} (2003) 400--420,
\href{http://arXiv.org/abs/hep-th/0205190}{{\tt hep-th/0205190}}.

\bibitem{Buchbinder:2002pr}
E.~I. Buchbinder, R.~Donagi, and B.~A. Ovrut, ``Vector bundle moduli
  superpotentials in heterotic superstrings and M-theory,'' {\em JHEP} {\bf 07}
  (2002) 066,
\href{http://arXiv.org/abs/hep-th/0206203}{{\tt hep-th/0206203}}.

\bibitem{Buchbinder:2002wz}
E.~Buchbinder and B.~A. Ovrut, ``Vector bundle moduli,'' {\em Russ. Phys. J.}
  {\bf 45} (2002)
662--669.

\bibitem{Buchbinder:2003pi}
E.~I. Buchbinder and B.~A. Ovrut, ``Vacuum stability in heterotic M-theory,''
  {\em Phys. Rev.} {\bf D69} (2004) 086010,
\href{http://arXiv.org/abs/hep-th/0310112}{{\tt hep-th/0310112}}.

\bibitem{Buchbinder:2002ji}
E.~Buchbinder, R.~Donagi, and B.~A. Ovrut, ``Vector bundle moduli and small
  instanton transitions,'' {\em JHEP} {\bf 06} (2002) 054,
\href{http://arXiv.org/abs/hep-th/0202084}{{\tt hep-th/0202084}}.

\bibitem{Braun:2005nv}
V.~Braun, Y.-H. He, B.~A. Ovrut, and T.~Pantev, ``The exact MSSM spectrum from
  string theory,'' {\em JHEP} {\bf 05} (2006) 043,
\href{http://arXiv.org/abs/hep-th/0512177}{{\tt hep-th/0512177}}.

\bibitem{Braun:2005ux}
V.~Braun, Y.-H. He, B.~A. Ovrut, and T.~Pantev, ``A heterotic standard model,''
  {\em Phys. Lett.} {\bf B618} (2005) 252--258,
\href{http://arXiv.org/abs/hep-th/0501070}{{\tt hep-th/0501070}}.

\bibitem{Braun:2005bw}
V.~Braun, Y.-H. He, B.~A. Ovrut, and T.~Pantev, ``A standard model from the
  E(8) x E(8) heterotic superstring,'' {\em JHEP} {\bf 06} (2005) 039,
\href{http://arXiv.org/abs/hep-th/0502155}{{\tt hep-th/0502155}}.

\bibitem{Braun:2005zv}
V.~Braun, Y.-H. He, B.~A. Ovrut, and T.~Pantev, ``Vector bundle extensions,
  sheaf cohomology, and the heterotic standard model,'' {\em Adv. Theor. Math.
  Phys.} {\bf 10} (2006) 4,
\href{http://arXiv.org/abs/hep-th/0505041}{{\tt hep-th/0505041}}.

\bibitem{Braun:2005fk}
V.~Braun, Y.-H. He, B.~A. Ovrut, and T.~Pantev, ``Heterotic standard model
  moduli,'' {\em JHEP} {\bf 01} (2006) 025,
\href{http://arXiv.org/abs/hep-th/0509051}{{\tt hep-th/0509051}}.

\bibitem{Braun:2005xp}
V.~Braun, Y.-H. He, B.~A. Ovrut, and T.~Pantev, ``Moduli dependent mu-terms in
  a heterotic standard model,''
\href{http://arXiv.org/abs/hep-th/0510142}{{\tt hep-th/0510142}}.

\bibitem{Braun:2006me}
V.~Braun, Y.-H. He, and B.~A. Ovrut, ``Yukawa couplings in heterotic standard
  models,'' {\em JHEP} {\bf 04} (2006) 019,
\href{http://arXiv.org/abs/hep-th/0601204}{{\tt hep-th/0601204}}.

\bibitem{Braun:2006ae}
V.~Braun, Y.-H. He, and B.~A. Ovrut, ``Stability of the minimal heterotic
  standard model bundle,'' {\em JHEP} {\bf 06} (2006) 032,
\href{http://arXiv.org/abs/hep-th/0602073}{{\tt hep-th/0602073}}.

\bibitem{Braun:2006th}
V.~Braun and B.~A. Ovrut, ``Stabilizing moduli with a positive cosmological
  constant in heterotic M-theory,'' {\em JHEP} {\bf 07} (2006) 035,
\href{http://arXiv.org/abs/hep-th/0603088}{{\tt hep-th/0603088}}.

\bibitem{Aspinwall:1994uj}
P.~S. Aspinwall and D.~R. Morrison, ``Chiral rings do not suffice: N=(2,2)
  theories with nonzero fundamental group,'' {\em Phys. Lett.} {\bf B334}
  (1994) 79--86,
\href{http://arXiv.org/abs/hep-th/9406032}{{\tt hep-th/9406032}}.

\bibitem{Braun:2007tp}
V.~Braun, M.~Kreuzer, B.~A. Ovrut, and E.~Scheidegger, ``Worldsheet instantons,
  torsion curves, and non-perturbative superpotentials,''
\href{http://arXiv.org/abs/hep-th/0703134}{{\tt hep-th/0703134}}.

\bibitem{Hosono:1997hp}
S.~Hosono, M.~Saito, and J.~Stienstra, ``On the mirror symmetry conjecture for
  Schoen's Calabi-Yau 3-folds,''
  \href{http://arXiv.org/abs/alg-geom/9709027}{{\tt alg-geom/9709027}}. Given
  at Taniguchi Symposium on Integrable Systems and Algebraic Geometry, Kyoto,
  Japan, 7-11 Jul 1997.

\bibitem{Lust:2006zh}
D.~Lust, S.~Reffert, E.~Scheidegger, and S.~Stieberger, ``Resolved toroidal
  orbifolds and their orientifolds,''
\href{http://arXiv.org/abs/hep-th/0609014}{{\tt hep-th/0609014}}.

\bibitem{Kreuzer:2006ax}
M.~Kreuzer, ``Toric geometry and Calabi-Yau compactifications,''
\href{http://arXiv.org/abs/hep-th/0612307}{{\tt hep-th/0612307}}.

\bibitem{Klemm:2004km}
A.~Klemm, M.~Kreuzer, E.~Riegler, and E.~Scheidegger, ``Topological string
  amplitudes, complete intersection Calabi-Yau spaces and threshold
  corrections,'' {\em JHEP} {\bf 05} (2005) 023,
\href{http://arXiv.org/abs/hep-th/0410018}{{\tt hep-th/0410018}}.

\bibitem{Batyrev:1994hm}
V.~V. Batyrev, ``Dual polyhedra and mirror symmetry for Calabi-Yau
  hypersurfaces in toric varieties,'' {\em J. Alg. Geom.} {\bf 3} (1994)
493--545.

\bibitem{Borisov:1993ab}
L.~A. Borisov, ``Towards the Mirror Symmetry for Calabi-Yau Complete
  Intersections in Gorenstein Toric Fano Varieties,''
  \href{http://arXiv.org/abs/arXiv:alg-geom/9310001}{{\tt
  arXiv:alg-geom/9310001}}.

\bibitem{Batyrev:1994pg}
V.~V. Batyrev and L.~A. Borisov, ``On {C}alabi-{Y}au complete intersections in
  toric varieties,'' in {\em Higher-dimensional complex varieties (Trento,
  1994)}, pp.~39--65.
\newblock de Gruyter, Berlin, 1996.
\newblock
\href{http://arXiv.org/abs/alg-geom/9412017}{{\tt alg-geom/9412017}}.
\newblock

\bibitem{Kreuzer:2002uu}
M.~Kreuzer and H.~Skarke, ``PALP: A Package for analyzing lattice polytopes
  with applications to toric geometry,'' {\em Comput. Phys. Commun.} {\bf 157}
  (2004) 87--106,
\href{http://arXiv.org/abs/math.na/0204356}{{\tt math.na/0204356}}.

\bibitem{Batyrev:1995ca}
V.~V. Batyrev and L.~A. Borisov, ``Mirror duality and string-theoretic Hodge
  numbers,'' {\em Invent. Math.} {\bf 126} (1996) 183,
\href{http://arXiv.org/abs/alg-geom/9509009}{{\tt alg-geom/9509009}}.

\bibitem{Batyrev:1991ab}
V.~V. Batyrev, ``On the classification of smooth projective toric varieties,''
  {\em Tohoku Math. J., II. Ser.} {\bf 43} (1991) 569 -- 585.

\bibitem{Stienstra:1998ab}
J.~Stienstra, ``Resonant Hypergeometric Systems and Mirror Symmetry,'' in {\em
  Integrable systems and algebraic geometry}, M.~H.~S. et~al., ed., pp.~412 --
  452.
\newblock World Scientific, Singapore, 1998.
\newblock \href{http://arXiv.org/abs/alg-geom/9711002}{{\tt alg-geom/9711002}}.

\bibitem{Danilov:1978ab}
V.~Danilov, ``Geometry of Toric Varieties,'' {\em Russ. Math. Surv.} {\bf 33}
  (1978) 97 -- 154.

\bibitem{Batyrev:1994rs}
V.~V. Batyrev, ``Quantum {C}ohomology {R}ings of {T}oric {Manifolds},'' {\em
  Ast\'erisque} (1993), no.~218, 9--34,
  \href{http://arXiv.org/abs/alg-geom/9310004}{{\tt alg-geom/9310004}}.

\bibitem{Wall:1966ab}
C.~T.~C. Wall, ``Classification Problems in Differential Topology. V: On
  Certain 6- manifolds,'' {\em Invent. Math.} {\bf 1} (1966) 355 -- 374.
  Corrigendum. Ibid. 2, 306 (1967).

\bibitem{Berglund:1995gd}
P.~Berglund, S.~H. Katz, and A.~Klemm, ``Mirror symmetry and the moduli space
  for generic hypersurfaces in toric varieties,'' {\em Nucl. Phys.} {\bf B456}
  (1995) 153--204,
\href{http://arXiv.org/abs/hep-th/9506091}{{\tt hep-th/9506091}}.

\bibitem{Cox:1999ab}
D.~A. Cox and S.~Katz, {\em Mirror symmetry and algebraic geometry}, vol.~68 of
  {\em Mathematical Surveys and Monographs}.
\newblock American Mathematical Society, Providence, RI, 1999.

\bibitem{Givental:1996ab}
A.~B. Givental, ``Equivariant {G}romov-{W}itten invariants,'' {\em Internat.
  Math. Res. Notices} (1996), no.~13, 613--663,
  \href{http://arXiv.org/abs/alg-geom/9603021}{{\tt alg-geom/9603021}}.

\bibitem{Givental:1998ab}
A.~B. Givental, ``A mirror theorem for toric complete intersections,'' in {\em
  Topological field theory, primitive forms and related topics (Kyoto, 1996)},
  vol.~160 of {\em Progr. Math.}, pp.~141--175.
\newblock Birkh\"auser Boston, Boston, MA, 1998.
\newblock \href{http://arXiv.org/abs/alg-geom/9701016}{{\tt alg-geom/9701016}}.

\bibitem{Hosono:1994ax}
S.~Hosono, A.~Klemm, S.~Theisen, and S.-T. Yau, ``Mirror symmetry, mirror map
  and applications to complete intersection Calabi-Yau spaces,'' {\em Nucl.
  Phys.} {\bf B433} (1995) 501--554,
\href{http://arXiv.org/abs/hep-th/9406055}{{\tt hep-th/9406055}}.

\bibitem{Cox:1993fz}
D.~A. Cox, ``The Homogeneous Coordinate Ring of a Toric Variety,'' {\em J.
  Algebr. Geom.} {\bf 4} (1995) 17 -- 50,
  \href{http://arXiv.org/abs/alg-geom/9210008}{{\tt alg-geom/9210008}}.

\bibitem{Kreuzer:1994uc}
M.~Kreuzer, ``The Mirror map for invertible LG models,'' {\em Phys. Lett.} {\bf
  B328} (1994) 312--318,
\href{http://arXiv.org/abs/hep-th/9402114}{{\tt hep-th/9402114}}.

\bibitem{Aspinwall:1990xe}
P.~S. Aspinwall, C.~A. Lutken, and G.~G. Ross, ``Construction and couplings of
  mirror manifolds,'' {\em Phys. Lett.} {\bf B241} (1990)
373--380.

\bibitem{MR1416344}
M.~Gross, ``The deformation space of {C}alabi-{Y}au {$n$}-folds with canonical
  singularities can be obstructed,'' in {\em Mirror symmetry, II}, vol.~1 of
  {\em AMS/IP Stud. Adv. Math.}, pp.~401--411.
\newblock Amer. Math. Soc., Providence, RI, 1997.

\bibitem{MR1390655}
Y.~Ruan, ``Topological sigma model and {D}onaldson-type invariants in {G}romov
  theory,'' {\em Duke Math. J.} {\bf 83} (1996), no.~2, 461--500.

\bibitem{Huang:2006hq}
M.-x. Huang, A.~Klemm, and S.~Quackenbush, ``Topological string theory on
  compact Calabi-Yau: Modularity and boundary conditions,''
\href{http://arXiv.org/abs/hep-th/0612125}{{\tt hep-th/0612125}}.

\bibitem{Grimm:2007tm}
T.~W. Grimm, A.~Klemm, M.~Marino, and M.~Weiss, ``Direct Integration of the
  Topological String,''
\href{http://arXiv.org/abs/hep-th/0702187}{{\tt hep-th/0702187}}.

\bibitem{Kreuzer:2000xy}
M.~Kreuzer and H.~Skarke, ``Complete classification of reflexive polyhedra in
  four dimensions,'' {\em Adv. Theor. Math. Phys.} {\bf 4} (2002) 1209--1230,
\href{http://arXiv.org/abs/hep-th/0002240}{{\tt hep-th/0002240}}.

\bibitem{Hosono:1999qc}
S.~Hosono, M.~H. Saito, and A.~Takahashi, ``Holomorphic anomaly equation and
  BPS state counting of rational elliptic surface,'' {\em Adv. Theor. Math.
  Phys.} {\bf 3} (1999) 177--208,
\href{http://arXiv.org/abs/hep-th/9901151}{{\tt hep-th/9901151}}.

\bibitem{GPS05}
G.-M. Greuel, G.~Pfister, and H.~Sch\"onemann, ``{\sc Singular} 3.0,'' a
  computer algebra system for polynomial computations, Centre for Computer
  Algebra, University of Kaiserslautern, 2005.
\newblock {\tt http://www.singular.uni-kl.de}.

\bibitem{MR1074022}
L.~J. Billera, P.~Filliman, and B.~Sturmfels, ``Constructions and complexity of
  secondary polytopes,'' {\em Adv. Math.} {\bf 83} (1990), no.~2, 155--179.

\bibitem{MR1264417}
I.~M. Gel{$'$}fand, M.~M. Kapranov, and A.~V. Zelevinsky, {\em Discriminants,
  resultants, and multidimensional determinants}.
\newblock Mathematics: Theory \& Applications. Birkh\"auser Boston Inc.,
  Boston, MA, 1994.

\bibitem{Wisniewski:2002ab}
J.~A. Wi\'sniewski, ``Toric Mori theory and Fano manifolds,'' {\em S\'eminaires
  \& Congr\`es} {\bf 6} (2002) 249.

\end{thebibliography}\endgroup
